\documentclass[aps,prl,reprint,superscriptaddress,longbibliography,floatfix]{revtex4-2}

\usepackage[T1]{fontenc}
\usepackage[utf8]{inputenc}
\usepackage{lmodern}
\usepackage{microtype}
\usepackage{amsmath,amssymb,bm,mathtools}
\usepackage{graphicx}
\usepackage{tikz}
\usetikzlibrary{arrows.meta,calc}
\definecolor{unmeasured}{RGB}{52,101,164}
\definecolor{measured}{RGB}{204,80,62}
\usepackage[colorlinks=true,citecolor=blue,urlcolor=blue,linkcolor=blue]{hyperref}

\newcommand{\ii}{\mathrm{i}}
\newcommand{\ee}{\mathrm{e}}
\newcommand{\dd}{\mathrm{d}}
\newcommand{\bbT}{\mathbb T}
\newcommand{\bbR}{\mathbb R}
\newcommand{\bbZ}{\mathbb Z}
\newcommand{\cE}{\mathcal E}
\newcommand{\cD}{\mathcal D}
\newcommand{\cR}{\mathcal R}
\newcommand{\Tr}{\operatorname{Tr}}
\newcommand{\doilink}[1]{\href{https://doi.org/#1}{doi:#1}}
\newcommand{\arxiv}[1]{\href{https://arxiv.org/abs/#1}{arXiv:#1}}

\begin{document}

\title{Quantum Snapshots Reveal a Compact Conformal Boundary Mode}

\author{M.~A.~Rajabpour}
\affiliation{Instituto de Física, Universidade Federal Fluminense, Av.~Gal.~Milton Tavares de Souza s/n, Gragoatá, 24210-346, Niterói, RJ, Brazil}

\date{\today}

\begin{abstract}
A projective measurement of a many-body state produces a microscopic snapshot, usually viewed as random classical data.  We show that partial occupation snapshots of the critical XX chain contain a universal angle with a precise conformal meaning.  Dividing the ring into two measured and two unmeasured arcs, we assign geometry-dependent conformal side weights to the observed occupations and obtain a compact variable $\delta_L$.  At every finite size, $\delta_L$ is fixed by the measured sites alone; the particular complete-configuration lift $X_L$ used in the proof additionally depends on unobserved particles.  This angle is an exact microscopic compact coordinate whose scaling-limit law is that of the relative Dirichlet phase of the associated conformal quadrilateral---the boundary coordinate conjugate to charge in continuum post-measurement descriptions.  Exact free-fermion determinants yield all of its Fourier moments.  We prove that the lift becomes Gaussian with variance $2h(\zeta)$, where $h(\zeta)$ is the rectangle modulus, and hence
$\langle e^{\ii q\delta_L}\rangle\to e^{-h(\zeta)q^2}$.
Thus raw quantum snapshots realize the heat kernel on a circle and provide an outcome-level microscopic foundation for the compact zero-mode sector of Born averages over fluctuating conformal boundary conditions.
\end{abstract}

\maketitle

Site-resolved measurements have changed how quantum many-body states are accessed.  Quantum-gas microscopes and quantum processors now return large ensembles of binary or spin-resolved configurations rather than only a few averaged correlators~\cite{Bakr2009,Sherson2010,Cheuk2015,Parsons2015,Brydges2019,Noel2022}.  Each run looks like a noisy microscopic string, but the distribution of strings is the Born distribution of the many-body wave function.  This raises a basic question with both experimental and conceptual force: \emph{can a universal field-theory coordinate be decoded directly from individual snapshots?}  At a critical point the long-distance theory is geometric and conformal, whereas a snapshot is defined in a local lattice basis.  A positive answer would make universality visible without reconstructing a density matrix, an entanglement spectrum, or even a conventional observable chosen in advance.

Measurements are not passive probes: they condition the state and can redistribute quantum information between the unmeasured degrees of freedom.  This principle underlies localizable entanglement~\cite{Verstraete2004,Popp2005}, measurement-induced transitions~\cite{Skinner2019,Li2019,Fisher2023} and their experimental realizations~\cite{Noel2022,GoogleAI2023,Koh2023}, measurement-altered and monitored critical states~\cite{Garratt2023,Murciano2023,Yang2023,Minoguchi2022,Buchhold2021,PatilLudwig2024,KhannaMurcianoVasseur2026}, and measurement-prepared long-range entanglement~\cite{Lu2022,Tantivasadakarn2024}.  For a ground state, one may either force a special outcome or average a post-measurement quantity over all outcomes with their exact Born weights.  The latter distinction is central to measurement-induced entanglement and information~\cite{Lin2023,Cheng2024,McGinley2025}.  Within boundary CFT~\cite{Cardy1984,Cardy1989}, recent results for Tomonaga--Luttinger liquids show that this Born average is naturally organized as an average over conformal Dirichlet boundary conditions of a compact boson~\cite{KhannaVasseur2026,KhannaVasseurStats2026}.

These developments expose a sharp missing link.  Continuum calculations introduce a fluctuating boundary phase~\cite{KhannaVasseur2026,KhannaVasseurStats2026}, but a generic microscopic measurement record is a jagged charge pattern that need not resemble a conformal boundary condition pointwise.  Which function of the observed bits is the compact phase?  Is it fixed by a partial record when the unmeasured regions contain an unknown number of particles?  And does its Born distribution really approach the CFT weight, rather than merely reproducing an averaged entropy?  Earlier work treated selected measurement outcomes through slit and conformal-boundary geometries~\cite{Rajabpour2015,Rajabpour2016,NajafiRajabpour2016,Hoshino2025} and analyzed fixed-configuration formation probabilities through slit/boundary free energies and Casimir interactions~\cite{Stephan2014,RajabpourFormation2015}.  Those results begin with a boundary condition or a distinguished outcome.  Here the boundary coordinate itself is random, and its microscopic decoder and probability law are the objects to be found.

The relative compact phase is also physically more than a formal label.  In a Dirac or compact-boson CFT it is a twist conjugate to the conserved charge.  Entanglement Hamiltonians translate conformal geometry into local inverse-temperature profiles~\cite{CardyTonni2016,DalmonteEHReview2022,EislerPeschel2017,EislerTonniPeschel2019,EislerTonniPeschel2022}, while BCFT decompositions organize entanglement spectra into boundary-condition and charge sectors~\cite{DiGiulio2023}.  They can now be reconstructed experimentally by entanglement-Hamiltonian tomography~\cite{Kokail2021}.  A record-level construction of the relative phase would therefore connect three levels that are presently separate: microscopic snapshots, the Born-weighted boundary ensemble~\cite{KhannaVasseur2026,KhannaVasseurStats2026}, and the charge sector of the post-measurement reduced state.  Existing entanglement-Hamiltonian results establish the continuum and lattice structures of the reduced state, while recent work directly connects partial measurements to the entanglement-Hamiltonian charge sector~\cite{EislerTonni2026}; these results do not provide the record-level decoder or its Born law established here.    This is experimentally attractive because, once the geometry is chosen, the decoder is fixed and requires only classical post-processing of the observed bits.

We establish this microscopic bridge exactly in the half-filled critical XX chain.  Four consecutive arcs $A,C_1,B,C_2$ form a conformal quadrilateral, with occupations measured only on $C=C_1\cup C_2$.  The conformal map to a rectangle supplies a conformal boundary coordinate $f$ that is constant on the unmeasured arcs and winds by one compact period along each measured arc.  From every partial record $m_C$ we construct an angle $\delta_L(m_C)$.  Our first result is finite-size and purely algebraic: although a real lift contains the unknown particle number in $B$, its class modulo $2\pi$ is determined exactly by $m_C$.  Thus all physical characters $e^{\ii q\delta_L}$, $q\in\mathbb Z$, are accessible from incomplete data.

The second result is an exact determinant bridge.  The XX Born measure is a projection determinantal process, so the Fourier moments of $\delta_L$ can be written equivalently as determinants on the full lattice, on the occupied one-particle subspace, or only on the measured sites.  Consecutive occupied momenta turn the smallest representation into a discrete Toeplitz determinant.  We control its thermodynamic limit despite the square-root corner singularities of $f$, proving that the associated real statistic converges to a Gaussian with quadratic form $\sum_{m\ge1}m|f_m|^2$.  Finally, conformal invariance identifies this abstract Fourier energy with the Dirichlet energy of the vertical rectangle coordinate and evaluates it exactly as the rectangle width $h(\zeta)$.  Therefore
\begin{equation}
 \lim_{L\to\infty}\mathbb E\!\left[e^{\ii q\delta_L}\right]
 =e^{-h(\zeta)q^2},\qquad q\in\mathbb Z,
\label{eq:headline}
\end{equation}
and the compact variable converges to a wrapped Gaussian, equivalently the heat kernel on a circle~\cite{KhannaVasseur2026,KhannaVasseurStats2026}.  No replica construction and no entanglement observable enter the proof: the universal boundary mode is already present in the classical probability distribution of the snapshots.

The result is deliberately narrower than assigning a smooth boundary profile to every microscopic outcome.  We isolate one collective compact coordinate---whose scaling-limit law is that of the relative Dirichlet zero mode---for which an exact record-only theorem is possible.  This distinction clarifies the relation to existing CFT: the continuum theory predicts why compact boundary data should matter; our lattice construction identifies which data are carried by each record and derives their full Born law.  It also places the problem between two familiar limits.  A sharp particle-counting window has jump singularities, logarithmically growing fluctuations, and Fisher--Hartwig branches~\cite{AbanovIvanov2011,IvanovLevkivskyi2016}.  Our conformal window is continuous but has a square-root divergent derivative at four corners, producing an order-one variance and a Szeg\H{o}-type Gaussian lift while retaining compact charge quantization.  The combination of exact partial-record measurability, determinantal asymptotics, and conformal geometry is the central result of this Letter.

\begin{figure}[t]
\centering
\begin{tikzpicture}[
  x=1cm,y=1cm,>=Latex,
  every node/.style={font=\footnotesize},
  arclabel/.style={fill=white,inner sep=1.1pt},
  sidelabel/.style={fill=white,inner sep=1.0pt}
]
  \node[font=\bfseries\footnotesize] at (-3.22,2.53) {(a)};
  \coordinate (O) at (-1.98,1.38);
  \def\R{1.08}
  \draw[line width=0.45pt,gray!55] (O) circle (\R);
  \draw[unmeasured,line width=2.15pt]
    ($(O)+(25:\R)$) arc[start angle=25,end angle=105,radius=\R];
  \draw[measured,line width=2.15pt,densely dashed]
    ($(O)+(105:\R)$) arc[start angle=105,end angle=205,radius=\R];
  \draw[unmeasured,line width=2.15pt]
    ($(O)+(205:\R)$) arc[start angle=205,end angle=285,radius=\R];
  \draw[measured,line width=2.15pt,densely dashed]
    ($(O)+(285:\R)$) arc[start angle=285,end angle=385,radius=\R];
  \foreach \ang in {25,105,205,285}
    \fill ($(O)+(\ang:\R)$) circle (1.2pt);

  \node[arclabel] at ($(O)+(65:1.39)$) {$A$};
  \node[arclabel] at ($(O)+(155:1.41)$) {$C_1$};
  \node[arclabel] at ($(O)+(245:1.39)$) {$B$};
  \node[arclabel] at ($(O)+(335:1.43)$) {$C_2$};

  \draw[->,line width=0.75pt] (-0.42,1.62)--(0.48,1.62)
    node[midway,above=2pt,font=\scriptsize] {$W_{\boldsymbol\theta}$};

  \node[font=\bfseries\footnotesize] at (0.64,2.53) {(b)};
  \draw[unmeasured,line width=2.15pt] (0.82,0.38)--(3.48,0.38);
  \draw[measured,line width=2.15pt,densely dashed] (3.48,0.38)--(3.48,2.38);
  \draw[unmeasured,line width=2.15pt] (3.48,2.38)--(0.82,2.38);
  \draw[measured,line width=2.15pt,densely dashed] (0.82,2.38)--(0.82,0.38);

  \node[sidelabel,below=2pt] at (2.15,0.38) {$A$};
  \node[sidelabel,above=2pt] at (2.15,2.38) {$B$};
  \node[sidelabel,left=2pt]  at (0.82,1.38) {$C_2$};
  \node[sidelabel,right=2pt] at (3.48,1.38) {$C_1$};

  \draw[->,line width=0.65pt] (2.15,0.70)--(2.15,2.06)
    node[midway,right=3pt,font=\scriptsize] {$f=2v$};
  \node[font=\scriptsize,anchor=east] at (2.03,0.68) {$0$};
  \node[font=\scriptsize,anchor=east] at (2.03,2.08) {$2\pi$};

  \draw[<->,line width=0.55pt] (0.82,0.06)--(3.48,0.06)
    node[midway,fill=white,inner sep=1pt,font=\scriptsize] {$h(\zeta)$};
\end{tikzpicture}
\caption{Geometry and conformal coordinate. The measured arcs $C_1,C_2$ (dashed red) and unmeasured arcs $A,B$ (solid blue) map to the vertical and horizontal sides of the rectangle $\cR_h$, respectively. The boundary coordinate $f=2\,\mathrm{Im}\,W=2v$ equals $0$ on $A$, $2\pi$ on $B$, and winds once with opposite orientations along the two measured sides.}
\label{fig:geometry}
\end{figure}
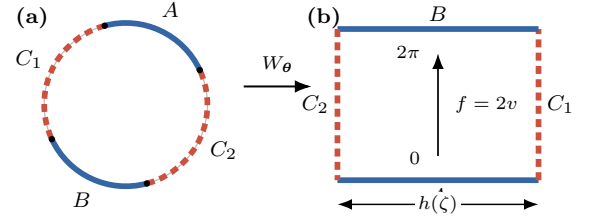

\emph{Microscopic setting.---}
Consider the periodic spin-$1/2$ XX chain at zero magnetization. The Jordan--Wigner transformation maps its ground state to a half-filled Fermi sea of $N=L/2$ consecutive momenta~\cite{JordanWigner1928,LiebSchultzMattis1961,Katsura1962}. For a configuration $X=\{x_1<\cdots<x_N\}$ of occupied sites, the Slater determinant reduces to a circular Vandermonde determinant,
\begin{equation}
 p_L(X)=L^{-N}\prod_{a<b}4\sin^2\!\left[\frac{\pi(x_b-x_a)}{L}\right].
\label{eq:born}
\end{equation}
Thus the Born measure is a discrete $\beta=2$ circular log gas and a projection determinantal process. This places the problem at the intersection of free-fermion full counting statistics~\cite{AbanovIvanov2011,IvanovLevkivskyi2016,LevitovLesovik1993,Klich2003,KlichLevitov2009,Song2012,CalabreseMintchevVicari2012,Gamayun2020}, determinantal processes and random-matrix fluctuation theory~\cite{Macchi1975,CostinLebowitz1995,Soshnikov2000,Soshnikov2002,Johansson1998,DiaconisEvans2001}, Toeplitz asymptotics~\cite{BasorTracy1991,DeiftItsKrasovsky2013}, and modern free-fermion universality and central-limit results~\cite{DeleporteLambert2025,DeleporteLambertJEMS2025}. Equation~\eqref{eq:born} is more than a calculational convenience. It means that every occupation snapshot is a configuration of a strongly correlated Coulomb gas, while every multiplicative deformation of the snapshot weight can be written as a determinant. The universal field fluctuations of a critical Fermi sea are therefore encoded in the asymptotics of a finite matrix whose entries remain directly tied to lattice sites. This exact bridge is what allows us to preserve the Born probabilities throughout the construction rather than replacing the measured string by an assumed continuum boundary profile. The physical boundary sector, normalization of Eq.~\eqref{eq:born}, and all conventions are collected in Supplemental Material (SM), Sec.~S1~\cite{SupplementalMaterial}.

Choose four ordered points $0\leq\theta_1<\theta_2<\theta_3<\theta_4<2\pi$, defining consecutive arcs $A,C_1,B,C_2$ as in Fig.~\ref{fig:geometry}. The measured region is $C=C_1\cup C_2$; $A$ and $B$ remain unobserved. Their conformal cross ratio is
\begin{equation}
 \zeta=\frac{\sin\frac{\theta_2-\theta_1}{2}\,\sin\frac{\theta_4-\theta_3}{2}}
 {\sin\frac{\theta_3-\theta_1}{2}\,\sin\frac{\theta_4-\theta_2}{2}}\in(0,1).
\label{eq:zeta}
\end{equation}
Let $W$ map the disk with these four marked boundary points to the rectangle
$\cR_h=\{0<u<h,\,0<v<\pi\}$. With $A,B$ mapped to the horizontal sides and $C_1,C_2$ to the vertical sides,
\begin{equation}
 h(\zeta)=\pi\frac{K(\zeta)}{K(1-\zeta)},\qquad
 f(\theta)=2\,\mathrm{Im}\,W(\ee^{\ii\theta}).
\label{eq:map}
\end{equation}
Here $K$ is the complete elliptic integral of the first kind. The function $f$ equals $0$ on $A$, $2\pi$ on $B$, rises from $0$ to $2\pi$ on $C_1$, and falls back on $C_2$. Its endpoint derivative has the integrable corner singularity $|f'(\theta)|\sim|\theta-\theta_r|^{-1/2}$. M\"obius transformations can move the four endpoints while preserving $\zeta$, so the detailed function $f(\theta)$ depends on the chosen coordinate on the microscopic circle, whereas its energy and limiting distribution can depend only on $\zeta$. This separation between a coordinate-dependent decoder and a coordinate-independent law is central to the construction. The modulus summarizes all conformally invariant information in the four endpoints. When one unmeasured arc is pinched, $\zeta\to0$ and $h\to0$; when one measured arc is pinched, $\zeta\to1$ and $h\to\infty$. These limits will become respectively the localized and uniform limits of the compact probability law. The complete Schwarz--Christoffel map, branch convention, and midpoint discretization are given in SM, Sec.~S2~\cite{SupplementalMaterial}; the complementary cross-ratio and factor-of-two cylinder conventions common in post-measurement and entanglement-Hamiltonian work are reconciled in SM, Sec.~S8~\cite{SupplementalMaterial}. The choice $f=2v$ is fixed by compactness. Across either measured arc the vertical rectangle coordinate changes by $\pi$, so $f$ changes by one full period $2\pi$.

Place sites at midpoint angles $\vartheta_j=2\pi(j+1/2)/L$, set $f_j=f(\vartheta_j)$, and define the centered linear statistic
\begin{equation}
 X_L=\sum_{j=0}^{L-1}f_j\left(n_j-\frac12\right).
\label{eq:XL}
\end{equation}
The central finite-size observation is elementary but decisive. Because $f=0$ on $A$ and $f=2\pi$ on $B$,
\begin{equation}
 X_L=\sum_{j\in C}f_j\left(n_j-\frac12\right)+2\pi N_B-\pi|B|,
\label{eq:decomp}
\end{equation}
where $|B|$ is the number of lattice sites in $B$ and $N_B=\sum_{j\in B}n_j\in\bbZ$. Hence the circle class
\begin{equation}
 \delta_L(m_C)=\left[\sum_{j\in C}f_j\left(m_C(j)-\frac12\right)-\pi|B|\right]_{2\pi}
\label{eq:delta}
\end{equation}
is determined entirely by the measured record $m_C$, even though the real lift $X_L$ is not. For every $q\in\bbZ$ and every complete configuration compatible with $m_C$,
\begin{equation}
 \ee^{\ii qX_L}=\ee^{\ii q\delta_L(m_C)}.
\label{eq:character}
\end{equation}
This is an exact information-theoretic statement, not a scaling-limit identification: the unobserved region can change only the integer winding of the lift. In particular, no reconstruction of $N_B$ is required to evaluate any physical character $\exp(\ii q\delta_L)$. Experimentally, one assigns to each measured site a geometry-dependent number $f_j$, forms the weighted centered sum in Eq.~\eqref{eq:delta}, and reduces it modulo $2\pi$. The weights are deterministic and can be tabulated once the four endpoints are chosen. The equivalent microscopic-height construction, including the precise edge weights at the four endpoints, is given in SM, Sec.~S2~\cite{SupplementalMaterial}. There the lattice heights satisfy $\Phi_{j+1}-\Phi_j=2\pi(n_j-1/2)$ and periodicity follows from half filling. The signed discrete derivative of $f$ has positive unit mass on one measured channel and negative unit mass on the other. Consequently, $X_L$ is exactly the difference of two normalized height averages. This makes the relative-height coordinate exact already before the continuum limit; its identification with the continuum relative zero mode is a scaling-limit statement.

\emph{Exact determinant bridge.---}
Let $U$ be the $L\times N$ matrix of occupied one-particle orbitals, $K=UU^\dagger$ the Fermi projector, and $D_t=\mathrm{diag}(\ee^{\ii t f_0},\ldots,\ee^{\ii t f_{L-1}})$. Slater overlap and the determinantal generating functional give, for every real $t$,
\begin{equation}
 \chi_L(t):=\langle\ee^{\ii tX_L}\rangle
 =\ee^{-\frac{\ii t}{2}\sum_j f_j}\det_N(U^\dagger D_tU).
\label{eq:detN}
\end{equation}
The same quantity is the $L\times L$ determinant $\ee^{-\frac{\ii t}{2}\sum_j f_j}\det[I+(D_t-I)K]$. More remarkably, for integer $q$ it reduces to the measured sites alone,
\begin{equation}
 \chi_L(q)=\ee^{-\ii q(\frac12\sum_{j\in C}f_j+\pi|B|)}
 \det_C[I_C+(D_{q,C}-I_C)K_C].
\label{eq:detC}
\end{equation}
Here $K_C$ and $D_{q,C}$ denote the restrictions of $K$ and $D_q$ to the measured sites. Equations~\eqref{eq:character} and \eqref{eq:detC} show that the complete free-fermion determinant computes the Fourier moments of a compact variable available from partial data. To see the origin of Eq.~\eqref{eq:detN}, note that the commuting projectors $n_j$ factorize the exponential as
$\exp(\ii t\sum_jf_jn_j)=\prod_j[1+(\ee^{\ii tf_j}-1)n_j]$. Wick's theorem converts this product to $\det[I+(D_t-I)K]$, and Sylvester's identity reduces the determinant to the occupied subspace. For integer $q$, $D_q-I$ vanishes identically on both unmeasured arcs because their $f$ values differ by exactly $2\pi$. Ordering the one-particle space as $C\oplus C^c$ makes the full determinant block triangular, yielding Eq.~\eqref{eq:detC}. Thus the measured-space formula is not a thermodynamic approximation or a consequence of tracing out a Gaussian state; it is an exact algebraic reduction tied to compactification. The full-space, occupied-space, measured-space, and direct Vandermonde representations are proved and cross-checked in SM, Sec.~S3~\cite{SupplementalMaterial}. Their equivalence clarifies different aspects of the result. The $N\times N$ form is the most economical for asymptotics, the $|C|\times|C|$ form makes partial-record sufficiency manifest, and the full-space determinant is the standard determinantal-process generating functional. Expanding the occupied-space determinant by Cauchy--Binet reproduces the Vandermonde sum over all complete configurations with the phase insertion $\prod_{x\in X}\exp(\ii t f_x)$. Thus no continuum assumption enters before the Toeplitz limit.

Because the occupied momenta are consecutive, $U^\dagger D_tU$ is a discrete Toeplitz matrix: its $(r,s)$ entry is the discrete Fourier coefficient of $\exp(\ii t f)$ at $r-s$. A constant shift of $f$ multiplies this matrix by a scalar, but the scalar is canceled exactly by the centering factor in Eq.~\eqref{eq:detN} because $N=L/2$. This lattice cancellation is the determinant counterpart of charge neutrality $\sum_j(n_j-1/2)=0$. Particle--hole symmetry also gives $X_L\overset{d}{=}-X_L$, so $\chi_L(t)$ is real and even at every finite size.

For a smooth real function $g$ on the circle, the strong Szeg\H{o} mechanism predicts a Gaussian limit with the homogeneous $H^{1/2}$ energy
\begin{equation}
 \cE[g]=\sum_{m=1}^{\infty}m|g_m|^2,
 \qquad
 g_m=\frac{1}{2\pi}\int_0^{2\pi}g(\theta)\ee^{-\ii m\theta}\dd\theta.
\label{eq:energy}
\end{equation}
Here $f$ is not smooth: the conformal corners imply $f_m=O(|m|^{-3/2})$. This is sufficient for $\cE[f]<\infty$ but lies at the point where a naive replacement of the discrete Toeplitz determinant by its continuum counterpart requires control. The endpoint expansion and Fourier regularity are derived in SM, Sec.~S4~\cite{SupplementalMaterial}.

We resolve this by combining Fourier truncation with an exact finite-size variance identity. For any real lattice function $u$,
\begin{equation}
 \mathrm{Var}\,X_L(u)=\Tr[K\mathsf U(I-K)\mathsf U]
 =\sum_{m=1}^{L-1}d_L(m)|\widehat u_L(m)|^2,
\label{eq:variance}
\end{equation}
where $\mathsf U=\mathrm{diag}(u_j)$, $\widehat u_L(m)=L^{-1}\sum_{j=0}^{L-1}u_j\ee^{-2\pi\ii mj/L}$, and $d_L(m)=\min(m,L-m)$. The weight $d_L(m)$ has a simple Fermi-surface interpretation: it counts the occupied momentum states that are shifted outside the Fermi sea by a lattice Fourier mode $m$. For $m\ll L$, it equals $m$, directly producing the $H^{1/2}$ norm in Eq.~\eqref{eq:energy}; for $m$ near $L$, periodicity replaces $m$ by $L-m$. At fixed Fourier cutoff, midpoint aliasing is exponentially small and the complex strong Szeg\H{o} theorem applies. Equation~\eqref{eq:variance} then removes the cutoff uniformly, including the square-root endpoint singularities. The result is the real-characteristic-function limit
\begin{equation}
 \lim_{L\to\infty}\chi_L(t)=\exp[-t^2\cE[f]],
 \qquad t\in\bbR.
\label{eq:CLT}
\end{equation}
The exact pair count in Eq.~\eqref{eq:variance}, aliasing estimates, and the three-limit argument are presented in SM, Sec.~S5~\cite{SupplementalMaterial}. The organization of the proof is useful. Write $f=f^{(M)}+r^{(M)}$, with $f^{(M)}$ a finite Fourier polynomial. For fixed $M$, midpoint sampling differs from the continuous Toeplitz coefficients only through aliases separated by multiples of $L$; analyticity of $\exp(\ii t f^{(M)})$ makes these contributions exponentially small. The complex strong Szeg\H{o} theorem then yields $\exp[-t^2\sum_{m\leq M}m|f_m|^2]$. The remainder is not controlled in the uniform norm, which would be too weak for an extensive fermion statistic. Instead, Eq.~\eqref{eq:variance} gives
$\mathrm{Var}\,X_L(r^{(M)})\leq c_1/M+c_2/L$. The inequality $|\ee^{\ii x}-\ee^{\ii y}|\leq|x-y|$ converts this into a uniform bound on characteristic functions, allowing $L\to\infty$ first and $M\to\infty$ afterward. Equation~\eqref{eq:CLT} is consistent with general Gaussian fluctuation theorems for determinantal fermions~\cite{Soshnikov2002,DeleporteLambert2025}, but the observable and the geometric evaluation below are special.

\emph{Why the variance is geometry.---}
The quadratic form in Eq.~\eqref{eq:energy} has a conformal meaning. Let $H_f$ be the harmonic extension of $f$ to the disk and let
$\cD_\Omega[F]=\int_\Omega|\nabla F|^2\dd^2x$. Fourier expansion gives
\begin{equation}
 \cD_{\mathbb D}[H_f]=4\pi\cE[f].
\label{eq:diskenergy}
\end{equation}
But $H_f(z)=2\,\mathrm{Im}\,W(z)$. Dirichlet energy is conformally invariant in two dimensions, so
\begin{equation}
 4\pi\cE[f]=\cD_{\mathbb D}[2\,\mathrm{Im}\,W]
 =\cD_{\cR_h}[2v].
\label{eq:confenergy}
\end{equation}
On the rectangle, $|\nabla(2v)|^2=4$ and $\mathrm{Area}(\cR_h)=\pi h$. Therefore
\begin{equation}
 \cE[f]=h(\zeta).
\label{eq:EequalsH}
\end{equation}
The Poisson-trace argument, weak Dirichlet principle, and real-space $H^{1/2}$ identity are supplied in SM, Sec.~S6~\cite{SupplementalMaterial}. This step is the geometric core: an abstract variance of a microscopic fermion statistic becomes exactly the conformal modulus of the four measurement regions. It also explains why the boundary weights cannot be replaced by uniform arc averages. The normalized conformal side measure is the weighting for which the boundary interpolation is the pullback of the linear potential $2v$ on the rectangle. In our $2\pi$-period normalization, $\cE[f]=h$ is the normalized Dirichlet energy of the potential $f=2v$; equivalently, it is proportional to the rectangle conductance with the normalization fixed by this potential drop. Consequently, the limiting Dirichlet energy depends on the geometry only through the cross ratio.

Combining Eqs.~\eqref{eq:CLT} and \eqref{eq:EequalsH}, the complete real statistic converges to
\begin{equation}
 X_L\Longrightarrow \mathsf N(0,2h).
\label{eq:realGaussian}
\end{equation}
For the partial record, only integer characters are intrinsic. Equations~\eqref{eq:character} and \eqref{eq:CLT} imply
\begin{equation}
 \lim_{L\to\infty}\mathbb E[\ee^{\ii q\delta_L}]=\ee^{-h q^2},
 \qquad q\in\bbZ.
\label{eq:fourierlimit}
\end{equation}
Fourier coefficients determine probability measures on the circle, giving the main result
\begin{equation}
 \operatorname{Law}(\delta_L)\Longrightarrow P_h(\delta)\dd\delta.
\label{eq:compactlimit}
\end{equation}
The limiting density is~\cite{KhannaVasseur2026,KhannaVasseurStats2026}
\begin{equation}
 P_h(\delta)=\frac{1}{\sqrt{4\pi h}}\sum_{\ell\in\bbZ}
 \exp\!\left[-\frac{(\delta+2\pi\ell)^2}{4h}\right].
\label{eq:wrapped}
\end{equation}
Equivalently,
$P_h(\delta)=(2\pi)^{-1}\sum_{q\in\bbZ}\ee^{-hq^2}\ee^{-\ii q\delta}$: it is the heat kernel on $\bbT$ at time $h$. This form makes the compact physics transparent. For a narrow rectangle, $h\to0$, the distribution is concentrated near a definite relative boundary value. For a long rectangle, $h\to\infty$, all nonzero Fourier modes are exponentially suppressed and the boundary angle becomes uniform. The compact law retains information that the Gaussian lift alone would obscure: shifts by $2\pi$ are physically identical, and the intrinsic characters are the integer harmonics $\ee^{\ii q\delta}$, $q\in\bbZ$. Conversely, the real Gaussian is not an additional measured variable; it is a convenient lift defined on complete configurations. The partial-record distribution can be established without choosing a branch by applying the Fourier convergence criterion for probability measures on the compact group $\bbT$. This point prevents a common ambiguity: ordinary means and variances of an angle depend on a branch cut, whereas the moments $\langle\ee^{\ii q\delta}\rangle$ are intrinsic. The weak-convergence proof, lift map, Fourier uniqueness, and heat-kernel forms are detailed in SM, Sec.~S6~\cite{SupplementalMaterial}.

\emph{Discussion.---}
Our result identifies an exact microscopic compact coordinate whose scaling-limit law is that of the relative Dirichlet phase, rather than only an observable averaged over boundary conditions.  In compact Dirac and bosonic descriptions the relative Dirichlet phase is conjugate to the conserved charge and, in the post-measurement entanglement Hamiltonian, maps to the chemical-potential parameter~\cite{EislerTonni2026}; conformal geometry fixes the inverse-temperature profile~\cite{CardyTonni2016,DalmonteEHReview2022,EislerTonniPeschel2019,EislerTonniPeschel2022}.  This realizes, at the free-fermion point and without replicas, the relative-zero-mode sector of the ``Born average over Dirichlet boundaries'' anticipated in compact-boson CFT~\cite{KhannaVasseur2026,KhannaVasseurStats2026}.  It is complementary to calculations for forced outcomes and formation probabilities~\cite{Rajabpour2015,Rajabpour2016,NajafiRajabpour2016,Hoshino2025,Stephan2014,RajabpourFormation2015}, which determine the cost or entanglement associated with prescribed boundaries rather than the random zero-mode law.

The free-fermion point makes the proof exact, but the geometric result suggests a broader principle. In an interacting Tomonaga--Luttinger liquid, the same conformal coordinate should survive with a stiffness set by the Luttinger parameter~\cite{KhannaVasseur2026,KhannaVasseurStats2026}; establishing this directly from XXZ snapshots would separate universal compactification data from lattice harmonics. Such an extension is nontrivial: the exact determinant disappears, irrelevant oscillatory operators can contaminate microscopic height variables, and the compactification radius changes continuously. Nevertheless, the modulo-$2\pi$ cancellation in Eq.~\eqref{eq:decomp} relies only on charge quantization and on choosing a boundary function that differs by one compact period between the unmeasured arcs. This makes the record-only compactification more robust than the Gaussian evaluation and provides a concrete target for bosonization, tensor-network sampling, and experiment. An experimental test would choose four endpoint positions, compute the deterministic conformal weights, and form $\delta_L$ for each snapshot. The entire prediction can then be tested through the low harmonics $\langle\cos(q\delta_L)\rangle=\exp[-h(\zeta)q^2]$, avoiding branch choices and density estimation on the circle. Varying the endpoints at fixed cross ratio provides an especially stringent universality check: the site weights change, but the limiting moments do not.

Multiple measured components are expected to produce several relative zero modes and a higher-dimensional compact Gaussian governed by an energy matrix playing the role of a capacitance or period matrix. In that setting, the present scalar conformal side measure is replaced by a basis of harmonic functions with prescribed values on the measured boundary components, and the Dirichlet energies form the covariance matrix. The determinant approach remains available for free fermions, suggesting a direct route from multicomponent snapshot statistics to the period geometry of multiply connected domains. A further mathematical question is whether the corner-regularity threshold used here is optimal, and how the compact law changes when endpoints merge on an $L$-dependent scale. Such crossover regimes should connect the finite-energy Szeg\H{o} problem to Fisher--Hartwig counting statistics and may reveal universal interpolating kernels. Changing the measurement basis may select Neumann, mixed, or symmetry-twisted boundary coordinates. It will also be useful to determine whether the compact heat kernel controls portions of measurement-induced entanglement distributions, rather than only their averaged partition functions. These extensions, along with robustness to endpoint motion and convention changes, are organized in SM, Sec.~S8~\cite{SupplementalMaterial}. More generally, the result shows that universal conformal geometry need not be reconstructed from correlation functions or entanglement entropies: it can already be present, in directly decodable form, in the randomness of individual many-body measurements.

{\it Acknowledgements:}
M.A.R. acknowledges partial support from CNPq and FAPERJ (grant number E-26/210.062/2023).

\clearpage
\onecolumngrid

\setcounter{section}{0}
\setcounter{subsection}{0}
\setcounter{equation}{0}
\setcounter{figure}{0}
\setcounter{table}{0}
\renewcommand{\thesection}{S\arabic{section}}
\renewcommand{\thesubsection}{\thesection.\arabic{subsection}}
\renewcommand{\theequation}{S\arabic{equation}}
\renewcommand{\thefigure}{S\arabic{figure}}
\renewcommand{\thetable}{S\arabic{table}}
\setcounter{secnumdepth}{3}

\renewcommand{\theHsection}{S.\arabic{section}}
\renewcommand{\theHsubsection}{S.\arabic{section}.\arabic{subsection}}
\renewcommand{\theHequation}{S.\arabic{equation}}
\renewcommand{\theHfigure}{S.\arabic{figure}}
\renewcommand{\theHtable}{S.\arabic{table}}

\begin{center}
{\large\bfseries Supplemental Material for \textit{Quantum Snapshots Reveal a Compact Conformal Boundary Mode}\par}
\vspace{1em}
{M.~A.~Rajabpour\par}
\vspace{0.4em}
{\small Instituto de F\'isica, Universidade Federal Fluminense, Av.~Gal.~Milton Tavares de Souza s/n, Gragoat\'a, 24210-346, Niter\'oi, RJ, Brazil\par}
\end{center}
\vspace{1em}
This Supplemental Material is organized in the nine sections cited in the Letter.  Sections~\ref{sec:S1}--\ref{sec:S9} provide the complete microscopic derivation, asymptotic proof, geometric evaluation, numerical verification, convention matching, and comparison with continuum boundary-phase and charge-sector descriptions used in the main text.  The notation follows the Letter: the ring contains an even number $L$ of sites, the ground state is at half filling $N=L/2$, and a complete occupation configuration is denoted by $X=\{x_1<\cdots<x_N\}$.

\setcounter{tocdepth}{2}
\tableofcontents
\clearpage

\section{Jordan--Wigner sector and exact Born measure}
\label{sec:S1}

This section derives the microscopic probability measure used in the Letter directly from the periodic spin-$1/2$ XX chain.  We keep the Jordan--Wigner boundary term explicit, identify the physical momentum grid for both congruence classes of even $L$, and then derive the exact circular Vandermonde probability.  The analysis also fixes the spin/occupation convention and proves that the final Born law is independent of the common momentum offset.  Standard background on the Jordan--Wigner solution of the XY family may be found in Refs.~\cite{JordanWigner1928,LiebSchultzMattis1961,Katsura1962}.

\subsection{Periodic XX chain and measurement convention}
\label{subsec:S1-spin-model}

Let $\sigma_j^{x,y,z}$ be Pauli matrices on the sites
\begin{equation}
 j\in\bbZ_L:=\{0,1,\ldots,L-1\},
 \qquad
 \sigma_{j+L}^{\alpha}=\sigma_j^{\alpha}.
\label{eq:S1-sites}
\end{equation}
We use the periodic XX Hamiltonian
\begin{align}
 H_{\mathrm{XX}}^{\mathrm{spin}}
 &=-\frac12\sum_{j=0}^{L-1}
 \left(\sigma_j^x\sigma_{j+1}^x+\sigma_j^y\sigma_{j+1}^y\right)
 \notag\\
 &=-\sum_{j=0}^{L-1}
 \left(\sigma_j^+\sigma_{j+1}^-+\sigma_j^-\sigma_{j+1}^+\right),
\label{eq:S1-spin-H}
\end{align}
where
\begin{equation}
 \sigma_j^{\pm}:=\frac{\sigma_j^x\pm\ii\sigma_j^y}{2}.
\label{eq:S1-ladder}
\end{equation}
The overall hopping scale in Eq.~\eqref{eq:S1-spin-H} only rescales energies and has no effect on the ground-state occupation probabilities.

We choose the Jordan--Wigner convention in which an occupied fermion corresponds to $\sigma_j^z=+1$.  Introduce
\begin{equation}
 p_j:=-\sigma_j^z
\label{eq:S1-local-parity}
\end{equation}
and define
\begin{equation}
 c_j=\left(\prod_{\ell=0}^{j-1}p_\ell\right)\sigma_j^- ,
 \qquad
 c_j^\dagger=\left(\prod_{\ell=0}^{j-1}p_\ell\right)\sigma_j^+,
\label{eq:S1-JW}
\end{equation}
with the empty product equal to the identity.  The strings in Eq.~\eqref{eq:S1-JW} give the canonical anticommutation relations
\begin{equation}
 \{c_j,c_k^\dagger\}=\delta_{jk},
 \qquad
 \{c_j,c_k\}=\{c_j^\dagger,c_k^\dagger\}=0.
\label{eq:S1-CAR}
\end{equation}
Indeed, for $j<k$ the string of the operator at $k$ contains the single factor $p_j=-\sigma_j^z$ that anticommutes with $\sigma_j^{\pm}$; all other factors act on distinct sites.  On the same site, the Pauli algebra gives $\{c_j,c_j^\dagger\}=1$ and $c_j^2=(c_j^\dagger)^2=0$.

The fermion occupation is
\begin{equation}
 n_j:=c_j^\dagger c_j
 =\sigma_j^+\sigma_j^-
 =\frac{1+\sigma_j^z}{2},
\label{eq:S1-number-spin}
\end{equation}
so that
\begin{equation}
 \sigma_j^z=2n_j-1,
 \qquad
 p_j=1-2n_j.
\label{eq:S1-z-number}
\end{equation}
Thus a projective occupation measurement on a set of sites is exactly a projective $\sigma^z$ measurement in the spin chain, with bit value $1$ assigned to $\sigma^z=+1$.

\subsection{Parity-dependent fermionic boundary condition}
\label{subsec:S1-boundary-sector}

For a bulk bond $0\leq j\leq L-2$, cancellation of the adjacent Jordan--Wigner strings gives
\begin{equation}
 \sigma_j^+\sigma_{j+1}^-=c_j^\dagger c_{j+1},
 \qquad
 \sigma_j^-\sigma_{j+1}^+=c_{j+1}^\dagger c_j.
\label{eq:S1-bulk-bonds}
\end{equation}
The bond crossing the periodic boundary depends on the total fermion parity
\begin{equation}
 \mathcal P:=\prod_{j=0}^{L-1}p_j=(-1)^{N_f},
 \qquad
 N_f:=\sum_{j=0}^{L-1}n_j.
\label{eq:S1-total-parity}
\end{equation}
Since $\prod_{\ell=0}^{L-2}p_\ell=\mathcal P p_{L-1}$, one obtains
\begin{equation}
 \sigma_{L-1}^+\sigma_0^-=-\mathcal P c_{L-1}^\dagger c_0,
 \qquad
 \sigma_{L-1}^-\sigma_0^+=-\mathcal P c_0^\dagger c_{L-1}.
\label{eq:S1-boundary-bonds}
\end{equation}
Consequently,
\begin{align}
 H_{\mathrm{XX}}^{\mathrm{spin}}
 ={}&-\sum_{j=0}^{L-2}
 \left(c_j^\dagger c_{j+1}+c_{j+1}^\dagger c_j\right)
 \notag\\
 &+\mathcal P
 \left(c_{L-1}^\dagger c_0+c_0^\dagger c_{L-1}\right).
\label{eq:S1-fermion-H-parity}
\end{align}
Within a fixed parity sector this is the translation-invariant hopping Hamiltonian
\begin{equation}
 H_{\mathrm{XX}}^{(\mathcal P)}
 =-\sum_{j=0}^{L-1}
 \left(c_j^\dagger c_{j+1}+c_{j+1}^\dagger c_j\right)
\label{eq:S1-uniform-H}
\end{equation}
provided the fermions obey the twisted boundary condition
\begin{equation}
 c_{j+L}=-\mathcal P c_j.
\label{eq:S1-fermion-BC}
\end{equation}
A plane wave $\ee^{\ii kj}$ is allowed when
\begin{equation}
 \ee^{\ii kL}=-\mathcal P.
\label{eq:S1-momentum-quantization}
\end{equation}
Therefore
\begin{equation}
\begin{array}{c|c|c}
 \mathcal P & c_{j+L} & \text{allowed momenta}\\
 \hline
 +1 & -c_j & k=\dfrac{2\pi}{L}\left(m+\dfrac12\right)\\[2mm]
 -1 & +c_j & k=\dfrac{2\pi m}{L}
\end{array}
\qquad (m\in\bbZ).
\label{eq:S1-sector-table}
\end{equation}
Hence even fermion parity corresponds to antiperiodic Jordan--Wigner fermions, while odd parity corresponds to periodic fermions.

\subsection{Physical half-filled momentum block}
\label{subsec:S1-ground-state}

Using
\begin{equation}
 c_j=\frac1{\sqrt L}\sum_{k\in\mathcal K_{\mathcal P}}
 \ee^{\ii kj}d_k,
\label{eq:S1-Fourier-transform}
\end{equation}
Eq.~\eqref{eq:S1-uniform-H} becomes
\begin{equation}
 H_{\mathrm{XX}}^{(\mathcal P)}
 =\sum_{k\in\mathcal K_{\mathcal P}}
 \varepsilon(k)d_k^\dagger d_k,
 \qquad
 \varepsilon(k)=-2\cos k.
\label{eq:S1-dispersion}
\end{equation}
At zero magnetization,
\begin{equation}
 N_f=N=\frac L2,
 \qquad
 \mathcal P=(-1)^{L/2}.
\label{eq:S1-half-filling-parity}
\end{equation}
The negative-energy region is $-\pi/2<k<\pi/2$ modulo $2\pi$.  The physical occupied modes are therefore
\begin{equation}
\begin{array}{c|c|c|c}
 L & N & \mathcal P & \text{occupied labels}\\
 \hline
 4\ell & 2\ell & +1 &
 k=\dfrac{2\pi}{L}\left(m+\dfrac12\right),\quad
 m=-\ell,\ldots,\ell-1\\[3mm]
 4\ell+2 & 2\ell+1 & -1 &
 k=\dfrac{2\pi m}{L},\quad
 m=-\ell,\ldots,\ell
\end{array}.
\label{eq:S1-Lmod4-table}
\end{equation}
Both rows are represented by the single consecutive block
\begin{equation}
 k_r=k_\star+\frac{2\pi r}{L},
 \qquad
 k_\star=-\frac\pi2+\frac\pi L,
 \qquad
 r=0,1,\ldots,N-1.
\label{eq:S1-occupied-block}
\end{equation}
No allowed momentum lies at $k=\pm\pi/2$, so the half-filled ground state is the unique Slater determinant obtained by occupying all negative-energy modes:
\begin{equation}
 |\Omega_L\rangle
 :=\prod_{r=0}^{N-1}d_{k_r}^\dagger|0\rangle,
 \qquad
 d_k^\dagger=\frac1{\sqrt L}\sum_{j=0}^{L-1}\ee^{\ii kj}c_j^\dagger.
\label{eq:S1-Fermi-sea}
\end{equation}
The product in Eq.~\eqref{eq:S1-Fermi-sea} is ordered in increasing $r$.  The orbitals are orthonormal because
\begin{equation}
 \{d_{k_r},d_{k_s}^\dagger\}
 =\frac1L\sum_{j=0}^{L-1}
 \ee^{2\pi\ii(s-r)j/L}
 =\delta_{rs},
\label{eq:S1-orbital-orthogonality}
\end{equation}
and hence $\langle\Omega_L|\Omega_L\rangle=1$.

\subsection{Coordinate-space Slater determinant}
\label{subsec:S1-Slater}

For an ordered set of occupied sites
\begin{equation}
 X=\{x_1<x_2<\cdots<x_N\}\subset\bbZ_L,
\label{eq:S1-configuration}
\end{equation}
define the occupation-basis state
\begin{equation}
 |X\rangle
 :=c_{x_1}^\dagger c_{x_2}^\dagger\cdots c_{x_N}^\dagger|0\rangle.
\label{eq:S1-occupation-state}
\end{equation}
The states $|X\rangle$ form an orthonormal basis of the $N$-particle sector.  Expanding the product in Eq.~\eqref{eq:S1-Fermi-sea}, the coefficient of the ordered monomial in Eq.~\eqref{eq:S1-occupation-state} is the antisymmetrized product of the occupied orbitals:
\begin{equation}
 \langle X|\Omega_L\rangle
 =\det_{\substack{0\leq r\leq N-1\\1\leq a\leq N}}
 \left[\frac{\ee^{\ii k_rx_a}}{\sqrt L}\right].
\label{eq:S1-Slater-amplitude}
\end{equation}
Equivalently, defining the $L\times N$ orbital matrix
\begin{equation}
 U_{jr}:=\frac1{\sqrt L}\ee^{\ii k_rj},
\label{eq:S1-U-matrix}
\end{equation}
Eq.~\eqref{eq:S1-Slater-amplitude} is $\langle X|\Omega_L\rangle=\det U_X$, where $U_X$ is the $N\times N$ submatrix formed from the rows indexed by $X$ in increasing order.

\subsection{Circular Vandermonde probability}
\label{subsec:S1-Vandermonde}

Introduce the $L$th roots of unity
\begin{equation}
 z_a:=\ee^{2\pi\ii x_a/L}.
\label{eq:S1-roots}
\end{equation}
Since the occupied momenta are consecutive,
\begin{equation}
 \ee^{\ii k_rx_a}=\ee^{\ii k_\star x_a}z_a^r.
\label{eq:S1-factorization}
\end{equation}
Extracting the common phase from each column of the Slater determinant gives
\begin{align}
 \langle X|\Omega_L\rangle
 &=L^{-N/2}\ee^{\ii k_\star\sum_{a=1}^Nx_a}
 \det[z_a^r]_{\substack{0\leq r\leq N-1\\1\leq a\leq N}}
 \notag\\
 &=L^{-N/2}\ee^{\ii k_\star\sum_ax_a}
 \prod_{1\leq a<b\leq N}(z_b-z_a),
\label{eq:S1-Vandermonde-amplitude}
\end{align}
where the second line is the Vandermonde identity.  Taking the absolute square removes the boundary-sector phase:
\begin{align}
 p_L(X)
 &:=|\langle X|\Omega_L\rangle|^2
 =L^{-N}\prod_{a<b}|z_b-z_a|^2
 \notag\\
 &=L^{-N}\prod_{a<b}
 4\sin^2\!\left[\frac{\pi(x_b-x_a)}{L}\right].
\label{eq:S1-Born-measure}
\end{align}
Equation~\eqref{eq:S1-Born-measure} is Eq.~(1) of the Letter.  Its normalization is exact:
\begin{equation}
 \sum_{\substack{X\subset\bbZ_L\\|X|=N}}p_L(X)
 =\sum_{|X|=N}|\langle X|\Omega_L\rangle|^2
 =\langle\Omega_L|\Omega_L\rangle
 =1.
\label{eq:S1-Born-normalization}
\end{equation}
Equivalently, Eq.~\eqref{eq:S1-Born-normalization} is the discrete circular Dyson identity
\begin{equation}
 \sum_{0\leq x_1<\cdots<x_N\leq L-1}
 \prod_{a<b}|\ee^{2\pi\ii x_b/L}-\ee^{2\pi\ii x_a/L}|^2
 =L^N.
\label{eq:S1-discrete-Dyson}
\end{equation}

The probability may be written in Gibbs form as
\begin{equation}
 p_L(X)=L^{-N}\ee^{-2E_L(X)},
 \qquad
 E_L(X):=-\sum_{a<b}
 \log\left|2\sin\frac{\pi(x_b-x_a)}{L}\right|.
\label{eq:S1-log-gas}
\end{equation}
Thus the snapshots form a discrete circular logarithmic gas at inverse temperature $\beta=2$.

The same law is a fixed-cardinality projection determinantal process.  Indeed, with
\begin{equation}
 K:=UU^\dagger,
 \qquad
 K_{j\ell}=\frac1L\sum_{r=0}^{N-1}
 \ee^{\ii k_r(j-\ell)},
\label{eq:S1-Fermi-projector}
\end{equation}
one has $K^2=K$, $K^\dagger=K$, and $\Tr K=N$.  For every $N$-site configuration,
\begin{equation}
 p_L(X)=|\det U_X|^2
 =\det(U_XU_X^\dagger)
 =\det K_X.
\label{eq:S1-projection-DPP}
\end{equation}
The correlation and generating-function identities associated with this projection process are derived in Sec.~\ref{sec:S3}.

\subsection{Partial occupation records}
\label{subsec:S1-partial-records}

Let $C=C_1\sqcup C_2$ be the measured set introduced in the Letter.  A measurement outcome is a bit string
\begin{equation}
 m_C:C\longrightarrow\{0,1\}.
\label{eq:S1-record}
\end{equation}
For a complete configuration $X$, write $\rho_C(X)=\mathbf1_X|_C$ for its restriction to the measured sites.  The Born probability of the partial record is the marginal
\begin{equation}
 \mathbb P_{L,C}(m_C)
 :=\sum_{\substack{X\subset\bbZ_L,\ |X|=N\\
 \rho_C(X)=m_C}}p_L(X).
\label{eq:S1-partial-marginal}
\end{equation}
The fibers of $\rho_C$ partition the set of complete configurations, and therefore
\begin{equation}
 \sum_{m_C\in\{0,1\}^{C}}\mathbb P_{L,C}(m_C)=1.
\label{eq:S1-partial-normalization}
\end{equation}
No postselection or continuum replacement enters Eq.~\eqref{eq:S1-partial-marginal}: it is the exact pushforward of the XX-chain Born measure under restriction to the observed sites.

\subsection{Independence of momentum-offset conventions}
\label{subsec:S1-convention-independence}

The parity analysis above identifies the physical momentum grid of the periodic spin chain.  The probability in Eq.~\eqref{eq:S1-Born-measure}, however, depends only on the fact that the occupied orbitals form a consecutive block.  Under a common shift
\begin{equation}
 k_r\longmapsto k_r+\kappa,
\label{eq:S1-common-shift}
\end{equation}
the coordinate amplitude transforms as
\begin{equation}
 \langle X|\Omega_L\rangle
 \longmapsto
 \ee^{\ii\kappa\sum_ax_a}\langle X|\Omega_L\rangle.
\label{eq:S1-common-shift-phase}
\end{equation}
The multiplier has unit modulus, so every complete probability $p_L(X)$ and every marginal probability $\mathbb P_{L,C}(m_C)$ is unchanged.  This establishes the convention independence used in the Letter: the Jordan--Wigner sector is required to identify the physical spin-chain ground state, but it leaves the occupation-basis Born law invariant once the consecutive Fermi sea is fixed.

\section{Schwarz--Christoffel map, microscopic height, and record-only compactification}
\label{sec:S2}

This section constructs the conformal coordinate used in the Letter and proves its exact finite-size interpretation.  We first fix the cross-ratio, branch, and orientation conventions for the map from the marked disk to a rectangle of height $\pi$.  We then define the compact boundary coordinate $f_{\boldsymbol\theta}=2\,\operatorname{Im}W_{\boldsymbol\theta}$, relate it to a relative conformal height, and discretize it at the lattice midpoints.  The final result is the exact identity
\begin{equation}
 [X_L(X)]_{2\pi}=\delta_L\!\left(\rho_{C,L}(X)\right),
\label{eq:S2-main-compactification}
\end{equation}
valid for every complete occupation configuration $X$ at every admissible even $L$.  Thus the particular complete-configuration statistic $X_L$ depends on unobserved occupations, but its compact class is a function of the partial record alone; the representative $r_L(m_C)$ introduced below is itself record-only.

\subsection{Marked circle and conformal cross ratio}
\label{subsec:S2-cross-ratio}

Write the spatial circle as $z=\ee^{\ii\theta}$ and choose four boundary points
\begin{equation}
 z_r=\ee^{\ii\theta_r},
 \qquad
 0\leq\theta_1<\theta_2<\theta_3<\theta_4<2\pi.
\label{eq:S2-marked-points}
\end{equation}
They define four consecutive open arcs in the positive angular orientation,
\begin{align}
 A&=(\theta_1,\theta_2),
 &C_1&=(\theta_2,\theta_3),
 \notag\\
 B&=(\theta_3,\theta_4),
 &C_2&=(\theta_4,\theta_1+2\pi).
\label{eq:S2-continuum-arcs}
\end{align}
The last interval is understood on the lifted angular coordinate; on $[0,2\pi)$ it is $(\theta_4,2\pi)\cup[0,\theta_1)$.  The measured arcs are $C_1$ and $C_2$, whereas $A$ and $B$ are unmeasured.

The M\"obius-invariant cross ratio adapted to this cyclic ordering is
\begin{align}
 \zeta
 &:=\frac{(z_2-z_1)(z_3-z_4)}{(z_2-z_4)(z_3-z_1)}
 \notag\\
 &=
 \frac{
 \sin\!\left[(\theta_2-\theta_1)/2\right]
 \sin\!\left[(\theta_4-\theta_3)/2\right]
 }{
 \sin\!\left[(\theta_3-\theta_1)/2\right]
 \sin\!\left[(\theta_4-\theta_2)/2\right]
 }.
\label{eq:S2-cross-ratio}
\end{align}
The phase factors in the four chord differences cancel, so the second line is real and positive.  Moreover,
\begin{align}
&\sin\!\left(\frac{\theta_3-\theta_1}{2}\right)
 \sin\!\left(\frac{\theta_4-\theta_2}{2}\right)
-
 \sin\!\left(\frac{\theta_2-\theta_1}{2}\right)
 \sin\!\left(\frac{\theta_4-\theta_3}{2}\right)
\notag\\
&\hspace{18mm}=
 \sin\!\left(\frac{\theta_3-\theta_2}{2}\right)
 \sin\!\left(\frac{\theta_4-\theta_1}{2}\right)>0,
\label{eq:S2-zeta-bound}
\end{align}
which proves
\begin{equation}
 0<\zeta<1.
\label{eq:S2-zeta-range}
\end{equation}
We write $\boldsymbol\theta=(\theta_1,\theta_2,\theta_3,\theta_4)$ and $\zeta=\zeta(\boldsymbol\theta)$.  The normalized boundary function constructed below depends on the full ordered quadruple in the fixed microscopic angular coordinate, while its conformal energy depends only on $\zeta$.

For the symmetric arrangement
\begin{equation}
 (\theta_1,\theta_2,\theta_3,\theta_4)
 =(0,\alpha,\pi,\pi+\alpha),
\label{eq:S2-symmetric-endpoints}
\end{equation}
one has
\begin{equation}
 \zeta=\sin^2(\alpha/2).
\label{eq:S2-symmetric-zeta}
\end{equation}
This is the convention used in the finite-size examples of Sec.~\ref{sec:S7}.

\subsection{Explicit map to the normalized rectangle}
\label{subsec:S2-SC-map}

The M\"obius map
\begin{equation}
 \xi(z)=
 \frac{(z-z_1)(z_3-z_4)}{(z-z_4)(z_3-z_1)}
\label{eq:S2-Mobius-map}
\end{equation}
sends the disk to the upper half-plane and the four marked points to
\begin{equation}
 \xi(z_1)=0,
 \qquad
 \xi(z_2)=\zeta,
 \qquad
 \xi(z_3)=1,
 \qquad
 \xi(z_4)=\infty.
\label{eq:S2-prevertices}
\end{equation}
Along the positively oriented boundary circle, the four arcs therefore map to the real intervals
\begin{equation}
 A:(0,\zeta),
 \quad
 C_1:(\zeta,1),
 \quad
 B:(1,\infty),
 \quad
 C_2:(-\infty,0).
\label{eq:S2-real-intervals}
\end{equation}

To fix the branch unambiguously, define in the upper half-plane
\begin{equation}
 R_\zeta(\xi)
 :=\sqrt{\xi}\,\sqrt{\xi-\zeta}\,\sqrt{\xi-1},
\label{eq:S2-root-branch}
\end{equation}
where each square root is analytic for $\operatorname{Im}\xi>0$ and is positive for a positive real argument.  Its boundary values from above are
\begin{equation}
 R_{\zeta,+}(x)=
 \begin{cases}
 -\ii\sqrt{(-x)(\zeta-x)(1-x)},&x<0,\\[1mm]
 -\sqrt{x(\zeta-x)(1-x)},&0<x<\zeta,\\[1mm]
 \ii\sqrt{x(x-\zeta)(1-x)},&\zeta<x<1,\\[1mm]
 \sqrt{x(x-\zeta)(x-1)},&x>1.
 \end{cases}
\label{eq:S2-root-boundary-values}
\end{equation}
These phases determine the orientation of all four rectangle sides.

We use the parameter convention
\begin{equation}
 K(m)=\int_0^{\pi/2}
 \frac{\dd\varphi}{\sqrt{1-m\sin^2\varphi}},
 \qquad 0<m<1,
\label{eq:S2-elliptic-K}
\end{equation}
and define the normalized Schwarz--Christoffel map by
\begin{equation}
 W_{\boldsymbol\theta}(z)
 =-\frac{\pi}{2K(1-\zeta)}
 \int_{0}^{\xi(z)}\frac{\dd t}{R_\zeta(t)}.
\label{eq:S2-explicit-W}
\end{equation}
The integral is taken in the upper half-plane, with the lower endpoint understood as an integrable boundary limit.  The integrand is analytic in the open upper half-plane, so the value is path independent.

The two adjacent absolute side integrals are
\begin{align}
 I_A
 &:=\int_0^\zeta
 \frac{\dd x}{\sqrt{x(\zeta-x)(1-x)}}
 =2K(\zeta),
\label{eq:S2-side-integral-A}\\
 I_C
 &:=\int_\zeta^1
 \frac{\dd x}{\sqrt{x(x-\zeta)(1-x)}}
 =2K(1-\zeta).
\label{eq:S2-side-integral-C}
\end{align}
The first equality follows from $x=\zeta\sin^2\varphi$ and the second from $x=\zeta+(1-\zeta)\sin^2\varphi$.  The branch phases in Eq.~\eqref{eq:S2-root-boundary-values} and the negative real prefactor in Eq.~\eqref{eq:S2-explicit-W} give the corner values
\begin{equation}
 W(z_1)=0,
 \quad
 W(z_2)=h,
 \quad
 W(z_3)=h+\ii\pi,
 \quad
 W(z_4)=\ii\pi,
\label{eq:S2-corner-values}
\end{equation}
where
\begin{equation}
 h(\zeta)=\pi\frac{K(\zeta)}{K(1-\zeta)}
 .
\label{eq:S2-rectangle-width}
\end{equation}
Thus $W_{\boldsymbol\theta}$ maps the disk conformally to
\begin{equation}
 \mathcal R_h
 =\{W=u+\ii v:0<u<h,\ 0<v<\pi\},
\label{eq:S2-rectangle}
\end{equation}
with the side assignment
\begin{equation}
 A\mapsto\{v=0\},
 \qquad
 C_1\mapsto\{u=h\},
 \qquad
 B\mapsto\{v=\pi\},
 \qquad
 C_2\mapsto\{u=0\}.
\label{eq:S2-side-assignment}
\end{equation}
Along the positively oriented circle, $v$ is constant on $A$ and $B$, rises from $0$ to $\pi$ on $C_1$, and falls from $\pi$ to $0$ on $C_2$.  Equation~\eqref{eq:S2-explicit-W} fixes the additive constant, overall scale, and orientation used throughout the Letter.  Alternative cross-ratio, elliptic-modulus, and doubled-cylinder conventions are reconciled in Sec.~\ref{sec:S8}.

\subsection{Boundary coordinate, conformal side measures, and corner behavior}
\label{subsec:S2-boundary-coordinate}

Define the real boundary coordinate
\begin{equation}
 f_{\boldsymbol\theta}(\theta)
 :=2\,\operatorname{Im}W_{\boldsymbol\theta}(\ee^{\ii\theta}).
\label{eq:S2-boundary-f}
\end{equation}
The normalization by the factor $2$ is fixed by compactness: the vertical change $\Delta v=\pi$ across a measured side becomes one full period $\Delta f=2\pi$.  From Eq.~\eqref{eq:S2-side-assignment},
\begin{equation}
 f_{\boldsymbol\theta}(\theta)=
 \begin{cases}
 0,&\theta\in A,\\
 2\pi,&\theta\in B,
 \end{cases}
\label{eq:S2-f-constant-arcs}
\end{equation}
and $f_{\boldsymbol\theta}$ rises monotonically on $C_1$ and falls monotonically on $C_2$.

The normalized positive conformal side measures on the measured arcs are
\begin{equation}
 \dd\omega_1=\frac{\dd v}{\pi}
 \quad\text{on }C_1,
 \qquad
 \dd\omega_2=-\frac{\dd v}{\pi}
 \quad\text{on }C_2.
\label{eq:S2-harmonic-measures}
\end{equation}
They obey
\begin{equation}
 \int_{C_1}\dd\omega_1
 =\int_{C_2}\dd\omega_2=1,
\label{eq:S2-measure-normalization}
\end{equation}
and, as a signed measure on the oriented boundary circle,
\begin{equation}
 \frac{\dd f_{\boldsymbol\theta}}{2\pi}
 =\dd\omega_1-\dd\omega_2.
\label{eq:S2-signed-measure}
\end{equation}
These are the normalized conformal side weights used by the decoder.  They are not uniform angular measures except in special geometries.

For later regularity estimates, it is useful to record the pullback of the Schwarz--Christoffel differential.  Algebraically, Eq.~\eqref{eq:S2-explicit-W} can be rewritten in the disk as
\begin{equation}
 \dd W_{\boldsymbol\theta}
 =C_{\boldsymbol\theta}
 \frac{\dd z}{\sqrt{\prod_{r=1}^4(z-z_r)}}
\label{eq:S2-disk-SC-form}
\end{equation}
for a nonzero constant $C_{\boldsymbol\theta}$.  Hence, on $z=\ee^{\ii\theta}$,
\begin{equation}
 \left|\frac{\dd W_{\boldsymbol\theta}}{\dd\theta}\right|
 =\mathcal N_{\boldsymbol\theta}
 \left|
 \prod_{r=1}^4
 \sin\!\left(\frac{\theta-\theta_r}{2}\right)
 \right|^{-1/2},
\label{eq:S2-boundary-Jacobian}
\end{equation}
where $\mathcal N_{\boldsymbol\theta}>0$ is fixed by the rectangle normalization.  Near either side of an endpoint $\theta_r$,
\begin{align}
 \frac{\dd W_{\boldsymbol\theta}}{\dd\theta}
 &=A_r^{\pm}(\theta-\theta_r)^{-1/2}
 +O(|\theta-\theta_r|^{1/2}),
\label{eq:S2-local-derivative}\\
 W_{\boldsymbol\theta}(\theta)-W_{\boldsymbol\theta}(\theta_r)
 &=2A_r^{\pm}(\theta-\theta_r)^{1/2}
 +O(|\theta-\theta_r|^{3/2}),
\label{eq:S2-local-map}
\end{align}
with consistent one-sided square-root branches and nonzero constants $A_r^{\pm}$.  In particular, $f_{\boldsymbol\theta}$ is continuous and $1/2$-H\"older at the corners, while its derivative has an integrable inverse-square-root singularity.  The sharper factorization needed to prove $|(f_{\boldsymbol\theta})_m|=O(|m|^{-3/2})$ is given in Sec.~\ref{sec:S4}.

\subsection{Continuum relative height}
\label{subsec:S2-continuum-height}

Let $\Phi$ be a periodic boundary height of bounded variation.  Its relative conformal zero mode is
\begin{equation}
 \delta[\Phi]
 :=\int_{C_2}\Phi\,\dd\omega_2
 -\int_{C_1}\Phi\,\dd\omega_1.
\label{eq:S2-continuum-zero-mode}
\end{equation}
The sign is chosen so that the associated charge statistic carries the positive coefficient $f_{\boldsymbol\theta}$.  If $\Phi$ is absolutely continuous and
\begin{equation}
 \dd\Phi=2\pi\rho(\theta)\,\dd\theta,
 \qquad \rho\in L^1(\mathbb T),
\label{eq:S2-continuum-height-charge}
\end{equation}
then Eqs.~\eqref{eq:S2-signed-measure} and periodic Stieltjes integration by parts give
\begin{align}
 \delta[\Phi]
 &=-\frac{1}{2\pi}\int_0^{2\pi}\Phi\,\dd f_{\boldsymbol\theta}
 =\frac{1}{2\pi}\int_0^{2\pi}f_{\boldsymbol\theta}\,\dd\Phi
 \notag\\
 &=\int_0^{2\pi}f_{\boldsymbol\theta}(\theta)
 \rho(\theta)\,\dd\theta.
\label{eq:S2-continuum-IBP}
\end{align}
Thus the conformally weighted difference of boundary heights is exactly the linear charge statistic generated by $f_{\boldsymbol\theta}$.

\subsection{Midpoint discretization and exact microscopic height identity}
\label{subsec:S2-lattice-height}

Place the $L$ sites at midpoint angles
\begin{equation}
 \vartheta_j=\frac{2\pi}{L}\left(j+\frac12\right),
 \qquad j\in\mathbb Z_L.
\label{eq:S2-midpoints}
\end{equation}
We call an even size $L$ \emph{admissible} when no $\vartheta_j$ coincides with an endpoint and each of the four induced site sets is nonempty.  Define
\begin{align}
 A_L&=\{j:\vartheta_j\in A\},
 &C_{1,L}&=\{j:\vartheta_j\in C_1\},
 \notag\\
 B_L&=\{j:\vartheta_j\in B\},
 &C_{2,L}&=\{j:\vartheta_j\in C_2\},
\label{eq:S2-lattice-regions}
\end{align}
and $C_L=C_{1,L}\cup C_{2,L}$.  The exact sampled coordinate is
\begin{equation}
 f_{\boldsymbol\theta,L}(j)
 :=f_{\boldsymbol\theta}(\vartheta_j).
\label{eq:S2-sampled-f}
\end{equation}

Introduce the centered occupation
\begin{equation}
 \eta_j:=n_j-\frac12.
\label{eq:S2-centered-charge}
\end{equation}
At half filling, Sec.~\ref{sec:S1} gives
\begin{equation}
 \sum_{j=0}^{L-1}\eta_j=N-\frac{L}{2}=0.
\label{eq:S2-charge-neutrality}
\end{equation}
Choose an arbitrary $\Phi_0\in\mathbb R$ and define the lifted microscopic height by
\begin{equation}
 \Phi_{j+1}-\Phi_j=2\pi\eta_j.
\label{eq:S2-height-increment}
\end{equation}
Equation~\eqref{eq:S2-charge-neutrality} implies
\begin{equation}
 \Phi_L=\Phi_0,
\label{eq:S2-height-periodicity}
\end{equation}
so the lift is periodic around the ring.

The real statistic of the Letter is
\begin{equation}
 X_L
 :=\sum_{j=0}^{L-1}f_{\boldsymbol\theta,L}(j)\eta_j.
\label{eq:S2-real-statistic}
\end{equation}
Define its signed discrete derivative on the oriented edge from $j-1$ to $j$ by
\begin{equation}
 \mu_j
 :=\frac{f_{\boldsymbol\theta,L}(j)-f_{\boldsymbol\theta,L}(j-1)}{2\pi},
 \qquad j\in\mathbb Z_L.
\label{eq:S2-edge-weight}
\end{equation}
Periodic telescoping gives
\begin{equation}
 \sum_{j=0}^{L-1}\mu_j=0.
\label{eq:S2-edge-zero-sum}
\end{equation}
Let
\begin{equation}
 \mathcal E_{1,L}:=\{j:\mu_j>0\},
 \qquad
 \mathcal E_{2,L}:=\{j:\mu_j<0\}.
\label{eq:S2-edge-sets}
\end{equation}
Because the sampled function rises only through the $C_1$ channel and falls only through the $C_2$ channel,
\begin{equation}
 \sum_{j\in\mathcal E_{1,L}}\mu_j=1,
 \qquad
 \sum_{j\in\mathcal E_{2,L}}(-\mu_j)=1.
\label{eq:S2-edge-unit-masses}
\end{equation}
The sets in Eq.~\eqref{eq:S2-edge-sets} are edge-index sets, not simply measured-site sets.  In general one endpoint-crossing edge is indexed by a neighboring unmeasured site; omitting it would destroy the exact unit normalization.

Using Eq.~\eqref{eq:S2-height-increment} and shifting the first sum cyclically,
\begin{align}
 X_L
 &=\frac1{2\pi}\sum_j
 f_{\boldsymbol\theta,L}(j)(\Phi_{j+1}-\Phi_j)
 \notag\\
 &=-\sum_j\mu_j\Phi_j
 \notag\\
 &=\sum_{j\in\mathcal E_{2,L}}(-\mu_j)\Phi_j
 -\sum_{j\in\mathcal E_{1,L}}\mu_j\Phi_j.
\label{eq:S2-discrete-height-identity}
\end{align}
Thus $X_L$ is exactly the difference of two normalized conformal height averages.  A global shift $\Phi_j\mapsto\Phi_j+\Phi_\star$ changes the right-hand side by $-\Phi_\star\sum_j\mu_j=0$, so only height differences enter.

\subsection{Exact compactification from the partial record}
\label{subsec:S2-record-only}

For a complete configuration $X$, let
\begin{equation}
 \rho_{C,L}(X)=\mathbf 1_X|_{C_L}
\label{eq:S2-restriction-map}
\end{equation}
be the measured record, consistently with Sec.~\ref{sec:S1}.  Since $f_{\boldsymbol\theta,L}=0$ on $A_L$ and $f_{\boldsymbol\theta,L}=2\pi$ on $B_L$, Eq.~\eqref{eq:S2-real-statistic} decomposes as
\begin{align}
 X_L(X)
 &=\sum_{j\in C_L}f_{\boldsymbol\theta,L}(j)
 \left(\mathbf1_X(j)-\frac12\right)
 +2\pi N_B(X)-\pi|B_L|,
\label{eq:S2-lift-decomposition}\\
 N_B(X)&:=\sum_{j\in B_L}\mathbf1_X(j)\in\mathbb Z.
\label{eq:S2-NB}
\end{align}
Define the record-dependent real representative
\begin{equation}
 r_L(m_C)
 :=\sum_{j\in C_L}f_{\boldsymbol\theta,L}(j)
 \left(m_C(j)-\frac12\right)-\pi|B_L|
\label{eq:S2-record-representative}
\end{equation}
and the compact variable
\begin{equation}
 \delta_L(m_C):=[r_L(m_C)]_{2\pi}
 \in\mathbb R/(2\pi\mathbb Z)
 .
\label{eq:S2-delta-definition}
\end{equation}
For every complete configuration,
\begin{equation}
 X_L(X)=r_L\!\left(\rho_{C,L}(X)\right)+2\pi N_B(X),
\label{eq:S2-exact-lift-relation}
\end{equation}
and therefore Eq.~\eqref{eq:S2-main-compactification} follows immediately.  The unobserved sites can change $X_L$ only by integer multiples of $2\pi$ through $N_B$, but cannot modify the circle class.

For $q\in\mathbb Z$, define the circle character
\begin{equation}
 \mathfrak e_q([x]_{2\pi})=\ee^{\ii qx}.
\label{eq:S2-circle-character}
\end{equation}
Equation~\eqref{eq:S2-exact-lift-relation} gives the configuration-wise identity
\begin{equation}
 \ee^{\ii qX_L(X)}
 =\mathfrak e_q\!\left(\delta_L(\rho_{C,L}(X))\right).
\label{eq:S2-character-identity}
\end{equation}
Grouping the complete Born average by the fibers of $\rho_{C,L}$ and using the marginal in Eq.~\eqref{eq:S1-partial-marginal}, one obtains the exact pushforward formula
\begin{equation}
 \left\langle\ee^{\ii qX_L}\right\rangle
 =\sum_{m_C\in\{0,1\}^{C_L}}
 \mathbb P_{L,C_L}(m_C)\,
 \mathfrak e_q(\delta_L(m_C)),
 \qquad q\in\mathbb Z.
\label{eq:S2-pushforward-character}
\end{equation}
No thermodynamic limit, postselection, reconstruction of $N_B$, or field-theory assumption enters Eqs.~\eqref{eq:S2-main-compactification}--\eqref{eq:S2-pushforward-character}.  This exact finite-size statement is the microscopic decoder used in the Letter.

\subsection{Convention checks}
\label{subsec:S2-convention-checks}

First, the real statistic is invariant under an additive shift $f_{\boldsymbol\theta,L}(j)\mapsto f_{\boldsymbol\theta,L}(j)+c$ because
\begin{equation}
 X_L\mapsto X_L+c\sum_j\eta_j=X_L.
\label{eq:S2-additive-shift}
\end{equation}
The convention $f=0$ on $A$ and $f=2\pi$ on $B$ is chosen because it makes Eq.~\eqref{eq:S2-lift-decomposition} and partial-record sufficiency manifest.

Second, reversing the rectangle orientation, $v\mapsto\pi-v$, sends
\begin{equation}
 f\mapsto2\pi-f,
 \qquad
 X_L\mapsto-X_L,
 \qquad
 \delta_L\mapsto-\delta_L.
\label{eq:S2-orientation-reversal}
\end{equation}
The limiting law derived in Sec.~\ref{sec:S6} is even, so this convention has no physical effect.

Finally, the coordinate function $f_{\boldsymbol\theta}(\theta)$ changes under a M\"obius reparametrization of the microscopic circle even when $\zeta$ is held fixed.  What is invariant is the Dirichlet energy evaluated in Sec.~\ref{sec:S6}, and hence the limiting compact law.  The decoder is therefore coordinate dependent, as any microscopic weighting must be, whereas its universal distribution depends only on the conformal quadrilateral.

\section{Four exact determinant representations}
\label{sec:S3}

This section derives the exact finite-size formulas used in the Letter.  The derivation is valid for an arbitrary real set of site weights $f_j$ and only at the final step is specialized to the conformal samples
\begin{equation}
 f_j=f_{\boldsymbol\theta,L}(j)
\label{eq:S3-conformal-samples}
\end{equation}
constructed in Sec.~\ref{sec:S2}.  Define the centered linear statistic and its complete-state characteristic function by
\begin{align}
 X_L(f)&:=\sum_{j=0}^{L-1}f_j\left(n_j-\frac12\right),
\label{eq:S3-linear-statistic}\\
 \Phi_L(t;f)&:=\langle\Omega_L|\ee^{\ii tX_L(f)}|\Omega_L\rangle,
 \qquad t\in\mathbb R.
\label{eq:S3-characteristic}
\end{align}
For the conformal function and an integer $q$, the exact compactification established in Eq.~\eqref{eq:S2-pushforward-character} gives
\begin{equation}
 \chi_L(q)
 :=\sum_{m_C}\mathbb P_{L,C_L}(m_C)\,
 \mathfrak e_q(\delta_L(m_C))
 =\Phi_L(q;f_{\boldsymbol\theta,L}),
 \qquad q\in\mathbb Z.
\label{eq:S3-compact-characteristic}
\end{equation}
We now express the same quantity as a determinant on: (i) the full $L$-dimensional site space, (ii) the $N$-dimensional occupied-orbital space, (iii) the measured site space $C_L$, and (iv) a discrete Toeplitz matrix.  A Cauchy--Binet expansion then reproduces the weighted Vandermonde sum over complete configurations.

\subsection{Fermi projector and determinantal occupation process}
\label{subsec:S3-DPP}

Let $U$ be the $L\times N$ occupied-orbital matrix from Sec.~\ref{sec:S1},
\begin{equation}
 U_{jr}=\frac1{\sqrt L}\ee^{\ii k_rj},
 \qquad
 j=0,\ldots,L-1,
 \quad
 r=0,\ldots,N-1,
\label{eq:S3-U}
\end{equation}
with $U^\dagger U=I_N$.  The Fermi projector is
\begin{equation}
 K:=UU^\dagger,
 \qquad
 K_{j\ell}=\frac1L\sum_{r=0}^{N-1}\ee^{\ii k_r(j-\ell)}.
\label{eq:S3-K}
\end{equation}
It is Hermitian, satisfies $K^2=K$, and has rank and trace $N$.

For a finite site set $Y\subset\mathbb Z_L$, write
\begin{equation}
 n_Y:=\prod_{y\in Y}n_y,
 \qquad
 K_Y:=(K_{y y'})_{y,y'\in Y},
\label{eq:S3-nY-KY}
\end{equation}
where site indices in a submatrix are ordered increasingly.  We first prove the correlation identity
\begin{equation}
 \langle n_Y\rangle=\det K_Y
 .
\label{eq:S3-DPP-correlations}
\end{equation}
Let $m=|Y|$.  If $m>N$, both sides vanish: an $N$-particle state cannot occupy all sites in $Y$, while $\operatorname{rank}K_Y\leq N<m$.  Suppose $m\leq N$ and list $Y=\{y_1<\cdots<y_m\}$.  Since the sites are distinct,
\begin{equation}
 n_Y=c_{y_1}^\dagger\cdots c_{y_m}^\dagger
 c_{y_m}\cdots c_{y_1}.
\label{eq:S3-nY-normal-order}
\end{equation}
On the Fermi sea, each site annihilator may be projected onto the occupied orbitals,
\begin{equation}
 c_y|\Omega_L\rangle
 =\sum_{r=0}^{N-1}U_{yr}\,d_{k_r}|\Omega_L\rangle.
\label{eq:S3-c-on-sea}
\end{equation}
Expanding $m$ annihilators and antisymmetrizing the removed orbital indices gives
\begin{equation}
 c_{y_m}\cdots c_{y_1}|\Omega_L\rangle
 =\sum_{\substack{R\subset\{0,\ldots,N-1\}\\|R|=m}}
 \varepsilon(R)\,
 \det (U_Y)_{:,R}\,|\Omega_{R^c}\rangle,
\label{eq:S3-contraction-minors}
\end{equation}
where $U_Y$ is the $m\times N$ row restriction of $U$, $|\Omega_{R^c}\rangle$ is the normalized Slater state obtained by removing the orbitals in $R$, and $\varepsilon(R)$ is an irrelevant sign.  States corresponding to distinct $R$ are orthogonal.  Taking the squared norm and using Cauchy--Binet therefore yields
\begin{align}
 \langle n_Y\rangle
 &=\sum_{|R|=m}|\det (U_Y)_{:,R}|^2
 \notag\\
 &=\det(U_YU_Y^\dagger)
 =\det K_Y,
\label{eq:S3-correlation-proof}
\end{align}
which proves Eq.~\eqref{eq:S3-DPP-correlations}.

Let $g_0,\ldots,g_{L-1}$ be arbitrary complex numbers and set
\begin{equation}
 G:=\operatorname{diag}(g_0,\ldots,g_{L-1}).
\label{eq:S3-G}
\end{equation}
Expanding the commuting product over subsets and using Eq.~\eqref{eq:S3-DPP-correlations},
\begin{align}
 \left\langle\prod_{j=0}^{L-1}(1+g_jn_j)\right\rangle
 &=\sum_{Y\subset\mathbb Z_L}
 \left(\prod_{y\in Y}g_y\right)\det K_Y
 \notag\\
 &=\sum_{Y\subset\mathbb Z_L}\det(GK)_Y.
\label{eq:S3-generating-subsets}
\end{align}
The principal-minor expansion $\det(I_L+M)=\sum_Y\det M_Y$ now gives the exact generating functional
\begin{equation}
 \left\langle\prod_{j=0}^{L-1}(1+g_jn_j)\right\rangle
 =\det(I_L+GK)
 .
\label{eq:S3-generating-functional}
\end{equation}
This identity is valid for arbitrary complex $g_j$ and at every finite size.

\subsection{Full-space and occupied-orbital determinants}
\label{subsec:S3-full-occupied}

Introduce the diagonal one-particle phase matrix
\begin{equation}
 D_t(f):=\operatorname{diag}
 \left(\ee^{\ii tf_0},\ldots,\ee^{\ii tf_{L-1}}\right).
\label{eq:S3-Dt}
\end{equation}
The number operators commute and satisfy $n_j^2=n_j$, hence
\begin{align}
 \ee^{\ii tX_L(f)}
 &=\ee^{-\frac{\ii t}{2}\sum_jf_j}
 \prod_{j=0}^{L-1}\ee^{\ii tf_jn_j}
 \notag\\
 &=\ee^{-\frac{\ii t}{2}\sum_jf_j}
 \prod_{j=0}^{L-1}
 \left[1+(\ee^{\ii tf_j}-1)n_j\right].
\label{eq:S3-exponential-product}
\end{align}
Applying Eq.~\eqref{eq:S3-generating-functional} with $g_j=\ee^{\ii tf_j}-1$ gives the full-space determinant
\begin{equation}
 \Phi_L(t;f)
 =\ee^{-\frac{\ii t}{2}\sum_jf_j}
 \det_L\!\left[I_L+(D_t(f)-I_L)K\right]
 .
\label{eq:S3-full-determinant}
\end{equation}
The equivalent orderings
\begin{equation}
 \det[I_L+(D_t-I_L)K]
 =\det[I_L+K(D_t-I_L)]
\label{eq:S3-ordering-equivalence}
\end{equation}
follow from Sylvester's identity $\det(I+AB)=\det(I+BA)$.

Using $K=UU^\dagger$ and applying the rectangular form of the same identity,
\begin{align}
 &\det_L\!\left[I_L+(D_t-I_L)UU^\dagger\right]
 \notag\\
 &\qquad=\det_N\!\left[I_N+U^\dagger(D_t-I_L)U\right]
 =\det_N(U^\dagger D_tU),
\label{eq:S3-Sylvester-reduction}
\end{align}
where $U^\dagger U=I_N$ was used in the last step.  Thus
\begin{equation}
 \Phi_L(t;f)
 =\ee^{-\frac{\ii t}{2}\sum_jf_j}
 \det_N\!\left[U^\dagger D_t(f)U\right]
 .
\label{eq:S3-occupied-determinant}
\end{equation}

Equation~\eqref{eq:S3-occupied-determinant} also has a direct Slater-overlap interpretation.  The many-body unitary
\begin{equation}
 \widehat D_t(f):=
 \exp\!\left(\ii t\sum_jf_jn_j\right)
\label{eq:S3-many-body-D}
\end{equation}
acts on a site creator as
\begin{equation}
 \widehat D_t(f)c_j^\dagger\widehat D_t(f)^{-1}
 =\ee^{\ii tf_j}c_j^\dagger.
\label{eq:S3-D-conjugation}
\end{equation}
Hence $\widehat D_t(f)|\Omega_L\rangle$ is the Slater determinant with orbital matrix $D_t(f)U$.  The overlap of Slater determinants with orbital matrices $A$ and $B$ is $\det(A^\dagger B)$, which immediately gives
\begin{equation}
 \langle\Omega_L|\widehat D_t(f)|\Omega_L\rangle
 =\det_N(U^\dagger D_t(f)U).
\label{eq:S3-Slater-overlap}
\end{equation}
The external centering phase in Eq.~\eqref{eq:S3-occupied-determinant} simply restores $n_j-1/2$.

\subsection{Exact determinant on the measured sites}
\label{subsec:S3-measured}

We now specialize to the conformal samples in Eq.~\eqref{eq:S3-conformal-samples}.  Order the sites of $C_L=C_{1,L}\cup C_{2,L}$ increasingly and define
\begin{equation}
 K_{C_L}:=(K_{j\ell})_{j,\ell\in C_L},
 \qquad
 D_{q,C_L}:=\operatorname{diag}_{j\in C_L}
 \left(\ee^{\ii qf_j}\right).
\label{eq:S3-restricted-matrices}
\end{equation}
For integer $q$, Eq.~\eqref{eq:S2-delta-definition} gives
\begin{equation}
 \mathfrak e_q(\delta_L(m_C))
 =\ee^{-\ii q\left(\frac12\sum_{j\in C_L}f_j+\pi|B_L|\right)}
 \prod_{j\in C_L}\ee^{\ii qf_jm_C(j)}.
\label{eq:S3-record-character-product}
\end{equation}
Marginalization preserves every inclusion probability on $C_L$:
\begin{equation}
 \sum_{m_C}\mathbb P_{L,C_L}(m_C)
 \prod_{y\in Y}m_C(y)
 =\langle n_Y\rangle
 =\det K_Y,
 \qquad Y\subset C_L.
\label{eq:S3-marginal-correlations}
\end{equation}
Therefore the restricted occupation process is determinantal with kernel $K_{C_L}$, and its generating functional is the restriction of Eq.~\eqref{eq:S3-generating-functional}.  Combining it with Eq.~\eqref{eq:S3-record-character-product} gives
\begin{equation}
 \chi_L(q)
 =\ee^{-\ii q\left(\frac12\sum_{j\in C_L}f_j+\pi|B_L|\right)}
 \det_{C_L}\!\left[
 I_{C_L}+(D_{q,C_L}-I_{C_L})K_{C_L}
 \right]
 ,
 \qquad q\in\mathbb Z.
\label{eq:S3-measured-determinant}
\end{equation}
Here $I_{C_L}$ is the identity on the $|C_L|$-dimensional measured-site space.

The equality with the full determinant is also visible directly.  Since $f_j=0$ on $A_L$ and $f_j=2\pi$ on $B_L$,
\begin{equation}
 \ee^{\ii qf_j}=1,
 \qquad j\in A_L\cup B_L,
 \quad q\in\mathbb Z.
\label{eq:S3-identity-unmeasured}
\end{equation}
Ordering the site basis as $C_L\oplus C_L^c$ gives the block upper-triangular matrix
\begin{equation}
 I_L+(D_q-I_L)K
 =\begin{pmatrix}
 I_{C_L}+(D_{q,C_L}-I_{C_L})K_{C_L}
 & (D_{q,C_L}-I_{C_L})K_{C_L,C_L^c}
 \\
 0&I_{C_L^c}
 \end{pmatrix}.
\label{eq:S3-block-triangular}
\end{equation}
Its determinant is the measured-site determinant.  The centering phases also agree because
\begin{equation}
 \frac12\sum_{j=0}^{L-1}f_j
 =\frac12\sum_{j\in C_L}f_j+\pi|B_L|.
\label{eq:S3-centering-split}
\end{equation}
The restriction to $q\in\mathbb Z$ is essential: only integer characters are single-valued functions of a circle class, and only then does $\ee^{2\pi\ii qN_B}=1$ for every possible unobserved particle number.

\subsection{Discrete Toeplitz representation and midpoint convention}
\label{subsec:S3-Toeplitz}

For the sampled weight $f_j$, define the lattice symbol
\begin{equation}
 a_{t,L}(j):=\ee^{\ii tf_j},
 \qquad j\in\mathbb Z_L,
\label{eq:S3-symbol}
\end{equation}
and its discrete Fourier coefficients
\begin{equation}
 \widehat a_{t,L}(m)
 :=\frac1L\sum_{j=0}^{L-1}
 a_{t,L}(j)\ee^{-2\pi\ii mj/L},
 \qquad m\in\mathbb Z.
\label{eq:S3-discrete-Fourier}
\end{equation}
Since the occupied momenta are consecutive,
\begin{align}
 [U^\dagger D_t(f)U]_{rs}
 &=\frac1L\sum_{j=0}^{L-1}
 \ee^{\ii tf_j}\ee^{2\pi\ii(s-r)j/L}
 \notag\\
 &=\widehat a_{t,L}(r-s),
 \qquad 0\leq r,s<N.
\label{eq:S3-Toeplitz-entry}
\end{align}
The common momentum offset cancels because only $k_s-k_r=2\pi(s-r)/L$ appears.  Define the discrete Toeplitz matrix and determinant
\begin{align}
 T_{N,L}^{\mathrm{disc}}[a]
 &:=\bigl[\widehat a_L(r-s)\bigr]_{r,s=0}^{N-1},
\label{eq:S3-discrete-Toeplitz-matrix}\\
 D_{N,L}^{\mathrm{disc}}[a]
 &:=\det T_{N,L}^{\mathrm{disc}}[a].
\label{eq:S3-discrete-Toeplitz-det}
\end{align}
Equation~\eqref{eq:S3-occupied-determinant} becomes
\begin{equation}
 \Phi_L(t;f)
 =\ee^{-\frac{\ii t}{2}\sum_jf_j}
 D_{N,L}^{\mathrm{disc}}[\ee^{\ii tf}]
 .
\label{eq:S3-exact-Toeplitz}
\end{equation}

The continuum function is sampled at the midpoint angles $\vartheta_j=2\pi(j+\tfrac12)/L$.  The corresponding midpoint Fourier sum is
\begin{equation}
 \widehat a_{t,L}^{\mathrm{mid}}(m)
 :=\frac1L\sum_{j=0}^{L-1}
 a_{t,L}(j)\ee^{-\ii m\vartheta_j}.
\label{eq:S3-midpoint-Fourier}
\end{equation}
It differs from Eq.~\eqref{eq:S3-discrete-Fourier} by the exact phase
\begin{equation}
 \widehat a_{t,L}(m)
 =\ee^{\ii\pi m/L}\widehat a_{t,L}^{\mathrm{mid}}(m).
\label{eq:S3-midpoint-phase}
\end{equation}
Let
\begin{equation}
 (R_L)_{rr}:=\ee^{\ii\pi r/L},
 \qquad r=0,\ldots,N-1.
\label{eq:S3-RL}
\end{equation}
Then
\begin{equation}
 \bigl[\widehat a_{t,L}(r-s)\bigr]
 =R_L\bigl[\widehat a_{t,L}^{\mathrm{mid}}(r-s)\bigr]R_L^{-1}.
\label{eq:S3-midpoint-conjugation}
\end{equation}
Thus the two determinants are identical, although the phase must be retained when comparing matrix entries to continuum Fourier coefficients in Sec.~\ref{sec:S5}.

A constant shift $f_j\mapsto f_j+c$ leaves $X_L(f)$ unchanged at half filling.  In Eq.~\eqref{eq:S3-exact-Toeplitz}, the Toeplitz matrix is multiplied by $\ee^{\ii tc}$, so its determinant gains $\ee^{\ii tcN}$, while the centering factor gains $\ee^{-\ii tcL/2}$.  These factors cancel because $N=L/2$.  Consequently, with
\begin{equation}
 \overline f_L:=\frac1L\sum_jf_j,
 \qquad
 \widetilde f_j:=f_j-\overline f_L,
\label{eq:S3-centered-samples}
\end{equation}
we may equivalently write
\begin{equation}
 \Phi_L(t;f)
 =D_{N,L}^{\mathrm{disc}}[\ee^{\ii t\widetilde f}].
\label{eq:S3-centered-Toeplitz}
\end{equation}
The measured-site reduction must, however, be performed before this shift: $\widetilde f$ is no longer $0$ on $A_L$ and $2\pi$ on $B_L$, even though it defines the same centered statistic.

\subsection{Weighted Vandermonde representation}
\label{subsec:S3-Vandermonde}

The Toeplitz determinant may also be expanded directly over complete occupation configurations.  Let $a_0,\ldots,a_{L-1}$ be arbitrary complex weights and $D_a=\operatorname{diag}(a_0,\ldots,a_{L-1})$.  Cauchy--Binet applied to $U^\dagger D_aU$ gives
\begin{align}
 \det_N(U^\dagger D_aU)
 &=\sum_{\substack{X\subset\mathbb Z_L\\|X|=N}}
 |\det U_X|^2\prod_{x\in X}a_x
 \notag\\
 &=\sum_{|X|=N}p_L(X)\prod_{x\in X}a_x,
\label{eq:S3-Heine}
\end{align}
where Eq.~\eqref{eq:S1-projection-DPP} was used in the second line.  Substituting the explicit Born measure from Sec.~\ref{sec:S1} yields the discrete Heine formula
\begin{align}
 \det_N(U^\dagger D_aU)
 &=\frac1{L^N}
 \sum_{0\leq x_1<\cdots<x_N<L}
 \prod_{a<b}
 \left|\ee^{2\pi\ii x_b/L}-\ee^{2\pi\ii x_a/L}\right|^2
 \prod_{a=1}^{N}a_{x_a}.
\label{eq:S3-discrete-Heine}
\end{align}
For $a_x=\ee^{\ii tf_x}$,
\begin{align}
 \Phi_L(t;f)
 &=\ee^{-\frac{\ii t}{2}\sum_jf_j}
 \sum_{|X|=N}p_L(X)
 \prod_{x\in X}\ee^{\ii tf_x}
 \notag\\
 &=\ee^{-\frac{\ii t}{2}\sum_jf_j}
 \frac1{L^N}
 \sum_{0\leq x_1<\cdots<x_N<L}
 \prod_{a<b}
 \left|\ee^{2\pi\ii x_b/L}-\ee^{2\pi\ii x_a/L}\right|^2
 \prod_{a=1}^{N}\ee^{\ii tf_{x_a}}.
\label{eq:S3-weighted-Vandermonde}
\end{align}
Thus the characteristic function is exactly the partition-function ratio of the same discrete $\beta=2$ circular log gas as in Sec.~\ref{sec:S1}, coupled to the complex one-body weight $\ee^{\ii tf_x}$.  No continuum or saddle-point approximation has been made.

\subsection{Particle--hole symmetry and exact checks}
\label{subsec:S3-checks}

At half filling the occupied and unoccupied momentum blocks differ by $\pi$.  Let
\begin{equation}
 G_{jj}:=(-1)^j.
\label{eq:S3-staggered-gauge}
\end{equation}
If $V$ is the orbital matrix of the unoccupied block, then $V=GU$.  Completeness of the full momentum basis gives
\begin{equation}
 I_L-K=VV^\dagger=GKG^\dagger.
\label{eq:S3-hole-gauge}
\end{equation}
The complement of a determinantal process with kernel $K$ has kernel $I_L-K$.  Since Eq.~\eqref{eq:S3-hole-gauge} is a diagonal gauge transformation, all principal minors of $I_L-K$ equal those of $K$.  Therefore the occupied set $X$ and its complement $X^c$ have the same finite-size distribution.  Under complementation,
\begin{equation}
 X_L(f;X^c)=-X_L(f;X),
\label{eq:S3-statistic-complement}
\end{equation}
so, for every real $t$,
\begin{equation}
 \Phi_L(t;f)=\Phi_L(-t;f)
 =\overline{\Phi_L(t;f)}\in\mathbb R
 .
\label{eq:S3-real-even}
\end{equation}
This symmetry holds for every real test function, not only for the symmetric geometries used in the numerical illustrations.

The exact formulas also imply
\begin{equation}
 \Phi_L(0;f)=1,
 \qquad
 |\Phi_L(t;f)|\leq1,
 \qquad
 \left.\frac{\partial\Phi_L(t;f)}{\partial t}\right|_{t=0}=0.
\label{eq:S3-elementary-checks}
\end{equation}
The last identity follows from $\langle n_j\rangle=K_{jj}=N/L=1/2$.  These relations provide useful analytic and numerical consistency checks.

\subsection{Master equivalence}
\label{subsec:S3-summary}

Combining the preceding results gives, for every real $t$,
\begin{equation}
 \begin{aligned}
 \Phi_L(t;f)
 &=\ee^{-\frac{\ii t}{2}\sum_jf_j}
 \det_L\!\left[I_L+(D_t-I_L)K\right]
 \\
 &=\ee^{-\frac{\ii t}{2}\sum_jf_j}
 \det_N(U^\dagger D_tU)
 \\
 &=\ee^{-\frac{\ii t}{2}\sum_jf_j}
 D_{N,L}^{\mathrm{disc}}[\ee^{\ii tf}]
 \\
 &=\ee^{-\frac{\ii t}{2}\sum_jf_j}
 \sum_{|X|=N}p_L(X)\prod_{x\in X}\ee^{\ii tf_x}.
 \end{aligned}
\label{eq:S3-master-real}
\end{equation}
For the conformal samples and integer $q$, the same quantity is the partial-record Fourier moment and admits the additional measured-space representation
\begin{equation}
 \begin{aligned}
 \chi_L(q)
 &=\Phi_L(q;f_{\boldsymbol\theta,L})
 \\
 &=\ee^{-\ii q\left(\frac12\sum_{j\in C_L}f_j+\pi|B_L|\right)}
 \det_{C_L}\!\left[
 I_{C_L}+(D_{q,C_L}-I_{C_L})K_{C_L}
 \right].
 \end{aligned}
\label{eq:S3-master-measured}
\end{equation}
All identities in Eqs.~\eqref{eq:S3-master-real} and \eqref{eq:S3-master-measured} are exact at finite $L$.  The occupied-orbital/Toeplitz form is the starting point for the asymptotic analysis in Secs.~\ref{sec:S4} and \ref{sec:S5}; the measured-space form makes record-only accessibility manifest; and the Vandermonde form preserves the original Born weights configuration by configuration.

\section{Endpoint Fourier regularity and exact finite-size variance}
\label{sec:S4}

This section supplies the two estimates that make the nonsmooth conformal coordinate accessible to ordinary strong Szeg\H{o} asymptotics.  First, we use the exact Schwarz--Christoffel derivative from Sec.~\ref{sec:S2} to prove
\begin{equation}
 |(f_{\boldsymbol\theta})_m|=O(|m|^{-3/2}).
\label{eq:S4-decay-summary}
\end{equation}
The square-root corners therefore have finite homogeneous $H^{1/2}$ energy even though $f_{\boldsymbol\theta}$ is not $C^1$.  Second, for an arbitrary real lattice weight $u_j$ we derive the exact half-filled variance
\begin{equation}
 \operatorname{Var}X_L(u)
 =\sum_{m=1}^{L-1}d_L(m)|\widehat u_L(m)|^2,
 \qquad d_L(m)=\min(m,L-m).
\label{eq:S4-variance-summary}
\end{equation}
The weight $d_L(m)$ counts particle--hole pairs across the two Fermi points.  Equation~\eqref{eq:S4-variance-summary} is the uniform control used in Sec.~\ref{sec:S5} to remove the Fourier cutoff.

\subsection{Fourier convention and exact boundary derivative}
\label{subsec:S4-Fourier-derivative}

For an integrable $2\pi$-periodic function $g$, we use the continuum Fourier coefficients
\begin{equation}
 g_m:=\frac{1}{2\pi}\int_0^{2\pi}
 g(\theta)\ee^{-\ii m\theta}\,\dd\theta,
 \qquad m\in\bbZ.
\label{eq:S4-continuum-Fourier}
\end{equation}
The conformal coordinate $f_{\boldsymbol\theta}=2\,\operatorname{Im}W_{\boldsymbol\theta}$ is constant on $A$ and $B$, increases from $0$ to $2\pi$ along $C_1$, and decreases from $2\pi$ to $0$ along $C_2$.  Combining this orientation with the boundary Jacobian in Eq.~\eqref{eq:S2-boundary-Jacobian} gives, away from the four endpoints,
\begin{equation}
 f_{\boldsymbol\theta}'(\theta)
 =\begin{cases}
 \displaystyle
 2\mathcal N_{\boldsymbol\theta}
 \left|
 \prod_{s=1}^{4}
 \sin\!\left(\frac{\theta-\theta_s}{2}\right)
 \right|^{-1/2},
 &\theta\in C_1,\\[4mm]
 \displaystyle
 -2\mathcal N_{\boldsymbol\theta}
 \left|
 \prod_{s=1}^{4}
 \sin\!\left(\frac{\theta-\theta_s}{2}\right)
 \right|^{-1/2},
 &\theta\in C_2,\\[4mm]
 0,&\theta\in A\cup B.
 \end{cases}
\label{eq:S4-fprime-explicit}
\end{equation}
The signs are fixed by the oriented boundary coordinate and not by a choice of square-root branch.

The endpoint analysis must retain more information than the big-$O$ expansion in Eq.~\eqref{eq:S2-local-derivative}.  For each endpoint $\theta_r$, let $\sigma_r\in\{+1,-1\}$ point from $\theta_r$ into the adjacent measured arc.  In the present cyclic convention,
\begin{equation}
 (\sigma_1,\sigma_2,\sigma_3,\sigma_4)=(-1,+1,-1,+1).
\label{eq:S4-sigma-values}
\end{equation}
When a neighborhood crosses $0\equiv2\pi$, the following local coordinate is understood periodically:
\begin{equation}
 \theta=\theta_r+\sigma_r x\pmod{2\pi},
 \qquad 0<x<\varepsilon_r.
\label{eq:S4-local-coordinate}
\end{equation}
Let $\tau_r=+1$ when the adjacent measured arc is $C_1$ and $\tau_r=-1$ when it is $C_2$; explicitly,
\begin{equation}
 (\tau_1,\tau_2,\tau_3,\tau_4)=(-1,+1,+1,-1).
\label{eq:S4-tau-values}
\end{equation}
Equation~\eqref{eq:S4-fprime-explicit} then factorizes exactly as
\begin{equation}
 f_{\boldsymbol\theta}'(\theta_r+\sigma_rx)
 =\tau_r x^{-1/2}B_r(x),
\label{eq:S4-exact-endpoint-factorization}
\end{equation}
where
\begin{align}
 B_r(x)
 &:=%
 2\sqrt{2}\,\mathcal N_{\boldsymbol\theta}
 \left[
 \frac{x/2}{|\sin(x/2)|}
 \right]^{1/2}
 \notag\\
 &\quad\times
 \prod_{\substack{s=1\\s\neq r}}^{4}
 \left|
 \sin\!\left(
 \frac{\theta_r+\sigma_rx-\theta_s}{2}
 \right)
 \right|^{-1/2}.
\label{eq:S4-Br}
\end{align}
For sufficiently small $\varepsilon_r$, none of the three factors with $s\neq r$ vanishes.  Moreover,
\begin{equation}
 \frac{x/2}{\sin(x/2)}=1+O(x^2),
 \qquad x\to0,
\label{eq:S4-sinc-expansion}
\end{equation}
so $B_r$ extends to a strictly positive $C^\infty$ function on $[0,\varepsilon_r]$.  In particular,
\begin{equation}
 f_{\boldsymbol\theta}'(\theta_r+\sigma_rx)
 =\tau_rB_r(0)x^{-1/2}+O(x^{1/2}).
\label{eq:S4-local-fprime}
\end{equation}
The inverse-square-root singularity is integrable.  Since the derivative is smooth away from the endpoints, $f_{\boldsymbol\theta}$ is periodic and absolutely continuous.

\subsection{Square-root oscillatory estimate}
\label{subsec:S4-oscillatory}

We isolate the elementary estimate needed for the Fourier transform of Eq.~\eqref{eq:S4-exact-endpoint-factorization}.

\paragraph{Lemma.}
Let $b\in C^1([0,\varepsilon])$ and suppose that $b$ vanishes in a neighborhood of $\varepsilon$.  There is a finite constant $C_b$ such that, for every integer $m\geq1$,
\begin{equation}
 \left|
 \int_0^{\varepsilon}x^{-1/2}b(x)\ee^{-\ii mx}\,\dd x
 \right|
 \leq C_bm^{-1/2}.
\label{eq:S4-square-root-lemma}
\end{equation}

\paragraph{Proof.}
If $m^{-1}\geq\varepsilon$, direct integration gives
\begin{equation}
 \left|
 \int_0^{\varepsilon}x^{-1/2}b(x)\ee^{-\ii mx}\,\dd x
 \right|
 \leq2\sqrt{\varepsilon}\,\|b\|_\infty
 \leq2\|b\|_\infty m^{-1/2}.
\label{eq:S4-small-m-bound}
\end{equation}
Now assume $m^{-1}<\varepsilon$ and split the integral at $x=m^{-1}$.  On the first interval,
\begin{equation}
 \left|
 \int_0^{1/m}x^{-1/2}b(x)\ee^{-\ii mx}\,\dd x
 \right|
 \leq2\|b\|_\infty m^{-1/2}.
\label{eq:S4-small-x-bound}
\end{equation}
On $[m^{-1},\varepsilon]$, set $h(x)=x^{-1/2}b(x)$.  Because $b$ vanishes near $\varepsilon$, integration by parts yields
\begin{equation}
 \int_{1/m}^{\varepsilon}h(x)\ee^{-\ii mx}\,\dd x
 =\frac{h(1/m)\ee^{-\ii}}{\ii m}
 +\frac{1}{\ii m}
 \int_{1/m}^{\varepsilon}h'(x)\ee^{-\ii mx}\,\dd x.
\label{eq:S4-IBP}
\end{equation}
The boundary term is at most $\|b\|_\infty m^{-1/2}$.  Furthermore,
\begin{equation}
 |h'(x)|
 \leq\frac12\|b\|_\infty x^{-3/2}
 +\|b'\|_\infty x^{-1/2},
\label{eq:S4-hprime-bound}
\end{equation}
so
\begin{equation}
 \frac1m\int_{1/m}^{\varepsilon}|h'(x)|\,\dd x
 \leq\|b\|_\infty m^{-1/2}
 +2\sqrt{\varepsilon}\,\|b'\|_\infty m^{-1}.
\label{eq:S4-hprime-integral}
\end{equation}
Combining Eqs.~\eqref{eq:S4-small-m-bound}--\eqref{eq:S4-hprime-integral} proves Eq.~\eqref{eq:S4-square-root-lemma}.  \hfill$\square$

\subsection{Fourier decay and finite conformal energy}
\label{subsec:S4-Fourier-decay}

Choose smooth one-sided cutoffs $\chi_r$ supported in the measured neighborhoods of Eq.~\eqref{eq:S4-local-coordinate}, equal to one sufficiently close to $x=0$, and vanishing near $x=\varepsilon_r$.  The localized singular pieces of $f_{\boldsymbol\theta}'$ have the form
\begin{equation}
 \mathbf1_{x>0}x^{-1/2}b_r(x),
 \qquad
 b_r(x):=\tau_r\chi_r(x)B_r(x)\in C^1([0,\varepsilon_r]).
\label{eq:S4-derivative-decomposition}
\end{equation}
After subtracting these four terms, the remainder is periodic, piecewise $C^1$, and supported away from the endpoints.  The localized terms have Fourier transforms $O(|m|^{-1/2})$ by Eq.~\eqref{eq:S4-square-root-lemma}; ordinary integration by parts on the finitely many smooth intervals gives $O(|m|^{-1})$ for the remainder.  Consequently,
\begin{equation}
 \left|
 \frac{1}{2\pi}\int_0^{2\pi}
 f_{\boldsymbol\theta}'(\theta)\ee^{-\ii m\theta}\,\dd\theta
 \right|
 \leq C_{\boldsymbol\theta}'|m|^{-1/2},
 \qquad m\neq0.
\label{eq:S4-fprime-Fourier-bound}
\end{equation}
Since $f_{\boldsymbol\theta}$ is periodic and absolutely continuous, integration by parts gives
\begin{align}
 \frac{1}{2\pi}\int_0^{2\pi}
 f_{\boldsymbol\theta}'(\theta)\ee^{-\ii m\theta}\,\dd\theta
 &=\frac{1}{2\pi}
 \left[f_{\boldsymbol\theta}(\theta)\ee^{-\ii m\theta}\right]_0^{2\pi}
 +\ii m(f_{\boldsymbol\theta})_m
 \notag\\
 &=\ii m(f_{\boldsymbol\theta})_m.
\label{eq:S4-derivative-Fourier-relation}
\end{align}
Therefore there is a finite constant $C_{\boldsymbol\theta}$ such that
\begin{equation}
 |(f_{\boldsymbol\theta})_m|
 \leq\frac{C_{\boldsymbol\theta}}{(1+|m|)^{3/2}},
 \qquad m\in\bbZ
 .
\label{eq:S4-Fourier-decay}
\end{equation}
Reality gives $(f_{\boldsymbol\theta})_{-m}=\overline{(f_{\boldsymbol\theta})_m}$, so the same bound holds for both signs of $m$; the zero mode is absorbed into the constant.  In particular, the quadratic form appearing in the Letter is finite:
\begin{equation}
 \mathcal E[f_{\boldsymbol\theta}]
 :=\sum_{m=1}^{\infty}m|(f_{\boldsymbol\theta})_m|^2
 <\infty
 .
\label{eq:S4-finite-energy}
\end{equation}
Indeed,
\begin{equation}
 \sum_{m=1}^{\infty}m|(f_{\boldsymbol\theta})_m|^2
 \leq C_{\boldsymbol\theta}^2
 \sum_{m=1}^{\infty}m^{-2}.
\label{eq:S4-energy-bound}
\end{equation}
Thus the conformal coordinate belongs to the natural homogeneous $H^{1/2}$ energy class.  The stronger pointwise decay in Eq.~\eqref{eq:S4-Fourier-decay}, rather than only the finiteness of Eq.~\eqref{eq:S4-finite-energy}, will control midpoint aliases uniformly when the Fourier cutoff is removed in Sec.~\ref{sec:S5}.

\subsection{Projection formula for the finite-size variance}
\label{subsec:S4-projection-variance}

Let $u_0,\ldots,u_{L-1}$ be arbitrary real numbers and define
\begin{equation}
 \mathsf U:=\operatorname{diag}(u_0,\ldots,u_{L-1}),
 \qquad
 X_L(u):=\sum_{j=0}^{L-1}u_j\left(n_j-\frac12\right).
\label{eq:S4-u-statistic}
\end{equation}
At half filling, translational invariance gives $K_{jj}=N/L=1/2$, and hence
\begin{equation}
 \langle X_L(u)\rangle=0.
\label{eq:S4-u-mean-zero}
\end{equation}
For $j\neq\ell$, the determinantal correlation formula in Eq.~\eqref{eq:S3-DPP-correlations} gives
\begin{equation}
 \langle n_jn_\ell\rangle
 =K_{jj}K_{\ell\ell}-K_{j\ell}K_{\ell j},
\label{eq:S4-two-point-offdiag}
\end{equation}
while $n_j^2=n_j$.  Both cases combine into
\begin{equation}
 \operatorname{Cov}(n_j,n_\ell)
 =\delta_{j\ell}K_{jj}-K_{j\ell}K_{\ell j}.
\label{eq:S4-number-covariance}
\end{equation}
Therefore
\begin{align}
 \operatorname{Var}X_L(u)
 &=\operatorname{Tr}(K\mathsf U^2)
 -\operatorname{Tr}(K\mathsf UK\mathsf U)
 \notag\\
 &=
 \operatorname{Tr}\!\left[K\mathsf U(I_L-K)\mathsf U\right]
 .
\label{eq:S4-projection-variance}
\end{align}
The positivity is manifest in Hilbert--Schmidt form.  With
\begin{equation}
 A:=(I_L-K)\mathsf UK,
\label{eq:S4-A-definition}
\end{equation}
cyclicity of the trace and $K^2=K$ give
\begin{equation}
 \operatorname{Var}X_L(u)
 =\operatorname{Tr}(A^\dagger A)
 =\left\|(I_L-K)\mathsf UK\right\|_{\mathrm{HS}}^2.
\label{eq:S4-HS-variance}
\end{equation}
Thus fluctuations are generated precisely by matrix elements of the multiplication operator $\mathsf U$ that connect occupied and empty one-particle states.

\subsection{Exact momentum-space pair count}
\label{subsec:S4-pair-count}

Define the lattice Fourier coefficients of the real weight $u$ by
\begin{equation}
 \widehat u_L(m)
 :=\frac1L\sum_{j=0}^{L-1}
 u_j\ee^{-2\pi\ii mj/L},
 \qquad m\in\bbZ.
\label{eq:S4-discrete-Fourier}
\end{equation}
They are $L$-periodic and satisfy
\begin{equation}
 \widehat u_L(L-m)=\overline{\widehat u_L(m)}.
\label{eq:S4-real-DFT-symmetry}
\end{equation}
In the momentum basis of Sec.~\ref{sec:S1}, the Fermi projector is diagonal.  We label the consecutive occupied and empty orbital indices by
\begin{equation}
 \mathcal I_{\mathrm{occ}}:=\{0,1,\ldots,N-1\},
 \qquad
 \mathcal I_{\mathrm{emp}}:=\{N,N+1,\ldots,L-1\}.
\label{eq:S4-index-blocks}
\end{equation}
The common momentum offset cancels from the matrix element
\begin{align}
 \langle k_r|\mathsf U|k_s\rangle
 &=\frac1L\sum_{j=0}^{L-1}
 u_j\ee^{\ii(k_s-k_r)j}
 \notag\\
 &=\widehat u_L(r-s).
\label{eq:S4-U-momentum-element}
\end{align}
Evaluating Eq.~\eqref{eq:S4-HS-variance} in this basis gives
\begin{equation}
 \operatorname{Var}X_L(u)
 =\sum_{r\in\mathcal I_{\mathrm{occ}}}
 \sum_{s\in\mathcal I_{\mathrm{emp}}}
 |\widehat u_L(r-s)|^2.
\label{eq:S4-occ-empty-sum}
\end{equation}
For a nonzero residue $m\in\{1,\ldots,L-1\}$, let
\begin{equation}
 \nu_L(m)
 :=\#\left\{r\in\mathcal I_{\mathrm{occ}}:
 r-m\pmod L\in\mathcal I_{\mathrm{emp}}
 \right\}.
\label{eq:S4-pair-count-definition}
\end{equation}
If $1\leq m\leq N$, the indices $r=0,\ldots,m-1$ wrap across the left Fermi point into the empty block, so $\nu_L(m)=m$.  If $N\leq m\leq L-1$, writing $p=L-m$ shows that the allowed occupied indices are $r=N-p,\ldots,N-1$, whose number is $p=L-m$.  At $m=N$ the two descriptions agree.  Hence
\begin{equation}
 \nu_L(m)=d_L(m):=\min(m,L-m).
\label{eq:S4-pair-count}
\end{equation}
Grouping Eq.~\eqref{eq:S4-occ-empty-sum} by the residue $m$ proves the exact identity
\begin{equation}
 \operatorname{Var}X_L(u)
 =\sum_{m=1}^{L-1}
 d_L(m)|\widehat u_L(m)|^2,
 \qquad
 d_L(m)=\min(m,L-m)
 .
\label{eq:S4-exact-momentum-variance}
\end{equation}
The zero Fourier mode is absent, consistently with invariance under $u_j\mapsto u_j+c$.  For $m\ll L$, the coefficient is $d_L(m)=m$, which is the lattice origin of the continuum $H^{1/2}$ weight.  The mode $L-m$ represents the corresponding negative continuum frequency and carries the same coefficient.

\subsection{Characteristic-function stability}
\label{subsec:S4-characteristic-stability}

The exact variance controls changes of the many-body characteristic function without using an operator-norm estimate on the extensive statistic.  For two real lattice weights $u$ and $v$, the elementary inequality $|\ee^{\ii x}-\ee^{\ii y}|\leq|x-y|$ and Cauchy--Schwarz imply
\begin{align}
 \left|
 \left\langle\ee^{\ii tX_L(u)}\right\rangle
 -\left\langle\ee^{\ii tX_L(v)}\right\rangle
 \right|
 &\leq |t|\,\left\langle|X_L(u-v)|\right\rangle
 \notag\\
 &\leq |t|\sqrt{\operatorname{Var}X_L(u-v)}.
\label{eq:S4-characteristic-stability}
\end{align}
In the last step we used Eq.~\eqref{eq:S4-u-mean-zero} for the difference weight.  Combining Eqs.~\eqref{eq:S4-exact-momentum-variance} and \eqref{eq:S4-characteristic-stability} gives
\begin{equation}
 \left|
 \left\langle\ee^{\ii tX_L(u)}\right\rangle
 -\left\langle\ee^{\ii tX_L(v)}\right\rangle
 \right|
 \leq |t|
 \left[
 \sum_{m=1}^{L-1}d_L(m)
 |\widehat{(u-v)}_L(m)|^2
 \right]^{1/2}.
\label{eq:S4-characteristic-DFT-bound}
\end{equation}
Section~\ref{sec:S5} applies this estimate with $u=f_{\boldsymbol\theta,L}$ and $v$ equal to a finite Fourier truncation sampled at the same midpoints.  The pointwise decay in Eq.~\eqref{eq:S4-Fourier-decay} then yields a variance tail of order $M^{-1}+L^{-1}$, allowing the thermodynamic limit to be taken before the cutoff is removed.

\section{Discrete-to-continuous Toeplitz limit and strong Szeg\H{o} asymptotics}
\label{sec:S5}

This section completes the probabilistic asymptotics used in the Letter.  We first prove the limit for a general real periodic function $g$ whose Fourier coefficients obey the decay established for $f_{\boldsymbol\theta}$ in Sec.~\ref{sec:S4}.  The proof keeps the order of limits explicit:
\begin{equation}
 \text{first }L\to\infty\text{ at fixed Fourier cutoff }M,
 \qquad
 \text{then }M\to\infty.
\label{eq:S5-order-of-limits}
\end{equation}
At fixed $M$, the symbol is analytic, midpoint aliases are exponentially small, and the ordinary complex strong Szeg\H{o} theorem applies.  The exact variance formula of Sec.~\ref{sec:S4} then removes the cutoff uniformly in $L$.  No Fisher--Hartwig theorem is needed.

\subsection{Statement of the asymptotic theorem}
\label{subsec:S5-main-theorem}

Let $g$ be a real, continuous, $2\pi$-periodic function, with Fourier coefficients
\begin{equation}
 g_m:=\frac{1}{2\pi}\int_0^{2\pi}
 g(\theta)\ee^{-\ii m\theta}\,\dd\theta,
 \qquad m\in\bbZ.
\label{eq:S5-continuous-Fourier}
\end{equation}
We assume that, for some finite $C_g$,
\begin{equation}
 |g_m|\leq\frac{C_g}{(1+|m|)^{3/2}},
 \qquad m\in\bbZ.
\label{eq:S5-decay-assumption}
\end{equation}
The midpoint sample and the corresponding centered occupation statistic are
\begin{equation}
 g_L(j):=g(\vartheta_j),
 \qquad
 \vartheta_j=\frac{2\pi}{L}\left(j+\frac12\right),
 \qquad
 X_L(g):=\sum_{j=0}^{L-1}g(\vartheta_j)
 \left(n_j-\frac12\right).
\label{eq:S5-midpoint-statistic}
\end{equation}
Define the finite quadratic form
\begin{equation}
 \mathcal E[g]
 :=\sum_{m=1}^{\infty}m|g_m|^2.
\label{eq:S5-energy-definition}
\end{equation}
Finiteness follows immediately from Eq.~\eqref{eq:S5-decay-assumption}.

We shall prove that, for every fixed $t\in\bbR$,
\begin{equation}
 \lim_{L\to\infty}
 \left\langle\ee^{\ii tX_L(g)}\right\rangle
 =\exp[-t^2\mathcal E[g]]
 .
\label{eq:S5-general-Gaussian-limit}
\end{equation}
Thus $X_L(g)$ converges in distribution to a centered Gaussian of variance $2\mathcal E[g]$.  Section~\ref{sec:S4} proves Eq.~\eqref{eq:S5-decay-assumption} for $g=f_{\boldsymbol\theta}$; Sec.~\ref{sec:S6} will evaluate the resulting energy geometrically.

\subsection{Continuous Toeplitz matrices and Fourier truncation}
\label{subsec:S5-Toeplitz-notation}

For an integrable symbol $a$ on the circle, write
\begin{equation}
 a_m:=\frac{1}{2\pi}\int_0^{2\pi}
 a(\theta)\ee^{-\ii m\theta}\,\dd\theta
\label{eq:S5-symbol-Fourier}
\end{equation}
and define the continuous Toeplitz matrix and determinant
\begin{equation}
 T_N[a]:=[a_{r-s}]_{r,s=0}^{N-1},
 \qquad
 D_N[a]:=\det T_N[a].
\label{eq:S5-continuous-Toeplitz}
\end{equation}
For midpoint samples of the same function, use the discrete coefficient already appearing in Sec.~\ref{sec:S3},
\begin{equation}
 \widehat a_L(m)
 :=\frac1L\sum_{j=0}^{L-1}
 a(\vartheta_j)\ee^{-2\pi\ii mj/L},
\label{eq:S5-sampled-Fourier}
\end{equation}
and
\begin{equation}
 T_{N,L}^{\mathrm{disc}}[a]
 :=[\widehat a_L(r-s)]_{r,s=0}^{N-1},
 \qquad
 D_{N,L}^{\mathrm{disc}}[a]
 :=\det T_{N,L}^{\mathrm{disc}}[a].
\label{eq:S5-discrete-Toeplitz}
\end{equation}
The exact determinant representation from Eq.~\eqref{eq:S3-exact-Toeplitz} reads
\begin{equation}
 \left\langle\ee^{\ii tX_L(g)}\right\rangle
 =\exp\!\left[-\frac{\ii t}{2}
 \sum_{j=0}^{L-1}g(\vartheta_j)\right]
 D_{N,L}^{\mathrm{disc}}[\ee^{\ii tg}],
 \qquad N=\frac L2.
\label{eq:S5-exact-characteristic-Toeplitz}
\end{equation}

For $M\geq0$, truncate the Fourier series of $g$ symmetrically:
\begin{equation}
 g^{(M)}(\theta)
 :=\sum_{|m|\leq M}g_m\ee^{\ii m\theta}.
\label{eq:S5-Fourier-truncation}
\end{equation}
Because $g_{-m}=\overline{g_m}$, the function $g^{(M)}$ is real.  For fixed $t\in\bbR$, set
\begin{equation}
 a_M(\theta):=\ee^{\ii t g^{(M)}(\theta)},
 \qquad
 V_M(\theta):=\ii t g^{(M)}(\theta).
\label{eq:S5-truncated-symbol}
\end{equation}
The symbol has unit modulus on the real circle and a globally defined smooth logarithm $V_M$; in particular, its winding number is zero.

\subsection{Exact midpoint aliasing}
\label{subsec:S5-aliasing}

Suppose that $a(\theta)=\sum_{k\in\bbZ}a_k\ee^{\ii k\theta}$ with $\sum_k|a_k|<\infty$.  Inserting this series into Eq.~\eqref{eq:S5-sampled-Fourier} gives
\begin{align}
 \widehat a_L(m)
 &=\sum_{k\in\bbZ}a_k\ee^{\ii\pi k/L}
 \left[\frac1L\sum_{j=0}^{L-1}
 \ee^{2\pi\ii(k-m)j/L}\right].
\label{eq:S5-aliasing-first-step}
\end{align}
The bracket equals one if $k-m$ is divisible by $L$ and vanishes otherwise.  Writing $k=m+\ell L$ yields the exact midpoint-aliasing identity
\begin{equation}
 \widehat a_L(m)
 =\ee^{\ii\pi m/L}
 \sum_{\ell\in\bbZ}(-1)^\ell a_{m+\ell L}
 .
\label{eq:S5-midpoint-aliasing}
\end{equation}
The prefactor is precisely the midpoint phase identified in Eq.~\eqref{eq:S3-midpoint-phase}.

For every fixed $M$ and $t$, the function $a_M$ extends analytically to the complex $\theta$ plane and is bounded on each closed strip $|\operatorname{Im}\theta|\leq\sigma$.  Shifting the Fourier contour downward for positive index and upward for negative index gives constants $C_{M,t}<\infty$ and $\sigma_{M,t}>0$ such that
\begin{equation}
 |(a_M)_k|
 \leq C_{M,t}\ee^{-\sigma_{M,t}|k|},
 \qquad k\in\bbZ.
\label{eq:S5-analytic-Fourier-decay}
\end{equation}
The constants may depend on the fixed cutoff and on $t$; no uniformity in $M$ is asserted or needed.

\subsection{Discrete-to-continuous comparison at fixed cutoff}
\label{subsec:S5-discrete-continuous}

Let $R_L$ be the diagonal unitary matrix from Eq.~\eqref{eq:S3-RL}.  Applying Eq.~\eqref{eq:S5-midpoint-aliasing} entry by entry gives
\begin{equation}
 T_{N,L}^{\mathrm{disc}}[a_M]
 =R_L\bigl(T_N[a_M]+E_{N,L}^{(M)}\bigr)R_L^{-1},
\label{eq:S5-matrix-decomposition}
\end{equation}
where
\begin{equation}
 [E_{N,L}^{(M)}]_{rs}
 :=\sum_{\ell\neq0}(-1)^\ell
 (a_M)_{r-s+\ell L}.
\label{eq:S5-alias-error}
\end{equation}
Since $N=L/2$ and $0\leq r,s<N$,
\begin{equation}
 |r-s|\leq\frac L2-1,
 \qquad
 |r-s+\ell L|
 \geq\left(|\ell|-\frac12\right)L
 \quad(\ell\neq0).
\label{eq:S5-alias-distance}
\end{equation}
The exponential decay in Eq.~\eqref{eq:S5-analytic-Fourier-decay} therefore implies
\begin{equation}
 |[E_{N,L}^{(M)}]_{rs}|
 \leq C_{M,t}'\ee^{-\sigma_{M,t}L/2}
\label{eq:S5-alias-entry-bound}
\end{equation}
and hence
\begin{equation}
 \|E_{N,L}^{(M)}\|_{\mathrm{op}}
 \leq\|E_{N,L}^{(M)}\|_{\mathrm F}
 \leq NC_{M,t}'\ee^{-\sigma_{M,t}L/2}.
\label{eq:S5-alias-operator-bound}
\end{equation}

We next record a determinant perturbation bound adapted to the present unitary symbols.  If $A$ and $B$ are $N\times N$ contractions, then
\begin{equation}
 |\det A-\det B|
 \leq N\|A-B\|_{\mathrm{op}}.
\label{eq:S5-determinant-Lipschitz}
\end{equation}
Indeed, replace the columns of $B$ by those of $A$ one at a time.  Multilinearity expresses the determinant difference as a sum of $N$ determinants, each containing one column of $A-B$ and $N-1$ columns of norm at most one.  Hadamard's inequality bounds each term by $\|A-B\|_{\mathrm{op}}$.

Both matrices to which we apply Eq.~\eqref{eq:S5-determinant-Lipschitz} are contractions.  The discrete matrix is the occupied-orbital compression
\begin{equation}
 T_{N,L}^{\mathrm{disc}}[a_M]
 =U^\dagger D_{t,L}^{(M)}U,
 \qquad
 D_{t,L}^{(M)}
 :=\operatorname{diag}\!\left(
 \ee^{\ii t g^{(M)}(\vartheta_0)},\ldots,
 \ee^{\ii t g^{(M)}(\vartheta_{L-1})}\right),
\label{eq:S5-discrete-compression}
\end{equation}
whose norm is at most one.  The continuous matrix $T_N[a_M]$ is the compression of multiplication by $a_M$ to the span of $1,z,\ldots,z^{N-1}$ in $L^2(\bbT)$, so
\begin{equation}
 \|T_N[a_M]\|_{\mathrm{op}}
 \leq\|a_M\|_\infty=1.
\label{eq:S5-continuous-contraction}
\end{equation}
Equation~\eqref{eq:S5-matrix-decomposition} shows that $T_N[a_M]+E_{N,L}^{(M)}$ is unitarily equivalent to the discrete contraction.  Applying Eq.~\eqref{eq:S5-determinant-Lipschitz} and then Eq.~\eqref{eq:S5-alias-operator-bound}, we obtain
\begin{equation}
 \left|
 D_{N,L}^{\mathrm{disc}}[a_M]-D_N[a_M]
 \right|
 \leq N^2C_{M,t}'\ee^{-\sigma_{M,t}L/2}
 \xrightarrow[L\to\infty]{}0.
\label{eq:S5-discrete-continuous-determinant}
\end{equation}
This proves the discrete-to-continuous reduction for each fixed cutoff.

\subsection{Strong Szeg\H{o} asymptotics and centering}
\label{subsec:S5-Szego}

We use the smooth complex form of the strong Szeg\H{o} theorem~\cite{DeiftItsKrasovsky2013}.  If $V$ is a smooth complex-valued function and $\ee^V$ has the globally defined logarithm $V$, then
\begin{equation}
 D_N[\ee^V]
 =\exp\!\left[
 NV_0+\sum_{m=1}^{\infty}mV_mV_{-m}
 \right]\,[1+o(1)]
 \qquad(N\to\infty).
\label{eq:S5-strong-Szego}
\end{equation}
All hypotheses are automatic for the finite Fourier polynomial $V_M=\ii t g^{(M)}$.  Its coefficients are
\begin{equation}
 (V_M)_m=
 \begin{cases}
 \ii t g_m,& |m|\leq M,\\
 0,& |m|>M,
 \end{cases}
\label{eq:S5-VM-coefficients}
\end{equation}
and reality of $g$ gives
\begin{equation}
 (V_M)_m(V_M)_{-m}
 =-t^2|g_m|^2.
\label{eq:S5-Szego-quadratic-term}
\end{equation}
Therefore
\begin{equation}
 D_N[a_M]
 =\exp\!\left[
 \ii tNg_0
 -t^2\sum_{m=1}^{M}m|g_m|^2
 \right]\,[1+o(1)].
\label{eq:S5-continuous-Szego-result}
\end{equation}

The extensive phase cancels exactly against the centering factor in the many-body statistic.  If $L>M$, midpoint quadrature is exact for the trigonometric polynomial $g^{(M)}$:
\begin{align}
 \frac1L\sum_{j=0}^{L-1}g^{(M)}(\vartheta_j)
 &=\sum_{|m|\leq M}g_m\ee^{\ii\pi m/L}
 \left[\frac1L\sum_{j=0}^{L-1}
 \ee^{2\pi\ii mj/L}\right]
 =g_0.
\label{eq:S5-midpoint-mean}
\end{align}
Since $N=L/2$,
\begin{equation}
 \exp\!\left[-\frac{\ii t}{2}
 \sum_{j=0}^{L-1}g^{(M)}(\vartheta_j)\right]
 =\ee^{-\ii tNg_0}.
\label{eq:S5-centering-cancellation}
\end{equation}
Combining Eqs.~\eqref{eq:S5-exact-characteristic-Toeplitz},
\eqref{eq:S5-discrete-continuous-determinant},
\eqref{eq:S5-continuous-Szego-result}, and
\eqref{eq:S5-centering-cancellation} yields, for every fixed $M$ and real $t$,
\begin{equation}
 \lim_{L\to\infty}
 \left\langle\ee^{\ii tX_L(g^{(M)})}\right\rangle
 =\exp\!\left[-t^2
 \sum_{m=1}^{M}m|g_m|^2\right]
 .
\label{eq:S5-fixed-cutoff-limit}
\end{equation}
The comparison in Eq.~\eqref{eq:S5-discrete-continuous-determinant} is additive.  This is sufficient because the right-hand side of Eq.~\eqref{eq:S5-continuous-Szego-result} has a nonzero limiting magnitude at every fixed $M$ and $t$.

\subsection{Uniform removal of the Fourier cutoff}
\label{subsec:S5-cutoff-removal}

Define the Fourier tail
\begin{equation}
 r^{(M)}(\theta)
 :=g(\theta)-g^{(M)}(\theta)
 =\sum_{|k|>M}g_k\ee^{\ii k\theta}.
\label{eq:S5-tail}
\end{equation}
The series is absolutely convergent by Eq.~\eqref{eq:S5-decay-assumption}, so the midpoint-aliasing identity applies.  For $m\in\{1,\ldots,L-1\}$, let
\begin{equation}
 d_L(m):=\min(m,L-m).
\label{eq:S5-dLm-recall}
\end{equation}
The residue class $m+L\bbZ$ has one representative of minimal absolute value $d_L(m)$, except at $m=L/2$, where the two representatives $\pm L/2$ have the same minimum.  If $d_L(m)\leq M$, these nearest representatives are absent from the tail.  Separating the at most two nearest representatives and summing the remaining $|k|^{-3/2}$ tail gives a constant $C_g'$ independent of $L,M,m$ such that
\begin{equation}
 \left|\widehat r_L^{(M)}(m)\right|
 \leq C_g'
 \left[
 \mathbf1_{\{d_L(m)>M\}}d_L(m)^{-3/2}
 +L^{-3/2}
 \right].
\label{eq:S5-tail-DFT-bound}
\end{equation}
For example, after the nearest representatives are removed, all remaining ones have magnitude at least $(p-\tfrac12)L$ for some $p\geq1$, and
\begin{equation}
 \sum_{p=1}^{\infty}
 [(p-\tfrac12)L]^{-3/2}
 \leq C L^{-3/2}.
\label{eq:S5-far-alias-sum}
\end{equation}

Using $(a+b)^2\leq2a^2+2b^2$ in Eq.~\eqref{eq:S5-tail-DFT-bound}, followed by the exact variance identity Eq.~\eqref{eq:S4-exact-momentum-variance}, gives
\begin{align}
 \operatorname{Var}X_L(r^{(M)})
 &\leq C
 \sum_{m=1}^{L-1}d_L(m)
 \left[
 \mathbf1_{\{d_L(m)>M\}}d_L(m)^{-3}
 +L^{-3}
 \right]
 \notag\\
 &\leq C_1
 \sum_{d=M+1}^{\lfloor L/2\rfloor}d^{-2}
 +C_2L^{-3}
 \sum_{d=1}^{\lfloor L/2\rfloor}d
 \notag\\
 &\leq\frac{C_3}{M}+\frac{C_4}{L}.
\label{eq:S5-tail-variance}
\end{align}
At most two residues have a prescribed value of $d_L(m)$; the constants absorb this multiplicity.

Applying the characteristic-function stability estimate Eq.~\eqref{eq:S4-characteristic-stability} with $u=g_L$ and $v=g_L^{(M)}$ yields
\begin{align}
 &\left|
 \left\langle\ee^{\ii tX_L(g)}\right\rangle
 -\left\langle\ee^{\ii tX_L(g^{(M)})}\right\rangle
 \right|
 \notag\\
 &\qquad\leq |t|
 \sqrt{\operatorname{Var}X_L(r^{(M)})}
 \leq |t|\left(\frac{C_3}{M}+\frac{C_4}{L}\right)^{1/2}.
\label{eq:S5-characteristic-tail}
\end{align}
Consequently,
\begin{equation}
 \lim_{M\to\infty}\limsup_{L\to\infty}
 \left|
 \left\langle\ee^{\ii tX_L(g)}\right\rangle
 -\left\langle\ee^{\ii tX_L(g^{(M)})}\right\rangle
 \right|=0.
\label{eq:S5-uniform-cutoff-removal}
\end{equation}
The nonnegative partial energies also satisfy
\begin{equation}
 \sum_{m=1}^{M}m|g_m|^2
 \xrightarrow[M\to\infty]{}\mathcal E[g].
\label{eq:S5-energy-partial-limit}
\end{equation}

\subsection{Completion of the Gaussian limit}
\label{subsec:S5-completion}

For every $L$ and $M$, the triangle inequality gives
\begin{align}
 &\left|
 \left\langle\ee^{\ii tX_L(g)}\right\rangle
 -\ee^{-t^2\mathcal E[g]}
 \right|
 \notag\\
 &\leq
 \left|
 \left\langle\ee^{\ii tX_L(g)}\right\rangle
 -\left\langle\ee^{\ii tX_L(g^{(M)})}\right\rangle
 \right|
 \notag\\
 &\quad+
 \left|
 \left\langle\ee^{\ii tX_L(g^{(M)})}\right\rangle
 -\exp\!\left[-t^2\sum_{m=1}^{M}m|g_m|^2\right]
 \right|
 \notag\\
 &\quad+
 \left|
 \exp\!\left[-t^2\sum_{m=1}^{M}m|g_m|^2\right]
 -\ee^{-t^2\mathcal E[g]}
 \right|.
\label{eq:S5-three-errors}
\end{align}
First let $L\to\infty$ at fixed $M$, using Eq.~\eqref{eq:S5-fixed-cutoff-limit}; then let $M\to\infty$, using Eqs.~\eqref{eq:S5-uniform-cutoff-removal} and \eqref{eq:S5-energy-partial-limit}.  All three terms vanish, proving Eq.~\eqref{eq:S5-general-Gaussian-limit}.  Since the limiting characteristic function is continuous at $t=0$, L\'evy's continuity theorem gives
\begin{equation}
 X_L(g)\ \Longrightarrow\
 \mathsf N\!\left(0,2\mathcal E[g]\right).
\label{eq:S5-real-Gaussian-convergence}
\end{equation}

For the conformal function, Eq.~\eqref{eq:S4-Fourier-decay} verifies the hypothesis.  Therefore, along the admissible sequence of even sizes defined in Sec.~\ref{sec:S2},
\begin{equation}
 \lim_{L\to\infty}
 \left\langle
 \ee^{\ii tX_L(f_{\boldsymbol\theta})}
 \right\rangle
 =\exp\!\left[-t^2
 \mathcal E[f_{\boldsymbol\theta}]\right]
 ,
 \qquad t\in\bbR.
\label{eq:S5-conformal-characteristic-limit}
\end{equation}
For integer $t=q\in\bbZ$, the left-hand side is exactly the $q$th Fourier moment of the compact partial-record variable by Eq.~\eqref{eq:S3-master-measured}.  For noninteger $t$, it is the characteristic function of the complete real statistic, not a single-valued character of the circle variable.  The exact particle--hole symmetry in Sec.~\ref{sec:S3} makes the finite-size characteristic function real and even, consistently with the limiting Gaussian.  Section~\ref{sec:S6} now proves
\begin{equation}
 \mathcal E[f_{\boldsymbol\theta}]
 =h\!\left(\zeta(\boldsymbol\theta)\right)
\label{eq:S5-energy-to-modulus-preview}
\end{equation}
and converts the integer moments into weak convergence of the compact law.

\section{Dirichlet energy and compact weak convergence}
\label{sec:S6}

Section~\ref{sec:S5} reduced the thermodynamic problem to the homogeneous Fourier energy
\begin{equation}
 \mathcal E[f]
 :=\sum_{m=1}^{\infty}m|f_m|^2
\label{eq:S6-Fourier-energy}
\end{equation}
and proved, for the conformal boundary coordinate of Sec.~\ref{sec:S2},
\begin{equation}
 \lim_{L\to\infty}
 \left\langle\ee^{\ii tX_L(f_{\boldsymbol\theta})}\right\rangle
 =\exp\!\left[-t^2\mathcal E[f_{\boldsymbol\theta}]\right],
 \qquad t\in\bbR.
\label{eq:S6-input-characteristic}
\end{equation}
The purpose of this section is twofold.  First, we evaluate the quadratic form exactly from the marked conformal geometry,
\begin{equation}
 \mathcal E[f_{\boldsymbol\theta}]
 =h\!\left(\zeta(\boldsymbol\theta)\right)
 ,
\label{eq:S6-energy-equals-modulus}
\end{equation}
where $h(\zeta)$ is the rectangle width in Eq.~\eqref{eq:S2-rectangle-width}.  Second, we convert the integer values of Eq.~\eqref{eq:S6-input-characteristic} into weak convergence of the compact partial-record law to the heat kernel on the circle.

\subsection{Poisson extension and disk Dirichlet energy}
\label{subsec:S6-Poisson-energy}

For a real function $U$ on a planar domain $\Omega$, define the Dirichlet energy
\begin{equation}
 \mathcal D_{\Omega}[U]
 :=\int_{\Omega}|\nabla U|^2\,\dd^2x.
\label{eq:S6-Dirichlet-definition}
\end{equation}
Let $f$ be a real continuous $2\pi$-periodic function with Fourier coefficients
\begin{equation}
 f_m=\frac{1}{2\pi}\int_0^{2\pi}
 f(\theta)\ee^{-\ii m\theta}\,\dd\theta
\label{eq:S6-Fourier-coefficients}
\end{equation}
and finite energy $\mathcal E[f]<\infty$.  Its Poisson extension to the unit disk $\mathbb D$ is
\begin{equation}
 H_f(r,\theta)
 :=\sum_{m\in\bbZ}f_m r^{|m|}\ee^{\ii m\theta}
 =\frac{1}{2\pi}\int_0^{2\pi}
 P_r(\theta-\varphi)f(\varphi)\,\dd\varphi,
\label{eq:S6-Poisson-extension}
\end{equation}
where
\begin{equation}
 P_r(t)=\frac{1-r^2}{1-2r\cos t+r^2}
 =\sum_{m\in\bbZ}r^{|m|}\ee^{\ii mt}.
\label{eq:S6-Poisson-kernel}
\end{equation}
The assumption $\mathcal E[f]<\infty$ implies absolute convergence on every closed subdisk.  Since $P_r/(2\pi)$ is an approximate identity and $f$ is uniformly continuous on the circle,
\begin{equation}
 \lim_{r\uparrow1}\sup_{\theta}
 |H_f(r,\theta)-f(\theta)|=0.
\label{eq:S6-Poisson-boundary-limit}
\end{equation}
Thus $H_f$ is the harmonic function in $\mathbb D$ that extends continuously to the boundary value $f$.

In polar coordinates,
\begin{equation}
 \mathcal D_{\mathbb D}[H_f]
 =\int_0^1r\,\dd r\int_0^{2\pi}\dd\theta
 \left(
 |\partial_rH_f|^2+
 \frac{1}{r^2}|\partial_\theta H_f|^2
 \right).
\label{eq:S6-disk-energy-polar}
\end{equation}
Termwise differentiation and Parseval's identity give, for every $r<1$,
\begin{align}
 \int_0^{2\pi}|\partial_rH_f|^2\,\dd\theta
 &=2\pi\sum_{m\neq0}m^2|f_m|^2r^{2|m|-2},
\label{eq:S6-radial-Parseval}\\
 \int_0^{2\pi}\frac{|\partial_\theta H_f|^2}{r^2}\,\dd\theta
 &=2\pi\sum_{m\neq0}m^2|f_m|^2r^{2|m|-2}.
\label{eq:S6-angular-Parseval}
\end{align}
Tonelli's theorem applies to the nonnegative summands.  Integrating over $r$ therefore yields
\begin{align}
 \mathcal D_{\mathbb D}[H_f]
 &=4\pi\sum_{m\neq0}m^2|f_m|^2
 \int_0^1r^{2|m|-1}\,\dd r
 \notag\\
 &=2\pi\sum_{m\neq0}|m||f_m|^2
 =4\pi\sum_{m=1}^{\infty}m|f_m|^2.
\label{eq:S6-Fourier-Dirichlet}
\end{align}
Hence
\begin{equation}
 \mathcal D_{\mathbb D}[H_f]=4\pi\mathcal E[f]
 .
\label{eq:S6-energy-normalization}
\end{equation}
The zero mode does not enter either side, consistently with the exact invariance of the lattice statistic under $f\mapsto f+c$ at half filling.

For completeness, the same quadratic form admits the real-space representation
\begin{equation}
 \mathcal E[f]
 =\frac{1}{8\pi^2}
 \int_0^{2\pi}\!\dd\theta
 \int_0^{2\pi}\!\dd\varphi\,
 \frac{|f(\theta)-f(\varphi)|^2}
 {4\sin^2[(\theta-\varphi)/2]}.
\label{eq:S6-Hhalf-real-space}
\end{equation}
Indeed, Parseval's identity for the translated difference gives
\begin{equation}
 \int_0^{2\pi}|f(\varphi+t)-f(\varphi)|^2\,\dd\varphi
 =2\pi\sum_{m\in\bbZ}|f_m|^2|\ee^{\ii mt}-1|^2.
\label{eq:S6-translation-Parseval}
\end{equation}
Using Tonelli's theorem and
\begin{equation}
 \int_0^{2\pi}
 \frac{|\ee^{\ii mt}-1|^2}{4\sin^2(t/2)}\,\dd t
 =2\pi|m|
\label{eq:S6-Fejer-integral}
\end{equation}
proves Eq.~\eqref{eq:S6-Hhalf-real-space}.  The kernel is singular on the diagonal; finiteness is precisely the homogeneous $H^{1/2}$ condition established for $f_{\boldsymbol\theta}$ in Sec.~\ref{sec:S4}.

\subsection{Conformal invariance and evaluation on the rectangle}
\label{subsec:S6-conformal-evaluation}

Let
\begin{equation}
 W_{\boldsymbol\theta}:\mathbb D\longrightarrow\mathcal R_h,
 \qquad
 \mathcal R_h=\{u+\ii v:0<u<h,\ 0<v<\pi\},
\label{eq:S6-map-recall}
\end{equation}
be the normalized conformal bijection of Sec.~\ref{sec:S2}, with
\begin{equation}
 h=h_{\boldsymbol\theta}
 :=h\!\left(\zeta(\boldsymbol\theta)\right).
\label{eq:S6-h-theta}
\end{equation}
The rectangle is a Jordan domain, so the conformal map extends continuously to the boundary.  Define
\begin{equation}
 F_{\boldsymbol\theta}(z)
 :=2\,\operatorname{Im}W_{\boldsymbol\theta}(z).
\label{eq:S6-bulk-coordinate}
\end{equation}
Because the imaginary part of a holomorphic function is harmonic, $F_{\boldsymbol\theta}$ is harmonic in $\mathbb D$.  Its boundary trace is exactly the function introduced in Eq.~\eqref{eq:S2-boundary-f},
\begin{equation}
 F_{\boldsymbol\theta}(\ee^{\ii\theta})
 =f_{\boldsymbol\theta}(\theta).
\label{eq:S6-boundary-trace}
\end{equation}
Both $F_{\boldsymbol\theta}$ and $H_{f_{\boldsymbol\theta}}$ are harmonic, continuous on the closed disk, and have the same boundary values.  The maximum principle therefore gives
\begin{equation}
 F_{\boldsymbol\theta}=H_{f_{\boldsymbol\theta}}
 \qquad\text{in }\mathbb D.
\label{eq:S6-harmonic-identification}
\end{equation}

We next recall the conformal invariance of the two-dimensional Dirichlet energy.  Let $W:\Omega\to\Omega'$ be a conformal bijection and let $U$ be a real $C^1$ function on $\Omega'$ with finite energy.  At every interior point the derivative has the form
\begin{equation}
 DW=\lambda R,
\label{eq:S6-conformal-derivative}
\end{equation}
where $\lambda>0$ and $R$ is a rotation.  Consequently,
\begin{equation}
 |\nabla(U\circ W)|^2
 =\lambda^2|\nabla U|^2\circ W,
 \qquad
 J_W=\lambda^2.
\label{eq:S6-gradient-Jacobian}
\end{equation}
The change-of-variables formula then gives
\begin{equation}
 \mathcal D_{\Omega}[U\circ W]
 =\mathcal D_{\Omega'}[U]
 .
\label{eq:S6-conformal-invariance}
\end{equation}
Boundary corners have zero area and do not affect this identity.

On the rectangle, the function corresponding to Eq.~\eqref{eq:S6-bulk-coordinate} is simply
\begin{equation}
 V(u,v)=2v.
\label{eq:S6-rectangle-coordinate}
\end{equation}
It satisfies
\begin{equation}
 \nabla V=(0,2),
 \qquad
 |\nabla V|^2=4.
\label{eq:S6-rectangle-gradient}
\end{equation}
Since the area of $\mathcal R_h$ is $\pi h$,
\begin{equation}
 \mathcal D_{\mathcal R_h}[2v]
 =4\pi h.
\label{eq:S6-rectangle-energy}
\end{equation}
Using Eqs.~\eqref{eq:S6-energy-normalization},
\eqref{eq:S6-harmonic-identification},
\eqref{eq:S6-conformal-invariance}, and
\eqref{eq:S6-rectangle-energy}, we obtain the exact chain
\begin{equation}
 4\pi\mathcal E[f_{\boldsymbol\theta}]
 =\mathcal D_{\mathbb D}[H_{f_{\boldsymbol\theta}}]
 =\mathcal D_{\mathcal R_h}[2v]
 =4\pi h_{\boldsymbol\theta}
 .
\label{eq:S6-energy-chain}
\end{equation}
This proves Eq.~\eqref{eq:S6-energy-equals-modulus}.  The coordinate function depends on the complete marked quadruple $\boldsymbol\theta$, but its energy depends only on the conformal cross ratio through the modulus $h(\zeta)$.

The normalization can also be checked without fixing the rectangle height.  On a rectangle $0<u<a$, $0<v<b$, the function with boundary range $0$ to $2\pi$ is $V_{a,b}=2\pi v/b$, and
\begin{equation}
 \mathcal D[V_{a,b}]
 =4\pi^2\frac{a}{b}.
\label{eq:S6-general-rectangle-energy}
\end{equation}
Rescaling the rectangle to height $\pi$ gives $h=\pi a/b$, so Eq.~\eqref{eq:S6-general-rectangle-energy} again becomes $4\pi h$.

\subsection{Real Gaussian limit}
\label{subsec:S6-real-Gaussian}

Substituting Eq.~\eqref{eq:S6-energy-equals-modulus} into the asymptotic result of Sec.~\ref{sec:S5} gives, for every fixed real $t$,
\begin{equation}
 \lim_{L\to\infty}
 \left\langle
 \ee^{\ii tX_L(f_{\boldsymbol\theta})}
 \right\rangle
 =\ee^{-h_{\boldsymbol\theta}t^2}
 .
\label{eq:S6-real-characteristic-final}
\end{equation}
The limiting function is the characteristic function of a centered normal random variable of variance $2h_{\boldsymbol\theta}$.  L\'evy's continuity theorem therefore implies
\begin{equation}
 X_L(f_{\boldsymbol\theta})
 \ \Longrightarrow\ 
 \mathsf N(0,2h_{\boldsymbol\theta})
 \qquad\text{on }\bbR
 .
\label{eq:S6-real-Gaussian-limit}
\end{equation}
For integer $t=q$, Eq.~\eqref{eq:S6-real-characteristic-final} is also the limiting $q$th Fourier moment of the compact variable accessible from the partial record.  For noninteger $t$, it concerns the complete-configuration statistic $X_L$ and is not an intrinsic circle character.

\subsection{Finite-size law on the circle}
\label{subsec:S6-circle-law}

Write
\begin{equation}
 \bbT:=\bbR/(2\pi\bbZ)
\label{eq:S6-circle}
\end{equation}
for the circle of circumference $2\pi$.  The finite-size compact law is the atomic pushforward measure
\begin{equation}
 \mu_L
 :=(\delta_L)_{\#}\mathbb P_{L,C_L}
 =\sum_{m_C\in\{0,1\}^{C_L}}
 \mathbb P_{L,C_L}(m_C)\,
 \delta^{\mathrm D}_{\delta_L(m_C)},
\label{eq:S6-muL}
\end{equation}
where $\delta_x^{\mathrm D}$ denotes the unit Dirac measure at $x\in\bbT$.  For the circle characters defined in Eq.~\eqref{eq:S2-circle-character},
\begin{align}
 \widehat\mu_L(q)
 &:=\int_{\bbT}\mathfrak e_q(\delta)\,\dd\mu_L(\delta)
 \notag\\
 &=\sum_{m_C}\mathbb P_{L,C_L}(m_C)
 \mathfrak e_q(\delta_L(m_C))
 =\left\langle\ee^{\ii qX_L(f_{\boldsymbol\theta})}\right\rangle,
 \qquad q\in\bbZ,
\label{eq:S6-muL-Fourier}
\end{align}
where the final equality is the exact finite-size pushforward identity of Secs.~\ref{sec:S2} and \ref{sec:S3}.  Equation~\eqref{eq:S6-real-characteristic-final} therefore gives
\begin{equation}
 \widehat\mu_L(q)
 \longrightarrow
 \ee^{-h_{\boldsymbol\theta}q^2},
 \qquad q\in\bbZ.
\label{eq:S6-Fourier-limit}
\end{equation}

For $h>0$, define the $2\pi$-periodic density
\begin{equation}
 P_h(\delta)
 :=\frac{1}{\sqrt{4\pi h}}
 \sum_{\ell\in\bbZ}
 \exp\!\left[-\frac{(\delta+2\pi\ell)^2}{4h}\right],
 \qquad 0\leq\delta<2\pi.
\label{eq:S6-wrapped-Gaussian}
\end{equation}
The defining series converges uniformly on $[0,2\pi]$.  Unfolding the real line gives normalization,
\begin{equation}
 \int_0^{2\pi}P_h(\delta)\,\dd\delta=1,
\label{eq:S6-wrapped-normalization}
\end{equation}
and, for $q\in\bbZ$,
\begin{align}
 \int_0^{2\pi}P_h(\delta)\ee^{\ii q\delta}\,\dd\delta
 &=\frac{1}{\sqrt{4\pi h}}
 \int_{-\infty}^{\infty}
 \ee^{-x^2/(4h)}\ee^{\ii qx}\,\dd x
 \notag\\
 &=\ee^{-hq^2}.
\label{eq:S6-wrapped-Fourier}
\end{align}
Thus $P_h(\delta)\,\dd\delta$ is the unique probability measure on $\bbT$ with Fourier coefficients $\ee^{-hq^2}$.

We now use the standard Fourier criterion for weak convergence on the circle.  Suppose $\nu_L$ and $\nu$ are probability measures on $\bbT$ and
\begin{equation}
 \int_{\bbT}\mathfrak e_q\,\dd\nu_L
 \longrightarrow
 \int_{\bbT}\mathfrak e_q\,\dd\nu
 \qquad\text{for every }q\in\bbZ.
\label{eq:S6-Fourier-criterion-assumption}
\end{equation}
For a trigonometric polynomial $p(\delta)=\sum_{|q|\leq M}c_q\mathfrak e_q(\delta)$, Eq.~\eqref{eq:S6-Fourier-criterion-assumption} gives convergence of $\int p\,\dd\nu_L$.  Given any $F\in C(\bbT)$, choose $p$ with $\|F-p\|_\infty<\varepsilon$.  Since both measures have unit mass,
\begin{align}
 \limsup_{L\to\infty}
 \left|\int F\,\dd\nu_L-\int F\,\dd\nu\right|
 \leq2\varepsilon.
\label{eq:S6-Fourier-criterion-proof}
\end{align}
Letting $\varepsilon\downarrow0$ proves $\nu_L\Longrightarrow\nu$.

Applying this criterion to Eqs.~\eqref{eq:S6-Fourier-limit} and \eqref{eq:S6-wrapped-Fourier} gives the compact theorem
\begin{equation}
 \mu_L
 \ \Longrightarrow\ 
 P_{h_{\boldsymbol\theta}}(\delta)\,\dd\delta
 \qquad\text{on }\bbT
 .
\label{eq:S6-compact-weak-limit}
\end{equation}
Equivalently, for every $F\in C(\bbT)$,
\begin{align}
 &\lim_{L\to\infty}
 \sum_{m_C\in\{0,1\}^{C_L}}
 \mathbb P_{L,C_L}(m_C)
 F(\delta_L(m_C))
 \notag\\
 &\qquad=
 \int_0^{2\pi}F(\delta)
 P_{h_{\boldsymbol\theta}}(\delta)\,\dd\delta.
\label{eq:S6-compact-weak-form}
\end{align}
No postselection is involved: the left-hand side averages over all partial records with their exact Born probabilities.

\subsection{Heat-kernel form and the role of the complete-configuration lift}
\label{subsec:S6-heat-kernel-lift}

The wrapped Gaussian has the absolutely and uniformly convergent Fourier series
\begin{equation}
 P_h(\delta)
 =\frac{1}{2\pi}
 \sum_{q\in\bbZ}\ee^{-hq^2}\ee^{-\ii q\delta}
 =\frac{1}{2\pi}
 \left[1+2\sum_{q=1}^{\infty}
 \ee^{-hq^2}\cos(q\delta)\right]
 .
\label{eq:S6-heat-kernel-series}
\end{equation}
It follows either from Eq.~\eqref{eq:S6-wrapped-Fourier} and uniqueness of finite measures with prescribed Fourier coefficients, or directly from Poisson summation.  Termwise differentiation gives
\begin{equation}
 \partial_hP_h(\delta)=\partial_\delta^2P_h(\delta),
\label{eq:S6-heat-equation}
\end{equation}
so $P_h$ is the heat kernel on a circle at time $h$ for the generator $\partial_\delta^2$.  It interpolates between a point mass at the circle origin as $h\downarrow0$ and the uniform density $1/(2\pi)$ as $h\to\infty$.

The second route to Eq.~\eqref{eq:S6-compact-weak-limit} is to apply the continuous mapping theorem to the quotient map $\pi_{\bbT}(x)=[x]_{2\pi}$.  Equation~\eqref{eq:S6-real-Gaussian-limit} implies
\begin{equation}
 \pi_{\bbT}(X_L)
 \Longrightarrow
 \pi_{\bbT}(G),
 \qquad
 G\sim\mathsf N(0,2h_{\boldsymbol\theta}),
\label{eq:S6-wrapping-continuous-map}
\end{equation}
and periodizing the Gaussian density gives exactly Eq.~\eqref{eq:S6-wrapped-Gaussian}.

At finite size, the distinction between the complete-configuration lift $X_L$ and the compact observable is exact.  With the representative $r_L(m_C)$ and the unmeasured occupation number $N_B(X)$ defined in Sec.~\ref{sec:S2},
\begin{equation}
 X_L(X)
 =r_L(\rho_{C,L}(X))+2\pi N_B(X).
\label{eq:S6-lift-decomposition}
\end{equation}
The partial record fixes $r_L$ and hence the class
\begin{equation}
 [X_L(X)]_{2\pi}
 =\delta_L(\rho_{C,L}(X)),
\label{eq:S6-record-class}
\end{equation}
but it need not fix the integer $N_B$ that labels which $2\pi$ lift is realized by $X_L$.  This is why the intrinsic observables of the compact variable are precisely the integer characters $\mathfrak e_q$, while noninteger exponentials probe additional lift information.  The microscopic coordinate relevant to the field-theory limit is therefore genuinely compact already at finite lattice size.

Combining the exact decoder of Sec.~\ref{sec:S2}, the determinant identities of Sec.~\ref{sec:S3}, the regularity and variance bounds of Sec.~\ref{sec:S4}, the Toeplitz limit of Sec.~\ref{sec:S5}, and the geometric evaluation above gives the complete chain
\begin{equation}
 m_C\ \longmapsto\ \delta_L(m_C),
 \qquad
 \left\langle\mathfrak e_q(\delta_L)\right\rangle
 \longrightarrow\ee^{-h(\zeta)q^2},
 \qquad
 \delta_L\Longrightarrow P_{h(\zeta)}(\delta)\,\dd\delta
 .
\label{eq:S6-final-chain}
\end{equation}

\section{Exact finite-size numerics}
\label{sec:S7}

This section verifies the finite-size identities derived in Secs.~\ref{sec:S2} and \ref{sec:S3}, and the asymptotic prediction of Secs.~\ref{sec:S5} and \ref{sec:S6}.  The calculation uses the exact midpoint samples
\begin{equation}
 f_j=f_{\boldsymbol\theta}(\vartheta_j),
 \qquad
 \vartheta_j=\frac{2\pi}{L}\left(j+\frac12\right),
\label{eq:S7-exact-samples}
\end{equation}
of the conformal boundary coordinate defined in Eq.~\eqref{eq:S2-boundary-f}.  In particular, we do not reconstruct the coordinate from discretized Jacobian weights.  The finite-size determinants therefore evaluate the same observable that appears in the theorem.

We perform three complementary tests.  First, at small size we enumerate every complete occupation configuration and compare the complete Born sum, the grouped partial-record sum, and all three determinant sizes.  Second, at larger sizes we evaluate the first three integer harmonics and compare the effective width with the conformal modulus.  Third, we replace the conformal side measure by uniform angular measure on the measured arcs.  The latter statistic still has a Gaussian limit, but its analytically calculable energy differs from the conformal modulus.  No Monte Carlo sampling or fitted parameter is used.

\subsection{Commensurate symmetric geometries}
\label{subsec:S7-geometries}

For the numerical scan we use the symmetric endpoint family
\begin{equation}
 \theta_1=0,
 \qquad
 \theta_2=\alpha,
 \qquad
 \theta_3=\pi,
 \qquad
 \theta_4=\pi+\alpha,
 \qquad
 \alpha=2\pi r,
 \quad 0<r<\frac12.
\label{eq:S7-symmetric-endpoints}
\end{equation}
For a commensurate size satisfying
\begin{equation}
 a:=rL\in\bbZ,
\label{eq:S7-commensurability}
\end{equation}
the midpoint partition of Eq.~\eqref{eq:S2-lattice-regions} is
\begin{equation}
 \begin{aligned}
 A_L&=\{0,\ldots,a-1\},
 &
 C_{1,L}&=\{a,\ldots,L/2-1\},
 \\
 B_L&=\{L/2,\ldots,L/2+a-1\},
 &
 C_{2,L}&=\{L/2+a,\ldots,L-1\}.
 \end{aligned}
\label{eq:S7-symmetric-regions}
\end{equation}
The cross ratio and conformal width are
\begin{equation}
 \zeta(r)=\sin^2(\pi r),
 \qquad
 h(r)=\pi\,
 \frac{K\!\left(\sin^2(\pi r)\right)}
      {K\!\left(\cos^2(\pi r)\right)}
 .
\label{eq:S7-symmetric-width}
\end{equation}
Here and throughout the Supplement, $K(m)$ uses the elliptic-parameter convention of Eq.~\eqref{eq:S2-elliptic-K}.  The numerical scan uses
\begin{equation}
 r\in\left\{\frac1{10},\frac16,\frac14,\frac13\right\},
 \qquad
 L\in\{120,240,360,480,600,720\},
\label{eq:S7-scan}
\end{equation}
retaining only commensurate pairs.  Thus the continuum geometry is held fixed exactly as $L$ changes.

\begin{table}[t]
\caption{Geometric parameters and the continuum energy of the uniform-measure control derived in Eq.~\eqref{eq:S7-uniform-energy}.}
\label{tab:S7-geometries}
\begin{ruledtabular}
\begin{tabular}{cccc}
$r$ & $\zeta(r)$ & $h(r)$ & $\mathcal E_{\rm unif}(r)$ \\
\hline
$1/10$ & $0.095491503$ & $1.946054521$ & $2.332699890$ \\
$1/6$  & $0.25$        & $2.455785997$ & $2.770809592$ \\
$1/4$  & $0.50$        & $3.141592654$ & $3.410227191$ \\
$1/3$  & $0.75$        & $4.018918754$ & $4.262783988$
\end{tabular}
\end{ruledtabular}
\end{table}

\subsection{Exact conformal samples}
\label{subsec:S7-conformal-samples}

For the geometry in Eq.~\eqref{eq:S7-symmetric-endpoints}, define
\begin{equation}
 \rho_r(\theta)
 :=
 \left|
 \prod_{s=1}^{4}
 \sin\!\left(\frac{\theta-\theta_s}{2}\right)
 \right|^{-1/2}.
\label{eq:S7-rho}
\end{equation}
Equation~\eqref{eq:S2-boundary-Jacobian} implies that on the first measured arc
\begin{equation}
 \frac{\dd f_{\boldsymbol\theta}}{\dd\theta}
 =2\mathcal N_{\boldsymbol\theta}\rho_r(\theta),
 \qquad
 \alpha<\theta<\pi.
\label{eq:S7-exact-derivative}
\end{equation}
The single global normalization constant can be eliminated by introducing
\begin{equation}
 J_r:=\int_{\alpha}^{\pi}\rho_r(s)\,\dd s.
\label{eq:S7-Jr}
\end{equation}
The exact boundary coordinate is then
\begin{equation}
 f_{\boldsymbol\theta}(\theta)=
 \begin{cases}
 0,
 &0<\theta<\alpha,
 \\[1mm]
 \displaystyle
 \frac{2\pi}{J_r}
 \int_{\alpha}^{\theta}\rho_r(s)\,\dd s,
 &\alpha<\theta<\pi,
 \\[4mm]
 2\pi-f_{\boldsymbol\theta}(\theta-\pi),
 &\pi<\theta<2\pi.
 \end{cases}
\label{eq:S7-exact-f}
\end{equation}
The last line follows from the half-turn symmetry
\begin{equation}
 f_{\boldsymbol\theta}(\theta+\pi)
 =2\pi-f_{\boldsymbol\theta}(\theta).
\label{eq:S7-half-turn}
\end{equation}
It gives $f=2\pi$ on $B$ and the required monotone decrease on $C_2$.

The integral in Eq.~\eqref{eq:S7-Jr} has inverse-square-root singularities at both endpoints.  They are removed by
\begin{equation}
 s=\alpha+(\pi-\alpha)\sin^2\varphi,
 \qquad
 0\leq\varphi\leq\frac\pi2,
\label{eq:S7-regularizing-substitution}
\end{equation}
which gives
\begin{align}
 J_r
 &=\int_0^{\pi/2}
 2(\pi-\alpha)\sin\varphi\cos\varphi
 \notag\\[-1mm]
 &\hspace{18mm}\times
 \rho_r\!\left(
 \alpha+(\pi-\alpha)\sin^2\varphi
 \right)\dd\varphi.
\label{eq:S7-Jr-regularized}
\end{align}
The transformed integrand has finite one-sided limits at both endpoints.  For a partial integral ending at $\theta\in(\alpha,\pi)$, the upper limit is
\begin{equation}
 \varphi(\theta)
 =\arcsin\sqrt{\frac{\theta-\alpha}{\pi-\alpha}}.
\label{eq:S7-partial-upper-limit}
\end{equation}
Adaptive quadrature of Eqs.~\eqref{eq:S7-Jr-regularized} and \eqref{eq:S7-partial-upper-limit}, followed by Eq.~\eqref{eq:S7-half-turn}, produces the samples in Eq.~\eqref{eq:S7-exact-samples}.  The symmetry is imposed algebraically for the second half of the circle, so independent quadrature noise is not introduced on $C_2$.  At every commensurate even size,
\begin{equation}
 f_{j+L/2}=2\pi-f_j,
 \qquad
 \sum_{j=0}^{L-1}f_j=L\pi.
\label{eq:S7-sampled-symmetry}
\end{equation}

\subsection{Determinant implementation}
\label{subsec:S7-determinants}

For an integer harmonic $q$, define
\begin{equation}
 a_j^{(q)}:=\ee^{\ii qf_j},
 \qquad
 \widehat a_L^{(q)}(m)
 :=\frac1L\sum_{j=0}^{L-1}
 a_j^{(q)}\ee^{-2\pi\ii mj/L}.
\label{eq:S7-symbol-DFT}
\end{equation}
With $N=L/2$, the occupied-orbital Toeplitz matrix of Eq.~\eqref{eq:S3-discrete-Toeplitz-matrix} is
\begin{equation}
 T_{rs}^{(L,q)}
 =\widehat a_L^{(q)}(r-s),
 \qquad
 0\leq r,s<N.
\label{eq:S7-Toeplitz-matrix}
\end{equation}
The exact finite-size characteristic function is
\begin{equation}
 \chi_{L,q}
 :=\left\langle
 \ee^{\ii qX_L(f_{\boldsymbol\theta})}
 \right\rangle
 =\ee^{-\frac{\ii q}{2}\sum_jf_j}
 \det T^{(L,q)}
 .
\label{eq:S7-characteristic}
\end{equation}
For integer $q$, this is simultaneously the Fourier coefficient of the compact partial-record law and the measured-site determinant in Eq.~\eqref{eq:S3-measured-determinant}.  We set the common momentum offset in the orbital matrix to zero; Sec.~\ref{sec:S3} proves that all determinants are invariant under the corresponding diagonal gauge transformation.

For $q\neq0$ and $\chi_{L,q}\neq0$, define
\begin{equation}
 h_{\rm eff}^{(q)}(L,r)
 :=-\frac1{q^2}\log|\chi_{L,q}|.
\label{eq:S7-heff}
\end{equation}
If $\chi_{L,q}=0$, we set $h_{\rm eff}^{(q)}=+\infty$.  Numerically, a logarithmic determinant returns
\begin{equation}
 \det T^{(L,q)}
 =\sigma_{L,q}\ee^{\lambda_{L,q}},
 \qquad
 |\sigma_{L,q}|=1,
\label{eq:S7-slogdet}
\end{equation}
for a nonsingular matrix.  Since the centering factor in Eq.~\eqref{eq:S7-characteristic} has unit modulus,
\begin{equation}
 \log|\chi_{L,q}|=\lambda_{L,q}.
\label{eq:S7-logabs}
\end{equation}
The discrete Fourier transform costs $O(L\log L)$, matrix assembly costs $O(L^2)$, and dense LU factorization costs $O(L^3)$ with $O(L^2)$ memory.  The characteristic functions are real and even at every finite size by Eq.~\eqref{eq:S3-real-even}; any residual imaginary part therefore directly measures floating-point error.

\subsection{Complete small-size verification}
\label{subsec:S7-small-size}

At $L=12$ and $r=1/6$, one has $a=2$, $N=6$, $|C_L|=8$, and
\begin{equation}
 \binom{12}{6}=924
\label{eq:S7-number-configurations}
\end{equation}
complete configurations.  We enumerate all of them with the exact Born weight
\begin{equation}
 p_L(X)
 =12^{-6}
 \prod_{\substack{x,y\in X\\x<y}}
 \left|
 \ee^{2\pi\ii x/12}-\ee^{2\pi\ii y/12}
 \right|^2.
\label{eq:S7-direct-probability}
\end{equation}
The complete-record sum is
\begin{equation}
 \chi_{12,q}^{\rm complete}
 =\sum_{|X|=6}p_L(X)
 \exp\!\left[
 \ii q\sum_{j=0}^{11}
 f_j\left(\mathbf1_X(j)-\frac12\right)
 \right].
\label{eq:S7-complete-enumeration}
\end{equation}
Independently, configurations are grouped by the record $m_C=\rho_{C,L}(X)$ to form
\begin{equation}
 \mathbb P_{12,C_L}(m_C)
 =\sum_{\substack{|X|=6\\\rho_{C,L}(X)=m_C}}p_L(X),
\label{eq:S7-enumerated-marginal}
\end{equation}
and the compact partial-record sum
\begin{equation}
 \chi_{12,q}^{\rm partial}
 =\sum_{m_C}
 \mathbb P_{12,C_L}(m_C)
 \mathfrak e_q\!\left(\delta_{12}(m_C)\right).
\label{eq:S7-partial-enumeration}
\end{equation}
These two sums are compared with the occupied-orbital determinant, the measured-site determinant, and the full $12\times12$ determinant from Sec.~\ref{sec:S3}.

The enumeration yields $238$ records with nonzero probability and
\begin{equation}
 \left|\sum_{|X|=6}p_L(X)-1\right|
 =3.1\times10^{-15}.
\label{eq:S7-enumeration-normalization}
\end{equation}
The five evaluations agree as shown in Table~\ref{tab:S7-small-check}.

\begin{table}[t]
\caption{Small-size verification at $L=12$, $r=1/6$.  The last column is the maximum absolute discrepancy among complete enumeration, partial-record enumeration, the occupied-orbital determinant, the measured-site determinant, and the full-space determinant.}
\label{tab:S7-small-check}
\begin{ruledtabular}
\begin{tabular}{ccc}
$q$ & common value of $\chi_{12,q}$ & maximum discrepancy \\
\hline
$1$ & $0.112455722032121$ & $6.0\times10^{-16}$ \\
$2$ & $0.0020638642968203$ & $1.2\times10^{-16}$
\end{tabular}
\end{ruledtabular}
\end{table}

This test verifies not only the Vandermonde/Toeplitz identity, but also the marginalization to partial records, the record-only compact phase, and the measured-site determinant.

\subsection{Finite-size approach to the conformal width}
\label{subsec:S7-convergence}

The theorem predicts
\begin{equation}
 h_{\rm eff}^{(q)}(L,r)\longrightarrow h(r)
 \qquad
 (L\to\infty)
\label{eq:S7-heff-limit}
\end{equation}
for every fixed nonzero integer $q$.  Table~\ref{tab:S7-L720} gives double-precision evaluations at $L=720$.  The final column is the spectral two-norm condition number of the occupied-orbital Toeplitz matrix.

\begin{table*}[t]
\caption{Finite-size determinant results from the exact conformal samples at $L=720$.  The relative deviation is $[h_{\rm eff}^{(q)}-h]/h$.}
\label{tab:S7-L720}
\resizebox{\textwidth}{!}{%
\begin{ruledtabular}
\begin{tabular}{ccccccc}
$r$ & $h(r)$ & $q$ & $|\chi_{720,q}|$ & $h_{\rm eff}^{(q)}$ & relative deviation & $\kappa_2(T^{(720,q)})$ \\
\hline
$1/10$ & $1.946055$ & $1$ & $1.433605\times10^{-1}$ & $1.942393$ & $-1.882\times10^{-3}$ & $2.22$ \\
       &            & $2$ & $4.224356\times10^{-4}$ & $1.942368$ & $-1.894\times10^{-3}$ & $12.5$ \\
       &            & $3$ & $2.594886\times10^{-8}$ & $1.940793$ & $-2.704\times10^{-3}$ & $83.9$ \\
$1/6$  & $2.455786$ & $1$ & $8.610638\times10^{-2}$ & $2.452172$ & $-1.472\times10^{-3}$ & $2.87$ \\
       &            & $2$ & $5.500112\times10^{-5}$ & $2.452039$ & $-1.526\times10^{-3}$ & $27.0$ \\
       &            & $3$ & $2.957665\times10^{-10}$& $2.437939$ & $-7.267\times10^{-3}$ & $286$ \\
$1/4$  & $3.141593$ & $1$ & $4.339739\times10^{-2}$ & $3.137356$ & $-1.349\times10^{-3}$ & $4.06$ \\
       &            & $2$ & $3.569092\times10^{-6}$ & $3.135800$ & $-1.844\times10^{-3}$ & $75.3$ \\
       &            & $3$ & $3.871962\times10^{-12}$& $2.919696$ & $-7.063\times10^{-2}$ & $680$ \\
$1/3$  & $4.018919$ & $1$ & $1.807934\times10^{-2}$ & $4.012986$ & $-1.476\times10^{-3}$ & $6.32$ \\
       &            & $2$ & $1.283558\times10^{-7}$ & $3.967115$ & $-1.289\times10^{-2}$ & $257$ \\
       &            & $3$ & $1.072065\times10^{-12}$& $3.062382$ & $-2.380\times10^{-1}$ & $508$
\end{tabular}
\end{ruledtabular}%
}
\end{table*}

For $q=1$, all four relative deviations at $L=720$ lie between $1.35\times10^{-3}$ and $1.88\times10^{-3}$ in magnitude.  The second harmonic behaves similarly for the first three geometries but converges more slowly for the widest rectangle.  The third harmonic is substantially more demanding because the limiting magnitude is $\ee^{-9h}$.  For example,
\begin{equation}
 \ee^{-9h(1/3)}=1.96\times10^{-16}\ldots,
 \qquad
 |\chi_{720,3}|=1.07\times10^{-12}.
\label{eq:S7-q3-scale}
\end{equation}
The corresponding Toeplitz matrix has condition number of order $5\times10^2$.  The data therefore display a large finite-size correction, but do not by themselves diagnose catastrophic numerical instability.  A small determinant is not synonymous with an inaccurate determinant; conditioning and higher-precision comparisons must be assessed separately.

\begin{figure}[t]
\centering
\includegraphics[width=0.52\textwidth]{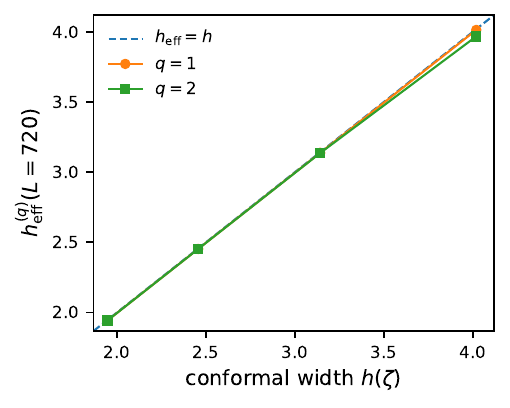}
\caption{Finite-size effective widths $h_{\mathrm{eff}}^{(q)}=-q^{-2}\log|\chi_L(q)|$ obtained from the exact determinant at $L=720$ for four fixed conformal geometries. The dashed line is the parameter-free prediction $h_{\mathrm{eff}}=h(\zeta)$.}
\label{fig:S7-numerics}
\end{figure}

Representative size dependence is shown in Table~\ref{tab:S7-size-convergence}.  The first harmonic converges regularly for every geometry.  The second harmonic also approaches $h(r)$, but the convergence slows as both $q$ and $h$ increase.  The $L=720$ comparison for the four geometries is shown in Fig.~\ref{fig:S7-numerics}.

\begin{table*}[t]
\caption{Finite-size effective widths obtained from exact conformal samples.}
\label{tab:S7-size-convergence}
\resizebox{\textwidth}{!}{%
\begin{ruledtabular}
\begin{tabular}{cccccccc}
$r$ & $h(r)$
& $h_{\rm eff}^{(1)}(120)$
& $h_{\rm eff}^{(1)}(360)$
& $h_{\rm eff}^{(1)}(720)$
& $h_{\rm eff}^{(2)}(120)$
& $h_{\rm eff}^{(2)}(360)$
& $h_{\rm eff}^{(2)}(720)$ \\
\hline
$1/10$ & $1.946055$ & $1.924192$ & $1.938744$ & $1.942393$ & $1.920601$ & $1.938543$ & $1.942368$ \\
$1/6$  & $2.455786$ & $2.434134$ & $2.448566$ & $2.452172$ & $2.421934$ & $2.447706$ & $2.452039$ \\
$1/4$  & $3.141593$ & $3.115932$ & $3.133118$ & $3.137356$ & $3.020669$ & $3.123969$ & $3.135800$ \\
$1/3$  & $4.018919$ & $3.981079$ & $4.006969$ & $4.012986$ & $3.347751$ & $3.819477$ & $3.967115$
\end{tabular}
\end{ruledtabular}%
}
\end{table*}

A useful magnitude diagnostic is
\begin{equation}
 R_{L,q}
 :=\log|\chi_{L,q}|-q^2\log|\chi_{L,1}|
 =-q^2\left[h_{\rm eff}^{(q)}-h_{\rm eff}^{(1)}\right].
\label{eq:S7-magnitude-residual}
\end{equation}
It vanishes for an exact Gaussian characteristic function but tests only magnitudes.  A stronger complex diagnostic is
\begin{equation}
 \Delta_{L,q}:=\chi_{L,q}-\chi_{L,1}^{\,q^2}.
\label{eq:S7-complex-residual}
\end{equation}
In the present problem $\chi_{L,q}$ is real by particle--hole symmetry, but Eq.~\eqref{eq:S7-complex-residual} remains the logically complete comparison.

\subsection{Uniform-measure control}
\label{subsec:S7-uniform-control}

To separate generic Gaussianity from the special geometric value of the variance, replace the conformal side measures of Eq.~\eqref{eq:S2-harmonic-measures} by normalized angular measure on the measured arcs.  Let
\begin{equation}
 \ell:=\pi-\alpha=\pi(1-2r).
\label{eq:S7-uniform-length}
\end{equation}
The resulting boundary coordinate is
\begin{equation}
 f_{\rm unif}(\theta)=
 \begin{cases}
 0,
 &0<\theta<\alpha,
 \\[1mm]
 \displaystyle
 \frac{2\pi}{\ell}(\theta-\alpha),
 &\alpha<\theta<\pi,
 \\[2mm]
 2\pi,
 &\pi<\theta<\pi+\alpha,
 \\[1mm]
 \displaystyle
 \frac{2\pi}{\ell}(2\pi-\theta),
 &\pi+\alpha<\theta<2\pi.
 \end{cases}
\label{eq:S7-uniform-function}
\end{equation}
For $m\neq0$, direct integration gives
\begin{equation}
 (f_{\rm unif})_m
 =-
 \frac{[1-(-1)^m][1+\ee^{-\ii m\alpha}]}
 {\ell m^2}.
\label{eq:S7-uniform-Fourier}
\end{equation}
Its exact continuum energy is therefore
\begin{equation}
 \mathcal E_{\rm unif}(r)
 =\frac{16}{\pi^2(1-2r)^2}
 \sum_{\substack{m\geq1\\m\ {
m odd}}}
 \frac{\cos^2(\pi rm)}{m^3}
 .
\label{eq:S7-uniform-energy}
\end{equation}
This value is listed in Table~\ref{tab:S7-geometries} and is different from $h(r)$.

\begin{table*}[t]
\caption{Uniform-measure control at $L=720$, $q=1$.  The finite-size determinant approaches $\mathcal E_{\rm unif}$ rather than the conformal width.}
\label{tab:S7-uniform-control}
\resizebox{\textwidth}{!}{%
\begin{ruledtabular}
\begin{tabular}{cccccc}
$r$ & $h(r)$ & $\mathcal E_{\rm unif}(r)$
& $h_{\rm eff,unif}^{(1)}(720)$
& relative error to $\mathcal E_{\rm unif}$
& relative shift from $h$ \\
\hline
$1/10$ & $1.946055$ & $2.332700$ & $2.332663$ & $-1.58\times10^{-5}$ & $0.1987$ \\
$1/6$  & $2.455786$ & $2.770810$ & $2.770760$ & $-1.79\times10^{-5}$ & $0.1283$ \\
$1/4$  & $3.141593$ & $3.410227$ & $3.410144$ & $-2.43\times10^{-5}$ & $0.0855$ \\
$1/3$  & $4.018919$ & $4.262784$ & $4.262611$ & $-4.07\times10^{-5}$ & $0.0606$
\end{tabular}
\end{ruledtabular}%
}
\end{table*}

The control demonstrates two distinct statements:
\begin{equation}
 \begin{aligned}
 &\text{XX linear statistics satisfying the regularity assumptions of Sec.~\ref{sec:S5} have Gaussian limits,}
 \\
 &\text{the conformal boundary coordinate has the special energy }h(r).
 \end{aligned}
\label{eq:S7-control-separation}
\end{equation}
The Toeplitz mechanism supplies Gaussianity, whereas the conformal Dirichlet-energy identity selects the variance $2h(r)$.

\section{Boundary/cylinder conventions, robustness, and extensions}
\label{sec:S8}

This section collects convention changes that frequently obscure comparisons between conformal calculations, states the precise sense in which the microscopic result is stable under finite-size perturbations, and separates direct consequences of the proof from natural extensions that remain to be established.  The central distinction is useful throughout: the decoder
\begin{equation}
 m_C\longmapsto \delta_L(m_C)
 \label{eq:S8-decoder-reminder}
\end{equation}
depends on the chosen coordinate and orientation of the microscopic circle, whereas the limiting law depends only on the conformal modulus of the marked quadrilateral.

\subsection{Cross-ratio and elliptic-integral conventions}
\label{subsec:S8-cross-ratio}

For the cyclic ordering used in the Letter and in Sec.~\ref{sec:S2},
\begin{equation}
 A=(\theta_1,\theta_2),\qquad
 C_1=(\theta_2,\theta_3),\qquad
 B=(\theta_3,\theta_4),\qquad
 C_2=(\theta_4,\theta_1+2\pi),
\label{eq:S8-arc-order}
\end{equation}
the cross ratio is
\begin{equation}
 \zeta
 =\frac{
 \sin\!\left(\frac{\theta_2-\theta_1}{2}\right)
 \sin\!\left(\frac{\theta_4-\theta_3}{2}\right)
 }{
 \sin\!\left(\frac{\theta_3-\theta_1}{2}\right)
 \sin\!\left(\frac{\theta_4-\theta_2}{2}\right)
 },
 \qquad 0<\zeta<1.
\label{eq:S8-our-cross-ratio}
\end{equation}
A one-step cyclic relabeling of the marked points,
\begin{equation}
 (\theta_1,\theta_2,\theta_3,\theta_4)
 \longmapsto
 (\theta_2,\theta_3,\theta_4,\theta_1+2\pi),
\label{eq:S8-cyclic-relabel}
\end{equation}
produces the complementary cross ratio
\begin{equation}
 \zeta_{\rm cyc}=1-\zeta.
\label{eq:S8-complementary-cross-ratio}
\end{equation}
This is the origin of many apparently reciprocal formulas in the literature: different authors begin the cyclic labeling on different physical arcs, or choose the measured rather than the unmeasured pair as the distinguished opposite sides.  More general permutations generate the standard six anharmonic values
\begin{equation}
 \zeta,
 \quad 1-\zeta,
 \quad \frac1\zeta,
 \quad \frac1{1-\zeta},
 \quad \frac{\zeta}{\zeta-1},
 \quad \frac{\zeta-1}{\zeta},
\label{eq:S8-six-cross-ratios}
\end{equation}
but only \(\zeta\) and \(1-\zeta\) lie in \((0,1)\) for the two cyclic choices relevant here.

Throughout this Supplement,
\begin{equation}
 K(m)=\int_0^{\pi/2}
 \frac{\dd\varphi}{\sqrt{1-m\sin^2\varphi}}
\label{eq:S8-parameter-K}
\end{equation}
is the complete elliptic integral in the \emph{parameter} convention.  If instead one uses the elliptic modulus \(k\), with
\begin{equation}
 K_{\rm mod}(k):=
 \int_0^{\pi/2}
 \frac{\dd\varphi}{\sqrt{1-k^2\sin^2\varphi}},
\label{eq:S8-modulus-K}
\end{equation}
then
\begin{equation}
 K_{\rm mod}(k)=K(k^2).
\label{eq:S8-modulus-parameter}
\end{equation}
The rectangle width can therefore be written equivalently as
\begin{equation}
 h(\zeta)
 =\pi\frac{K(\zeta)}{K(1-\zeta)}
 =\pi\frac{K_{\rm mod}(\sqrt\zeta)}
 {K_{\rm mod}(\sqrt{1-\zeta})}
 .
\label{eq:S8-width-conventions}
\end{equation}
With the complementary cross ratio \(\bar\zeta=1-\zeta\), the same expression becomes
\begin{equation}
 h
 =\pi\frac{K(1-\bar\zeta)}{K(\bar\zeta)}.
\label{eq:S8-width-complement}
\end{equation}
The complementary modulus obeys the exact relation
\begin{equation}
 h(\zeta)h(1-\zeta)=\pi^2
 .
\label{eq:S8-modulus-duality}
\end{equation}
Geometrically, replacing \(\zeta\) by \(1-\zeta\) exchanges the two pairs of opposite sides and rotates the conformal rectangle.  After the transverse side is renormalized to length \(\pi\), the width is consequently inverted as \(h\mapsto\pi^2/h\).  This is a change of which physical channel is designated longitudinal, not a discrepancy between conformal maps.

\subsection{Rectangle, doubled cylinder, and compact-phase normalization}
\label{subsec:S8-cylinder}

The single-sheet quadrilateral is mapped in Sec.~\ref{sec:S2} to
\begin{equation}
 \mathcal R_h
 =\{W=u+\ii v:0<u<h,\ 0<v<\pi\}.
\label{eq:S8-rectangle}
\end{equation}
Doubling this rectangle across its two horizontal sides produces a cylinder of longitudinal length \(h\) and circumference \(2\pi\),
\begin{equation}
 \widehat{\mathcal R}_h
 =\{0<u<h,\ 0\leq\widetilde v<2\pi\},
 \qquad
 \widetilde v\sim\widetilde v+2\pi.
\label{eq:S8-doubled-cylinder}
\end{equation}
Thus the factor of two between a rectangle of height \(\pi\) and a cylinder of circumference \(2\pi\) is simply the usual doubling construction employed in boundary CFT and entanglement-Hamiltonian calculations~\cite{Rajabpour2016,CardyTonni2016,KhannaVasseur2026}.  The longitudinal scale is unchanged: the cylinder length is the same \(h\) that appears in the compact heat kernel.

The normalization of the microscopic boundary coordinate is a separate issue.  On the original rectangle, \(v\) changes by \(\pi\) between the horizontal sides.  We define
\begin{equation}
 f_{\boldsymbol\theta}=2v
\label{eq:S8-f-normalization}
\end{equation}
so that
\begin{equation}
 f_{\boldsymbol\theta}|_A=0,
 \qquad
 f_{\boldsymbol\theta}|_B=2\pi.
\label{eq:S8-f-boundary-values}
\end{equation}
This normalization is fixed by the microscopic compactification: the height increment is \(2\pi(n_j-1/2)\), and the contribution of every occupation in the unmeasured arc \(B\) is therefore an integer multiple of one full period.  It is precisely Eq.~\eqref{eq:S8-f-boundary-values} that makes the class \([X_L]_{2\pi}\) record-only at finite size.

If one merely rescales the same compact coordinate to period \(2\pi R\), one uses
\begin{equation}
 f^{(R)}_{\boldsymbol\theta}=R f_{\boldsymbol\theta},
 \qquad
 \delta^{(R)}\in\bbR/(2\pi R\bbZ),
\label{eq:S8-radius-rescaling}
\end{equation}
and the intrinsic characters are \(\exp(\ii q\delta^{(R)}/R)\), \(q\in\bbZ\).  This is only a reparametrization of the same compact variable; changing the physical compactification radius or Luttinger stiffness of a different theory is not exhausted by this rescaling.  Our choice corresponds to \(R=1\).  The continuum relative Dirichlet or fermionic boundary phase is matched by the scaling-limit law and integer characters of \(\delta_L\), up to the orientation convention; the sign reversal leaves the limiting law invariant.

\subsection{A quantitative robustness criterion}
\label{subsec:S8-robustness}

The exact variance formula of Sec.~\ref{sec:S4} provides a natural finite-size seminorm.  For a real lattice vector \(u=(u_0,\ldots,u_{L-1})\), define
\begin{equation}
 \|u\|_{L,1/2}^2
 :=\sum_{m=1}^{L-1}
 d_L(m)|\widehat u_L(m)|^2,
 \qquad
 d_L(m)=\min(m,L-m).
\label{eq:S8-discrete-Hhalf}
\end{equation}
Then
\begin{equation}
 \|u\|_{L,1/2}^2
 =\operatorname{Var}X_L(u).
\label{eq:S8-norm-variance}
\end{equation}
For any two real lattice functions \(u_L\) and \(v_L\), Eq.~\eqref{eq:S4-characteristic-stability} gives
\begin{equation}
 \left|
 \left\langle\ee^{\ii tX_L(u_L)}\right\rangle
 -
 \left\langle\ee^{\ii tX_L(v_L)}\right\rangle
 \right|
 \leq
 |t|\,\|u_L-v_L\|_{L,1/2}.
\label{eq:S8-robust-characteristic}
\end{equation}
Consequently,
\begin{equation}
 \|u_L-v_L\|_{L,1/2}\longrightarrow0
 \quad\Longrightarrow\quad
 X_L(u_L)\text{ and }X_L(v_L)
 \text{ have the same limiting characteristic function}
 .
\label{eq:S8-robustness-criterion}
\end{equation}
For integer harmonics, the same statement applies to the compact laws whenever the two statistics admit compatible record-only reductions and those reductions agree modulo \(2\pi\).

Equation~\eqref{eq:S8-robustness-criterion} is the precise test for alternative discretizations, numerical approximations, weak smoothing of the corners, or an \(L\)-dependent family of endpoint positions.  Pointwise convergence alone is not sufficient; the relevant notion is convergence in the discrete \(H^{1/2}\) variance norm.  The exact conformal samples used in the theorem and in Sec.~\ref{sec:S7} avoid this issue entirely.

For the main scaling limit, the continuum endpoints are fixed and only the midpoint partition changes with \(L\).  Choosing the nearest admissible lattice bonds therefore does not move the conformal function itself.  If instead one recomputes the conformal map for endpoints \(\boldsymbol\theta^{(L)}\), the result remains valid provided
\begin{equation}
 \left\|
 f_{\boldsymbol\theta^{(L)},L}
 -f_{\boldsymbol\theta,L}
 \right\|_{L,1/2}
 \longrightarrow0.
\label{eq:S8-moving-endpoint-condition}
\end{equation}
For endpoint quadruples confined to a compact nondegenerate subset of cyclic configurations, this is the natural continuity requirement in the Dirichlet-energy topology; the statement above applies whenever the corresponding continuity estimate holds.  When endpoint separations vanish with \(L\), uniformity is no longer automatic and a different crossover problem emerges.

\subsection{Degenerate geometries and crossover regimes}
\label{subsec:S8-degenerate}

The theorem assumes a fixed cross ratio \(0<\zeta<1\).  The standard elliptic asymptotics give
\begin{align}
 K(\zeta)&=\frac\pi2+O(\zeta),
 &
 K(1-\zeta)&=\frac12\log\frac{16}{\zeta}
 +O\!\left(\zeta\log\frac1\zeta\right)
\label{eq:S8-K-small-zeta}
\end{align}
as \(\zeta\downarrow0\), and therefore
\begin{equation}
 h(\zeta)
 =\frac{\pi^2}{\log(16/\zeta)}[1+o(1)].
\label{eq:S8-h-small-zeta}
\end{equation}
At the complementary degeneration, \(\varepsilon=1-\zeta\downarrow0\),
\begin{equation}
 h(1-\varepsilon)
 =\log\frac{16}{\varepsilon}+o(1).
\label{eq:S8-h-large-zeta}
\end{equation}
Thus the compact law approaches
\begin{equation}
 P_{h(\zeta)}\Longrightarrow
 \begin{cases}
 \delta^{\mathrm D}_{[0]_{2\pi}},&\zeta\downarrow0,\\[1mm]
 \dd\delta/(2\pi),&\zeta\uparrow1.
 \end{cases}
\label{eq:S8-degenerate-laws}
\end{equation}
These are respectively the localized and completely delocalized limits of the relative compact phase.

If \(\zeta=\zeta_L\) approaches an endpoint simultaneously with \(L\to\infty\), the order of limits matters.  The Fourier cutoff and the corner constants used in Sec.~\ref{sec:S5} are then no longer uniform.  When a measured interval shrinks to a microscopic number of sites, the smooth finite-energy Szeg\H{o} problem should cross over to sharp-counting Fisher--Hartwig asymptotics~\cite{BasorTracy1991,AbanovIvanov2011}.  Determining the universal double-scaling kernels in this regime is a separate problem and is not part of the theorem proved here.

\subsection{Several measured components and a compact Gaussian vector}
\label{subsec:S8-multiple-modes}

The scalar result suggests a direct multicomponent extension.  Suppose a geometry admits \(d\) independent real boundary coordinates \(f_1,\ldots,f_d\), each constant in integer multiples of \(2\pi\) on every unmeasured component.  Define
\begin{equation}
 \delta_{L,a}
 :=[X_L(f_a)]_{2\pi},
 \qquad a=1,\ldots,d.
\label{eq:S8-vector-variable}
\end{equation}
The same finite-size argument as in Sec.~\ref{sec:S2} makes the vector
\begin{equation}
 \boldsymbol\delta_L
 =(\delta_{L,1},\ldots,\delta_{L,d})
 \in\bbT^d
\label{eq:S8-vector-torus}
\end{equation}
record-only.  For \(\boldsymbol q\in\bbZ^d\), the free-fermion determinant contains the single symbol
\begin{equation}
 \exp\!\left(\ii\sum_{a=1}^dq_af_a\right).
\label{eq:S8-vector-symbol}
\end{equation}
The strong Szeg\H{o} exponent is governed by the symmetric matrix
\begin{align}
 \mathsf H_{ab}
 &:=\sum_{m=1}^{\infty}m\,
 \operatorname{Re}\!\left[(f_a)_m\overline{(f_b)_m}\right]
\notag\\
 &=\frac{1}{4\pi}
 \int_{\mathbb D}
 \nabla H_{f_a}\cdot\nabla H_{f_b}\,\dd^2z.
\label{eq:S8-capacitance-matrix}
\end{align}
Formally, and rigorously under the same regularity assumptions used in Sec.~\ref{sec:S5}, one obtains
\begin{equation}
 \left\langle
 \exp\!\left(\ii\boldsymbol q\cdot\boldsymbol\delta_L\right)
 \right\rangle
 \longrightarrow
 \exp\!\left(-\boldsymbol q^{\mathsf T}
 \mathsf H\boldsymbol q\right).
\label{eq:S8-vector-characteristic}
\end{equation}
If \(\mathsf H\) is positive definite, the corresponding density on \(\bbT^d\) is the periodized Gaussian
\begin{equation}
 P_{\mathsf H}(\boldsymbol\delta)
 =\frac{1}{(4\pi)^{d/2}\sqrt{\det\mathsf H}}
 \sum_{\boldsymbol n\in\bbZ^d}
 \exp\!\left[
 -\frac14
 (\boldsymbol\delta+2\pi\boldsymbol n)^{\mathsf T}
 \mathsf H^{-1}
 (\boldsymbol\delta+2\pi\boldsymbol n)
 \right]
 .
\label{eq:S8-vector-wrapped-Gaussian}
\end{equation}
The matrix \(\mathsf H\) plays the role of a Dirichlet capacitance matrix for these boundary coordinates.  In multiply connected geometries one expects a relation to the corresponding period matrix.  Establishing that relation, together with the appropriate microscopic decoder for arbitrary collections of measured intervals, is a natural continuation of the present work; the determinant and Gaussian parts of the argument are already in place.

\subsection{Interacting liquids and other measurement bases}
\label{subsec:S8-interacting-bases}

The XX chain realizes the free-fermion point of a compact-boson theory.  In an interacting Tomonaga--Luttinger liquid, continuum calculations predict Born averaging over compact conformal boundary values, with the zero-mode action controlled by the Luttinger stiffness and compactification radius~\cite{KhannaVasseur2026,KhannaVasseurStats2026}.  The expected heat-kernel time is therefore rescaled by the appropriate stiffness factor.  Its exact numerical coefficient depends on the normalization of the bosonic field and of the compact phase.

What does not follow from the present proof is an outcome-level interacting decoder.  Away from the free point there is no projection determinant, and it is not known whether a simple linear functional of the microscopic record remains exact at finite size.  The natural conjecture is that the weighted height difference of Sec.~\ref{sec:S2} flows to the same relative Dirichlet mode, with interaction-dependent renormalization.  Demonstrating this statement requires either an interacting lattice argument or controlled numerical scaling and is beyond the scope of this Letter.

Occupation measurements select a density field and hence a Dirichlet-type compact coordinate in the present convention.  Measurements in other local bases may instead couple to the dual boson or to boundary-condition-changing operators.  Depending on the basis and preserved symmetry, one may expect Neumann, mixed, symmetry-twisted, or more general conformal boundary conditions~\cite{Cardy1984,Cardy1989,DiGiulio2023}.  The microscopic question is then to identify which record functionals become the corresponding boundary zero modes.  The exact logic exposed here suggests a useful diagnostic: search for a compact statistic whose unmeasured contribution is quantized, derive its finite-size generating function, and evaluate its continuum quadratic form by conformal energy.

The result proved in this work is therefore both specific and structural.  It is specific because the exact decoder and determinant representation rely on the half-filled XX chain.  It is structural because the final chain
\begin{equation}
 \text{record-only compactification}
 \ \longrightarrow\ 
 \text{Gaussian determinant asymptotics}
 \ \longrightarrow\ 
 \text{conformal Dirichlet energy}
\label{eq:S8-structural-chain}
\end{equation}
identifies the three ingredients required for analogous microscopic realizations in broader critical systems.

\section{Matching to continuum boundary ensembles and charge sectors}
\label{sec:S9}

The preceding sections establish the compact variable and its probability law directly from the XX-chain Born measure.  This final section matches those microscopic statements to the continuum language used in boundary conformal field theory and in entanglement-Hamiltonian descriptions.  The purpose is not to claim that the heat-kernel form was absent from continuum theory.  Rather, it is to identify precisely which continuum zero-mode law and charge characters are matched by the record-dependent variable of Sec.~\ref{sec:S2}, to reconcile the common cylinder and elliptic-integral conventions, and to separate the previously known effective boundary ensemble from the outcome-level lattice theorem proved here.

\subsection{Relative compact phase on the doubled cylinder}
\label{subsec:S9-cylinder-phase}

Doubling the normalized rectangle
\begin{equation}
 \mathcal R_h=\{0<u<h,\ 0<v<\pi\}
\label{eq:S9-rectangle}
\end{equation}
across either horizontal side produces a cylinder of height $h$ and circumference $2\pi$, with $v\sim v+2\pi$.  In a free Dirac or compact-boson description, conformal boundary conditions at $u=0$ and $u=h$ carry compact phases $\alpha_1$ and $\alpha_2$.  Only their difference~\cite{KhannaVasseur2026,KhannaVasseurStats2026}
\begin{equation}
 \delta:=[\alpha_2-\alpha_1]_{2\pi}\in\bbT
\label{eq:S9-relative-phase}
\end{equation}
is invariant under a common boundary shift.  This is the continuum relative zero mode whose probability law and integer characters are matched by the scaling limit of the microscopic variable $\delta_L$.  In the compact-boson convention $S=(g/4\pi)\int(\partial\phi)^2$, the XX point corresponds to $g=1/2$; with the dimensionless phase normalized to period $2\pi$, this gives the zero-mode normalization used below~\cite{KhannaVasseur2026,KhannaVasseurStats2026}.

The zero-mode dependence can be written in a form that makes the comparison exact~\cite{KhannaVasseur2026,KhannaVasseurStats2026}.  Let
\begin{equation}
 \nu:=\frac{\delta}{2\pi},
 \qquad
 \tau:=\frac{\ii\pi}{h}.
\label{eq:S9-nu-tau}
\end{equation}
Using the convention
\begin{equation}
 \vartheta_3(z\mid\tau)
 :=\sum_{n\in\bbZ}
 \exp\!\left(\ii\pi\tau n^2+2\pi\ii n z\right),
\label{eq:S9-theta-definition}
\end{equation}
one has the identity
\begin{align}
 \mathcal Z_{0}(\delta\mid h)
 &:=\exp\!\left(\ii\pi\tau\nu^2\right)
 \vartheta_3(\tau\nu\mid\tau)
 \notag\\
 &=\sum_{n\in\bbZ}
 \exp\!\left[\ii\pi\tau(n+\nu)^2\right]
 \notag\\
 &=\sum_{n\in\bbZ}
 \exp\!\left[-\frac{(\delta+2\pi n)^2}{4h}\right].
\label{eq:S9-zero-mode-partition}
\end{align}
The oscillator determinant is independent of $\delta$ and multiplies Eq.~\eqref{eq:S9-zero-mode-partition} by a common factor.  After normalization over one compact period,
\begin{equation}
 \int_0^{2\pi}\mathcal Z_0(\delta\mid h)\,\dd\delta
 =\sqrt{4\pi h},
\label{eq:S9-zero-mode-normalization}
\end{equation}
Eq.~\eqref{eq:S9-zero-mode-partition} becomes exactly the density $P_h$ of Eq.~\eqref{eq:S6-wrapped-Gaussian}.  Thus the continuum compact-boundary ensemble and the limiting microscopic law are the same normalized zero-mode object~\cite{KhannaVasseur2026,KhannaVasseurStats2026},
\begin{equation}
 P_h(\delta)
 =\frac{\mathcal Z_0(\delta\mid h)}{\sqrt{4\pi h}}
 .
\label{eq:S9-Ph-Z0}
\end{equation}
Poisson summation gives the dual charge-sector representation
\begin{equation}
 P_h(\delta)
 =\frac{1}{2\pi}
 \sum_{q\in\bbZ}\ee^{-hq^2}\ee^{-\ii q\delta},
\label{eq:S9-dual-charge-expansion}
\end{equation}
which is the Fourier form already obtained microscopically in Sec.~\ref{sec:S6}.  The integer $q$ is simultaneously a character of the compact phase and the quantum number dual to the Dirichlet zero mode.

\subsection{Twist, conserved charge, and effective chemical potential}
\label{subsec:S9-charge-chemical-potential}

The charge interpretation is particularly transparent for a massless Dirac fermion on the doubled cylinder.  Boundary gluing conditions may be written schematically as
\begin{equation}
 \psi_-(0,v)=\ee^{\ii\alpha_1}\psi_+(0,v),
 \qquad
 \psi_-(h,v)=-\ee^{\ii\alpha_2}\psi_+(h,v),
\label{eq:S9-boundary-gluing}
\end{equation}
where the displayed minus sign fixes the Neveu--Schwarz convention.  After the image doubling, the relative phase $\delta=\alpha_2-\alpha_1$ becomes a twist around the spatial cycle.  The allowed chiral momenta are shifted to
\begin{equation}
 p_k=\frac{\pi}{h}\left(k+\frac12-\nu\right),
 \qquad k\in\bbZ,
 \qquad \nu=\frac{\delta}{2\pi}.
\label{eq:S9-twisted-momenta}
\end{equation}
The corresponding partition function has the standard form~\cite{KhannaVasseur2026,KhannaVasseurStats2026}
\begin{equation}
 Z_{\delta}(\tau)
 =\frac{\vartheta_3(\tau\nu\mid\tau)}{\eta(\tau)}
 \exp\!\left(\ii\pi\tau\nu^2\right),
\label{eq:S9-twisted-partition}
\end{equation}
where $\eta$ is the Dedekind eta function.  A grand-canonical Dirac partition function with a homogeneous chemical potential $\mu$ has the same theta-function dependence after the replacement
\begin{equation}
 \nu=\frac{h\mu}{\pi}.
\label{eq:S9-mu-matching-condition}
\end{equation}
Therefore the twist is thermodynamically represented by
\begin{equation}
 \mu_0=\frac{\delta}{2h}
 .
\label{eq:S9-effective-mu}
\end{equation}
The full $\delta$-dependent combination $\exp(\ii\pi\tau\nu^2)\vartheta_3(\tau\nu\mid\tau)$ in Eq.~\eqref{eq:S9-twisted-partition} is precisely the compact zero-mode factor $\mathcal Z_0(\delta\mid h)$ of Eq.~\eqref{eq:S9-zero-mode-partition}; the factor $\eta(\tau)^{-1}$ is the $\delta$-independent oscillator contribution.  Thus the compact Born weight is supplied by the normalized full zero-mode factor, not by the Gaussian prefactor alone.  At the same time, the phase appears as the charge-sector twist, with the twist/chemical-potential matching in Eqs.~\eqref{eq:S9-mu-matching-condition} and \eqref{eq:S9-effective-mu} derived directly for the measurement-induced entanglement Hamiltonian in Ref.~\cite{EislerTonni2026}.  This distinction is consistent with the general CFT structure of local entanglement Hamiltonians and their lattice continuum limits~\cite{CardyTonni2016,DalmonteEHReview2022,EislerPeschel2017,EislerTonniPeschel2019,EislerTonniPeschel2022}.

Equation~\eqref{eq:S9-effective-mu} also explains why the compact phase is physically consequential.  It is not merely a coordinate used to label conformal boundary conditions: it is conjugate to the conserved $U(1)$ charge and organizes the charge sector seen by the post-measurement reduced density matrix.  At the same time, the zero mode cannot encode all microscopic information in a generic outcome.  Outcome-dependent nonzero modes may modify local charge profiles or other details while leaving the compact relative phase unchanged.  The theorem of this work isolates the universal record-dependent coordinate, not a complete continuum representation of every bit string.

\subsection{Exact microscopic decoder versus an effective continuum integration variable}
\label{subsec:S9-decoder-comparison}

Continuum Born-averaged boundary CFT introduces $\delta$ as an integration variable and predicts the normalized weight $P_h(\delta)$~\cite{KhannaVasseur2026,KhannaVasseurStats2026}.  The lattice construction supplies the missing outcome-level map.  For every measured record,
\begin{equation}
 \delta_L(m_C)
 =\left[
 \sum_{j\in C_L}f_{\boldsymbol\theta,L}(j)
 \left(m_C(j)-\frac12\right)
 -\pi|B_L|
 \right]_{2\pi},
\label{eq:S9-microscopic-decoder}
\end{equation}
and for every compatible complete configuration $X$,
\begin{equation}
 [X_L(X)]_{2\pi}
 =\delta_L\!\left(\rho_{C,L}(X)\right).
\label{eq:S9-decoder-exactness}
\end{equation}
Equations~\eqref{eq:S9-microscopic-decoder} and \eqref{eq:S9-decoder-exactness} hold at finite $L$ and require no coarse graining.  Compatible completions change the particular complete-configuration statistic $X_L$ only through the term $2\pi N_B(X)$; the representative $r_L(m_C)$ and the compact class are fixed by the partial record.

The complete comparison is summarized in Table~\ref{tab:S9-continuum-lattice}.  The central advance is therefore not the first appearance of a Gaussian compact-boundary weight.  It is the simultaneous construction of the microscopic random variable, the proof that it is measurable from a partial record, and the derivation of its continuum law from exact Born probabilities.

\begin{table}[t]
\caption{Continuum and microscopic descriptions of the same compact zero mode.}
\label{tab:S9-continuum-lattice}
\begin{ruledtabular}
\begin{tabular}{p{0.28\linewidth}p{0.31\linewidth}p{0.31\linewidth}}
Object & Continuum description & Microscopic statement proved here \\
\hline
Compact coordinate
& Relative Dirichlet or fermionic boundary phase $\delta\in\bbT$
& Exact decoder $m_C\mapsto\delta_L(m_C)$ in Eq.~\eqref{eq:S9-microscopic-decoder}, whose scaling-limit law and integer characters match the continuum relative phase \\
Geometry
& Cylinder/rectangle modulus $h$
& $h=\mathcal E[f_{\boldsymbol\theta}]$ from the exact conformal Dirichlet energy \\
Probability
& Zero-mode heat kernel $P_h(\delta)$
& Pushforward of the exact XX Born measure converging weakly to $P_h\,\dd\delta$ \\
Dual sector
& Integer charge/winding label $q$
& Record-accessible character $\mathfrak e_q(\delta_L)$ with moment $\ee^{-hq^2}$ \\
Finite size
& Effective long-distance boundary ensemble
& Exact full-, occupied-, and measured-space determinants for every $L$ \\
\end{tabular}
\end{ruledtabular}
\end{table}

The logical relation between continuum expectation and lattice theorem can be written compactly as
\begin{equation}
 \underbrace{\int_0^{2\pi}P_h(\delta)F(\delta)\,\dd\delta}_{\text{continuum boundary ensemble}}
 \quad\longleftarrow\quad
 \underbrace{\sum_{m_C}\mathbb P_{L,C_L}(m_C)
 F\!\left(\delta_L(m_C)\right)}_{\text{microscopic Born average}},
\label{eq:S9-observable-correspondence}
\end{equation}
where the arrow denotes the $L\to\infty$ limit for every continuous function $F$ on $\bbT$.  This follows directly from Eq.~\eqref{eq:S6-compact-weak-form}.  It applies to any continuum quantity that depends only and continuously on the relative zero mode.  If an observable also depends on outcome-specific nonzero modes, Eq.~\eqref{eq:S9-observable-correspondence} controls only its zero-mode component.

\end{document}